\documentclass[eqsecnum]{revtex4}
\usepackage{caption, subfigure,tocvsec2}
\usepackage{amsmath,amssymb}
\usepackage[pdftex]{graphicx}
\usepackage{epstopdf}
\usepackage{hyperref} 

\renewcommand{\subsubsection}[1]{{\bigskip\centering\it #1 \par\medskip}}
\newcommand{\be}{\begin{equation}}
\newcommand{\ee}{\end{equation}}
\newcommand{\bea}{\begin{eqnarray}}
\newcommand{\eea}{\end{eqnarray}}

\newcommand{\bp}{\ensuremath{\mathbf{p}}}

\newcommand{\br}{\ensuremath{\mathbf{r}}}

\def\bbn{$\beta\beta$-0$\nu$}
\def\bb{$\beta\beta$-2$\nu$}
\def\Qbb{$Q_{\beta\beta}$}

\def\Te{$^{130}$Te}

\def\Cd{$^{116}$Cd}

\def\Se{$^{82}$Se}
\def\Mo{$^{100}$Mo}

\def\teo{TeO$_2$}

\def\mee{$m_{ee}$}

\begin{document}

\title{{\Large The Future of Neutrino Mass Measurements: Terrestrial, Astrophysical, and Cosmological Measurements in the Next Decade} \\   
\medskip

Highlights of the $\nu$Mass 2013 Workshop \\
\smallskip
\small{Milano, Italy, February 4$^{th}$ -- 7$^{th}$ , 2013}

}

\author{   
G.J. Barker$^{(a)}$
M.~Biassoni$^{(b,c)}$
A. De Rujula$^{(d,e)}$ 
H.~J.~De Vega$^{(f)}$
J. W. Engle$^{(g)}$
M.~Faverzani$^{(b,c,*)}$
E.~Ferri$^{(b,c,*)}$
J.~Formaggio$^{(h)}$
L. Gastaldo$^{(i)}$
F.~Gatti$^{(l,m)}$
P.~Gorla$^{(n)}$
U. K\"oster$^{(o)}$
S.~Lahiri$^{(p)}$
M. Lusignoli$^{(q,r)}$
A. Nucciotti$^{(b,c,*)}$,    
T.~Ota$^{(s)}$
M.~Sisti$^{(b,c)}$
M.~Sorel$^{(t)}$
F. Terranova$^{(b,c,*)}$
F. Vissani$^{(n)}$
N. Wandkowsky$^{(u)}$
M. Yoshimura$^{(v)}$
}

\affiliation{ 
\medskip
$^{(a)}$Dept. of Physics, University of Warwick, Coventry, UK\\
$^{(b)}$Dipartimento di Fisica ``G. Occhialini'', Universit\`a di
  Milano-Bicocca, Milano, Italy\\
$^{(c)}$Istituto Nazionale di Fisica Nucleare, Sezione di
  Milano-Bicocca, Milano, Italy\\
$^{(d)}$CERN, Geneva, Switzerland\\
$^{(e)}$Instituto de F\'isica Te\'orica, CSIC and Univ. Aut\'onoma de Madrid, Madrid, Spain\\
$^{(f)}$LPTHE, Univers\'ite Pierre et Marie Curie (Paris VI) and Observatoire de
Paris, LERMA. Paris, France\\
$^{(g)}$Los Alamos National Laboratory, Los Alamos, NM, USA\\
$^{(h)}$Massachusetts Institute of Technology, Cambridge, MA, USA\\
$^{(i)}$Kirchhoff Institute for Physics, Heidelberg University, Germany\\
$^{(l)}$Dip. di Fisica, Universit\`a di Genova, Genova, Italy \\
$^{(m)}$INFN Sezione di Genova, Genova, Italy\\
$^{(n)}$Laboratori Nazionali del Gran Sasso dell'INFN, Assergi (AQ), Italy\\
$^{(o)}$Institut Laue Langevin, Grenoble, France\\
$^{(p)}$Chemical Sciences Division, Saha Institute of Nuclear Physics, Kolkata, India\\
$^{(q)}$Dip. di Fisica, Universit\`a di Roma ``La Sapienza'', Roma, Italy\\
$^{(r)}$ INFN, Sezione di Roma, Roma, Italy\\
$^{(s)}$Department of Physics, Saitama University, Saitama-Sakura, Japan\\
$^{(t)}$Instituto de F\'isica Corpuscular (IFIC), CSIC and Univ. de Valencia, Valencia, Spain\\
$^{(u)}$Institute for Nuclear Physics, Karlsruhe Institute of Technology, Germany\\
$^{(v)}$Okayama University, Okayama, Japan\\
$^{(*)}$ Editors
}

\maketitle
\newpage

\setcounter{secnumdepth}{3}
\setcounter{tocdepth}{0}
\tableofcontents
\newpage\section{Foreword} 

In the last decade, experimental neutrino physics has changed
dramatically. Precision measurements of neutrino oscillations show
unambiguously that neutrinos are massive particles and all leptonic
mixing angles - including $\theta_{13}$ - are significantly larger
than their quark counterparts.  This new information has profound
implications in particle physics, cosmology and astrophysics and will
drive the design of neutrino experiments for the decade to come. In
fact, oscillation data set the scale of sensitivity to be matched by
novel experiments in order to address neutrino mass issues with a
reasonable chance of success. These issues are firstly the measurement
of the absolute mass of neutrinos and the determination of its
Dirac/Majorana nature.

The $\nu$Mass 2013 Workshop on ``The Future of Neutrino Mass
Measurements: Terrestrial, Astrophysical, and Cosmological
Measurements in the Next Decade'' was mostly focused on this
challenge: re-consider the experimental strategies for absolute mass
measurements in the light of the oscillation data and the sought-for
precisions both in the normal and inverted hierarchy scenario. Such
re-consideration is entangled with the determination of the mass
hierarchy with terrestrial and astrophysics experiments and the
clarification of the anomalies pointing to additional
sterile states. Hence, together with the classical sections of the
$\nu$Mass Series on direct measurements and neutrinoless double beta
decay, this Workshop edition also hosted dedicated sessions on sterile
neutrinos and the neutrino mass pattern.

This paper collects most of the contributions presented in the
Workshop.  It witnesses a fresh-looking and interdisciplinary field of
research, gathering nuclear and particle physicists,
experts of observational cosmology and model builders.
We take the opportunity of this Foreword to recall the lively discussions and
the enthusiasm of the speakers on the new ideas 
that will shape our field in the years to come. 

Finally, we are greatly indebted with the Department of Physics of the
University of Milano Bicocca and with INFN for hospitality and
support. A special thank goes to Dr. Marco Faverzani and Dr. Elena
Ferri for their invaluable help during the Workshop
and the editing of these proceedings.
 
\vspace{1cm} 
\hspace{8cm} 
\noindent
Angelo Nucciotti and Francesco Terranova\\

\newpage\section{Workshop full program}
\begin{center}
\begin{tabular}{p{0.2\textwidth}p{0.6\textwidth}} 
{\bf Author} & {\bf Title}\\
\hline
G. Senjanovic & Neutrino mass models\\
E. Lisi & Neutrino masses from oscillation experiments\\
W. Rodejohann & Neutrinoless Double Beta Decay and Neutrino Mass\\
A. Melchiorri & Constraints on Neutrino Physics from Cosmology\\
H. De Vega & Galaxy and cosmological motivations to the search of warm dark matter sterile neutrinos\\
C. Yang & Perspectives for the measurement of mass hierarchy with the Daya Bay II experiment\\
G. Barker & Establishing the neutrino mass hierarchy at accelerators\\
Th. Lasserre & Sterile neutrinos\\
E. Figueroa-Feliciano & Ricochet: Coherent Neutrino Scattering with Cryogenic Athermal Detectors\\
A. De Rujula & Progress towards measurements of the neutrino mass in single and double electron-capture beta-decays \\
Ch. Weinheimer & Direct neutrino mass measurements\\
N. Wandkowsky & KATRIN - Status of spectrometer commissioning\\
L. Gastaldo & The ECHO Experiment\\
F. Gatti & MARE Experiment\\
J. Formaggio & Project 8\\
M. Yoshimura & Neutrino mass spectroscopy using atoms and molecules\\
T. Ota & Collider-testable neutrino mass generation mechanisms\\
J. Wilkerson & Neutrinoless Double Beta Decay and Neutrino Mass\\
P. Gorla & First CUORE-0 measurement on the way to CUORE\\
M. Marino & EXO\\
C. Cattadori & Results and status report of the GERDA experiment at LNGS\\
F. Avignone & Excited State Search for Neutrinoless Double Beta Decay with CUORE \\
M. Sisti & The future of neutrinoless double beta decay searches with thermal detectors \\
M. Sorel & The NEXT Experiment\\
P. Ranitzsch & Calorimetric measurement of the 163Ho electron   capture spectrum\\
M. Faverzani & Status of the MARE-1 experiment\\
J. Fowler & Large Microcalorimeter Arrays for Beta Decay Spectroscopy\\
U. Koester & Production and separation of 163Ho\\
L. Lahiri & Alternative Production routes and new separation methods for no-carrier-added 163-Ho\\
M. Rabin & Chemistry and materials issues for embedding radioisotopes into low-temperature microcalorimeters\\
J. Engle & Comparison of methods for 163Ho production for neutrino mass measurement\\
P. De Bernardis & Neutrino mass from cosmological observations\\
F. Vissani & Neutrino properties from Supernovae\\
M. Biassoni & The CUORE experiment potential as a supernova neutrinos observatory\\
D. Fargion & Timing extragalactic  Supernova neutrino burst delay with Gravity wave\\
\end{tabular}
\end{center}
The slides of all presentations are available on the workshop web site (\url{http://artico.mib.infn.it/numass2013})
\newpage
\section{Highlights from the presentations}
%
 \subsection{ H. J. de Vega: ``Galaxy and cosmological motivations to the search of keV sterile neutrinos''}
\begin{description}
\it\small 
\setlength{\parskip}{-1mm}
\item[\small\it  H. J. de Vega:] devega\@lpthe.jussieu.fr, LPTHE, Universit\'e Pierre et Marie Curie (Paris VI), Laboratoire Associ\'e au CNRS UMR 7589, Tour 13,
4\`eme. et 5\`eme. \'etages, Boite 126, 4, Place Jussieu, 75252 Paris, Cedex 05, France, and Observatoire de Paris, LERMA. Laboratoire Associ\'e au CNRS
UMR 8112. 61, Avenue de l'Observatoire, 75014 Paris, France.
\end{description}
\subsubsection{Abstract}
Warm dark matter (WDM) means DM particles with mass $ m $ in the keV scale.
For large scales, for structures beyond $ \sim 100$ kpc, WDM and CDM yield identical results 
which agree with observations. For intermediate scales, WDM gives the correct abundance of substructures.
Inside galaxy cores, below $ \sim 100$ pc, $N$-body classical physics simulations 
are incorrect for WDM because at such scales quantum effects are important for WDM.
Quantum calculations (Thomas-Fermi approach) provide galaxy cores, 
galaxy masses, velocity dispersions and density profiles in agreement with the observations.
All evidences point to a dark matter particle mass around 2 keV.
Baryons, which represent 16\% of DM, are expected to give a correction to pure WDM results.
The detection of the DM particle depends upon the particle physics model.
Sterile neutrinos with keV scale mass (the main WDM candidate) can be detected in 
beta decay for Tritium and Renium and in the electron capture in Holmiun.
The sterile neutrino decay into X rays can be detected observing DM
dominated galaxies and through the distortion of the black-body CMB spectrum.
The effective number of neutrinos, N$_{\rm eff}$ measured by WMAP9 and Planck satellites
is compatible with two Majorana sterile neutrinos with mass much smaller than the electron mass.
One of them can be a WDM sterile neutrino.
So far, {\bf not a single valid} objection arose against WDM.

\subsubsection{Introduction}

81 \% of the matter of the universe is {\bf dark}.
Dark matter (DM) is the dominant component of galaxies. 
DM interacts through gravity. 
DM interactions other than gravitational have been {\bf so far unobserved}, such possible couplings must be very weak:
much weaker than weak interactions in particle physics. DM is outside the standard model of particle physics.

The main proposed candidates for DM are:
Neutrinos (hot dark matter) back in the 1980's with particle mass $ m \sim 1 $ eV
already ruled out, 
Cold Dark Matter (CDM), weak interacting massive particles (WIMPS) 
in supersymmetric models with R-parity, with particle mass $ m \sim 10-1000 $ GeV 
seriously disfavoured by galaxy observations, and finally
Warm Dark Matter (WDM), mainly sterile neutrinos with particle mass $ m \sim 1 $ keV.

DM particles decouple due to the universe expansion, their distribution function 
{\bf freezes out} at decoupling. The characteristic length scale after decoupling
is the {\bf free streaming scale (or Jeans' scale)}. Following the DM evolution since
ultrarelativistic decoupling by solving the linear Boltzmann-Vlasov equations
yields (see for example [1]),
\be\label{fs}
r_{Jeans} = 57.2 \, {\rm kpc}
\; \frac{\rm keV}{m} \; \left(\frac{100}{g_d}\right)^{\! \frac13} \; , 
\ee
where $ g_d $ equals the number of UR degrees of freedom at decoupling. 

DM particles can {\bf freely} propagate over distances of the order of the free streaming scale.
Therefore, structures at scales smaller or of the order of $ r_{Jeans} $ are {\bf erased}
for a given value of $ m $.

The observed size of the DM galaxy substructures is in the 
$ \sim 1 - 100 $ kpc scale. Therefore, eq.(\ref{fs}) indicates that $ m $ 
should be in the keV scale. That is, Warm Dark Matter particles.
This indication is confirmed by phase-space density observations [6] and 
relevant further evidence [3,5,7,8,9,11,12].

For CDM particles with $ m \sim 100$ GeV we have $ r_{Jeans} \sim 0.1 $ pc.
Hence CDM structures keep forming till scales as small as the solar system.
This result from the linear regime is confirmed  as a {\bf robust result} by
$N$-body CDM simulations. However, it has {\bf never been observed} in the sky. 

Adding baryons to CDM does not cure this serious problem. 
There is {\bf over abundance} of small structures in CDM and in CDM+baryons
(also called the satellite problem). 

CDM has {\bf many serious} conflicts with observations as:
\begin{itemize}
\item{Galaxies naturally grow through merging in CDM models.
Observations show that galaxy mergers are {\bf rare} ($ <10 \% $).}
\item{Pure-disk galaxies (bulgeless) are observed whose formation 
through CDM is unexplained}.
\item{CDM predicts {\bf cusped} density profiles: $ \rho(r) \sim 1/r $ for small $ r $.
Observations show {\bf cored } profiles: $ \rho(r) $  bounded for small $ r $.
Adding by hand strong enough feedback in the CDM from baryons can eliminate cusps but spoils the star formation rate.}
\end{itemize}
Structures in the Universe as galaxies and cluster of galaxies
form out of the small primordial quantum fluctuations originated by inflation
just after the big-bang.

These linear small primordial  fluctuations grow due to gravitational unstabilities
(Jeans) and then classicalize. Structures form through non-linear gravitational evolution.
Hierarchical formation starts from small scales first.

$N$-body CDM simulations {\bf fail} to produce the observed structures 
for {\bf small} scales less than some kpc.

Both $N$-body WDM and CDM simulations yield {\bf identical and correct} structures 
for scales larger than some kpc.

At intermediate scales WDM give the {\bf correct abundance} of substructures [9].

Inside galaxy cores, below  $ \sim 100$ pc, $N$-body classical physics simulations 
are incorrect for WDM because quantum effects are important in WDM at these scales.
WDM predicts correct structures for small scales (below kpc) when its {\bf quantum} nature is
taken into account [3].

\subsubsection{Quantum physics in Galaxies}

To determine whether a physical system has a classical or quantum nature
one has to compare the average distance between particles with their
de Broglie wavelength.

The de Broglie wavelength of DM particles in a galaxy can be expressed as
$ \lambda_{dB}  = \hbar/(m \; v) $,
where $ v $ is the velocity dispersion, while the average interparticle distance $ d $ can be estimated as
$ d = \left( m/\rho_h \right)^{\! \! \frac13} \; , $
where $ \rho_h $ is the average density in the  galaxy core.
We can measure the classical or quantum character of the system by considering the ratio
$$ 
{\cal R} \equiv \frac{\lambda_{dB}}{d}  \; .
$$
$ \cal R $ can then be expressed as
\be
{\cal R} = \hbar \; \left( \frac{Q_h}{m^4}\right)^{\! \! \frac13} \quad {\rm where} ~ 
 Q_h \equiv \frac{\rho_h}{\sigma^3} \quad {\rm is ~ the ~ phase-space ~ density} \; .
\ee
Notice that $ \cal R $ as well as  $ Q_h $ are invariant under the expansion of the universe
because both $ \lambda_{dB} $ and $ d $ scale with the expansion scale factor.  $ \cal R $ and $ Q_h $
evolve by nonlinear gravitational relaxation.

Using now the observed  values of $ Q_h $ from Table \ref{pgal} yields $ \cal R $ in the range
\be\label{quant}
 2 \times 10^{-3}  \; \left( \displaystyle \frac{\rm keV}{m}\right)^{\! \frac43}
< {\cal R} < 1.4 \; \left( \displaystyle \frac{\rm keV}{m}\right)^{\! \frac43}
\ee
The larger value of $ \cal R $ is for ultracompact dwarfs while the smaller value of $ \cal R $ 
is for big spirals.

The ratio $ \cal R $ around unity clearly implies a macroscopic quantum object.
Notice that $ \cal R $ expresses solely in terms of $ Q $ and hence 
$ (\hbar^3 \; Q/m^4) $ measures how quantum or classical is the system, here, the galaxy. 
Therefore, we conclude {\bf solely from observations} 
that compact dwarf galaxies are natural macroscopic quantum objects for WDM [3].

We see from eq.(\ref{quant}) that for CDM with $ m\gtrsim $ GeV that
$ {\cal R} \lesssim 10^{-8} $
and hence there are no quantum effects in CDM.

\begin{table}
\begin{tabular}{|c|c|c|c|c|c|} \hline  
 Galaxy  & $ \displaystyle \frac{r_h}{\rm pc} $ & $  \displaystyle \frac{v}{\frac{\rm km}{\rm s}} $
& $ \displaystyle  \frac{\hbar^{\frac32} \;\sqrt{Q_h}}{({\rm keV})^2} $ & 
$ \rho(0)/\displaystyle \frac{M_\odot}{({\rm pc})^3} $ & $ \displaystyle \frac{M_h}{10^6 \; M_\odot} $
\\ \hline 
Willman 1 & 19 & $ 4 $ & $ 0.85 $ & $ 6.3 $ & $ 0.029 $
\\ \hline  
 Segue 1 & 48 & $ 4 $ & $ 1.3 $ & $ 2.5 $ & $ 1.93 $ \\ \hline  
  Leo IV & 400 & $ 3.3 $ & $ 0.2 $ & $ .19 $ & $ 200 $ \\ \hline  
Canis Venatici II & 245 & $ 4.6 $ & $ 0.2 $   & $ 0.49 $ & $ 4.8 $
\\ \hline  
Coma-Berenices & 123 & $ 4.6 $  & $ 0.42 $   & $ 2.09 $  & $ 0.14 $
\\ \hline  
 Leo II & 320 & $ 6.6 $ & $ 0.093 $  & $ 0.34 $ & $ 36.6 $
\\  \hline  
 Leo T & 170 & $ 7.8 $ &  $ 0.12 $  & $ 0.79 $ & $ 12.9 $
\\ \hline  
 Hercules & 387 & $ 5.1 $ &  $ 0.078 $  & $ 0.1 $ & $ 25.1 $
\\ \hline  
 Carina & 424 & $ 6.4 $ & $ 0.075 $  & $ 0.15 $ & $ 32.2 $
\\ \hline 
 Ursa Major I & 504 & 7.6  &  $ 0.066 $  & $ 0.25 $ & $ 33.2 $
\\ \hline  
 Draco & 305 & $ 10.1 $ &  $ 0.06 $  & $ 0.5 $ & $ 26.5 $
\\ \hline  
 Leo I & 518  & $ 9 $ &  $ 0.048 $  & $ 0.22 $ & $ 96 $
\\ \hline  
 Sculptor & 480  & $ 9 $ & $ 0.05 $  & $ 0.25  $ & $ 78.8 $
\\ \hline 
 Bo\"otes I & 362 & $ 9 $ & $ 0.058 $  & $ 0.38 $ & $ 43.2 $
\\ \hline  
 Canis Venatici I & 1220  & $ 7.6 $ & $ 0.037 $ & $ 0.08 $ & $ 344 $
\\ \hline  
Sextans & 1290 & $ 7.1 $ & $ 0.021 $ & $ 0.02 $ & $ 116 $
\\ \hline 
 Ursa Minor & 750 & $ 11.5 $ & $ 0.028 $  & $ 0.16 $ & $ 193 $
\\ \hline  
 Fornax  & 1730 & $ 10.7 $ & $ 0.016 $  & $ 0.053  $ & $ 1750 $
\\  \hline  
 NGC 185  & 450 & $ 31 $ & $ 0.033 $ & $ 4.09 $ & $ 975 $
\\ \hline  
 NGC 855  & 1063 & $ 58 $ & $ 0.01 $ & $ 2.64 $ & $ 8340 $
\\ \hline  
  Small Spiral  & 5100  & $ 40.7 $ & $ 0.0018 $ & $ 0.029 $ & $ 6900 $
\\ \hline  
NGC 4478 & 1890 & $ 147 $ & $ 0.003 $ & $ 3.7 $ & $ 6.55 \times 10^4 $
\\ \hline  
 Medium Spiral & $ 1.9 \times 10^4 $ & $ 76.2 $ & $ 3.7 \times 10^{-4} $ & $ 0.0076 $ & $ 1.01 \times 10^5 $
\\ \hline  
 NGC 731 & 6160 & $ 163 $ & $ 9.27 \times 10^{-4} $ & $ 0.47 $ & $ 2.87 \times 10^5 $
\\ \hline 
 NGC 3853   & 5220 & $ 198 $ & $ 8.8 \times 10^{-4} $ & $ 0.77 $  
& $ 2.87 \times 10^5 $ \\ \hline 
NGC 499  & 7700 &  $ 274 $ & $ 5.9 \times 10^{-4} $ & 
$ 0.91 $ & $ 1.09 \times 10^6 $ \\   \hline 
Large Spiral & $ 5.9 \times 10^4 $ & $ 125 $ & $ 0.96 \times 10^{-4} $ & $ 2.3 \times 10^{-3} $ & 
$ 1. \times 10^6 $ \\ \hline  
\end{tabular}
\caption{Observed values $ r_h $, velocity dispersion $ v, \;  \sqrt{Q_h}, \; \rho(0)$ 
and $ M_h $ covering from ultracompact galaxies to large spiral galaxies
from refs.[4,5]. The phase space density is larger
for smaller galaxies, both in mass and size.
Notice that the phase space density is obtained
from the stars velocity dispersion which is expected to be smaller than the DM  velocity dispersion.
Therefore, the reported $ Q_h $ are in fact upper bounds to the true values [4].}
\label{pgal}
\end{table}


We consider a single DM halo in the late stages of structure formation when DM
particles composing it are non--relativistic and their phase--space distribution
function $ f(t, \br,\bp) $ is relaxing to a time--independent form, at least for
$\br$ not too far from the halo center. In the Thomas--Fermi approach such a
time--independent form is taken to be a energy distribution function $f(E)$ of
the conserved single--particle energy $E = p^2/(2m) - \mu $, where $m$ is the
mass of the DM particle and $\mu$ is the chemical potential
$   \mu(\br) =  \mu_0 - m \, \phi(\br) $
with $\phi(\br)$ the gravitational potential and  $ \mu_0 $ some constant.

We consider the spherical symmetric case. 
The Poisson equation for $ \phi(r) $ is a nonlinear and selfconsistent equation
\be \label{pois}
  \frac{d^2 \mu}{dr^2} + \frac2{r} \; \frac{d \mu}{dr} = - 4\pi \, G \, m \, \rho(r)\; , 
\ee
where the mass density $ \rho(r) $ is a function of $ \mu(r) $ and
$ G $ is Newton's constant. $ \rho(r) $ is expressed here as a function of $ \mu(r) $ through the
standard integral of the DM phase--space distribution function over the momentum
for Dirac fermions as
\be \label{den}
  \rho(r) = \frac{m}{\pi^2 \, \hbar^3} \int_0^{\infty} dp\;p^2 
  \; f\left(\displaystyle \frac{p^2}{2m}-\mu(r)\right)\; , 
\ee
We see that $\mu(r)$ fully characterizes the fermionic DM halo in this
Thomas--Fermi framework [3].
Eqs.(\ref{pois}) and (\ref{den}) provide an ordinary nonlinear
differential equation that determines selfconsistently the chemical potential $ \mu(r) $ and
constitutes the Thomas--Fermi semi-classical approach. We obtain a family of solutions 
parametrized by the value of  $ \mu_0 \equiv \mu(0) $ [3].

We integrate the Thomas-Fermi nonlinear differential
equations (\ref{pois})-(\ref{den}) from $ r = 0 $ till the 
boundary $ r = R = R _{200} \sim R_{vir} $ defined as the radius where the 
mass density equals $ 200 $ times the mean DM density [3].

We define the core size $ r_h $ of the halo by analogy with the Burkert density profile as
\be\label{onequarter}
  \frac{\rho(r_h)}{\rho_0} = \frac14 \quad , \quad  r_h = l_0 \; \xi_h \; .
\ee
To explicitly solve eq. (\ref{pois})-(\ref{den}) we need to specify the distribution function
$ \Psi(E/E_0) $. But many important properties of the Thomas--Fermi semi-classical
approximation do not depend on the detailed form of the distribution function
$ \Psi(E/E_0) $. Indeed, a generic
feature of a physically sensible one--parameter form $ \Psi(E/E_0) $ is that it should
describe degenerate fermions for $ E_0 \to 0 $. That is, $ \Psi(E/E_0) $ should behave as
the step function $ \theta(-E) $ in such limit. In the opposite limit, $ \Psi(E/E_0) $ 
describes classical particles for $ \mu/E_0 \to -\infty $.
As an example of distribution function, we consider the Fermi--Dirac distribution 
\be\label{FD}
  \Psi_{\rm FD}(E/E_0) = \frac1{e^{E/E_0} + 1} \; .
\ee
We define the dimensionless chemical potential $ \nu(r) $ as  $ \nu(r) \equiv \mu(r)/E_0 $
and  $ \nu_0 \equiv \mu(0)/E_0 $. 
Large positive values of the chemical potential at the origin $ \nu_0 \gg 1 $ 
correspond to the degenerate 
fermions limit which is the extreme quantum case and oppositely, $ \nu_0 \ll -1 $ gives 
 the diluted limit which is the classical limit. In this classical limit the Thomas-Fermi equations
(\ref{pois})-(\ref{den}) become the equations for a self-gravitating Boltzmann gas.

The obtained fermion profiles are always cored. 
The sizes of the cores $ r_h $ defined by eq.(\ref{onequarter}) 
are in agreement with the observations, from the compact galaxies where $ r_h \sim 35 $ pc till
the spiral and elliptical galaxies where $ r_h \sim 0.2 - 60 $ kpc. The larger and positive is 
$ \nu_0 $, the smaller is the core. The minimal core size arises in
the degenerate case  $ \nu_0 \to +\infty $ (compact dwarf galaxies) [3].

\begin{figure}[h]
\begin{center}
\includegraphics[width=14.cm]{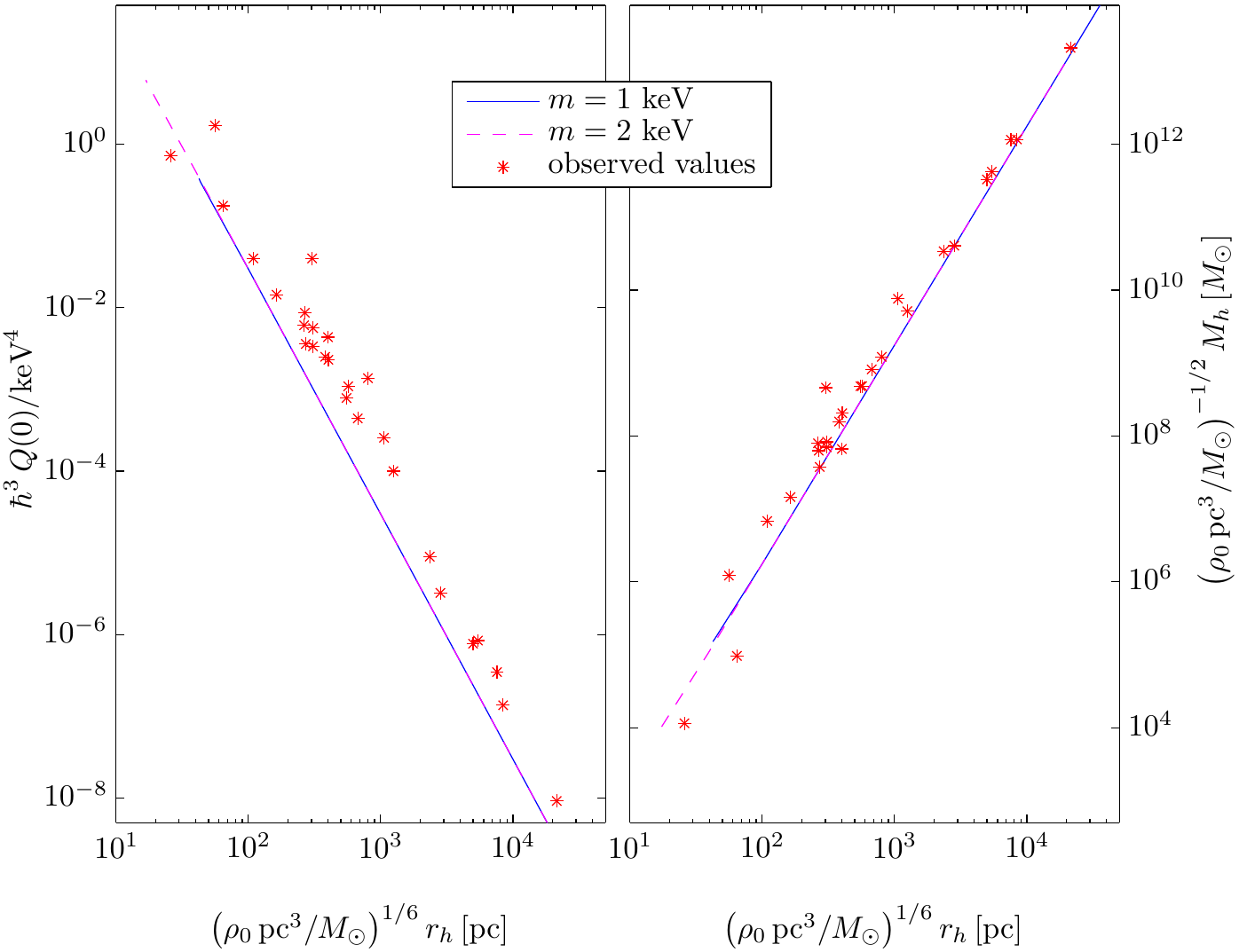}
\caption{In the left panel we display the galaxy phase-space density 
$ \hbar^3 \; Q(0)/({\rm keV})^4 $
obtained from the numerical resolution of the Thomas-Fermi eqs. (\ref{pois})-(\ref{den})
for WDM fermions of mass $ m = 1 $ and $ 2 $ keV 
versus the ordinary logarithm of the product $ \log_{10}\{r_h \; 
[{\rm pc}^3 \; \rho_0/ M_\odot]^{\frac16} \} $ in parsecs. 
The red stars $ * $ are the observed values of $ \hbar^3 \; Q(0)/({\rm keV})^4 $ from Table \ref{pgal}.  
Notice that the observed values $ Q_h $ from the stars' velocity 
dispersion are in fact upper bounds for the DM $ Q_h $ and therefore the theoretical curve is slightly below them.
In the right panel we display the galaxy mass 
$ (M / M_\odot) \sqrt{M_\odot / [\rho_0 \; {\rm pc}^3]} $ obtained from the numerical resolution of the 
Thomas-Fermi eqs.(\ref{pois})-(\ref{den}) for WDM fermions of mass $ m = 1 $ and $ 2 $ keV 
versus the product $ r_h \; [{\rm pc}^3 \; \rho_0/ M_\odot]^{\frac16} $
in parsecs.  The red stars $ * $ are the observed values of 
$ (M / M_\odot) \sqrt{M_\odot / [\rho_0 \; {\rm pc}^3]} $ 
from Table \ref{pgal}. Notice that the error bars 
of the observational data are not reported here but they are at least about $ 10-20 \%$.}
\label{halo}
\end{center}
\end{figure}

In the left panel of fig. \ref{halo} we plot the dimensionless quantity  [3]
$ \hbar^3 \; Q(0) /  ({\rm keV})^4 $. 
In the right panel of fig. \ref{halo}, we plot instead the dimensionless product
$ M_h/M_\odot \; \sqrt{M_\odot/[\rho_0 \; {\rm pc}^3 ]} $,
where $ M_h $ is the halo mass, namely the galaxy mass inside the core radius $ r_h
$ defined by eq.(\ref{onequarter}). 
In both cases we consider the two values $ m = 1 $ and $ 2 $ keV and we put in
the abscissa the product
$ r_h  \; \left([{\rm pc^3} / M_\odot] \; \rho_0\right)^{\! \! \frac16} $
in  parsecs, where $ r_h $ is the core radius. The phase-space density $ Q(0) $ and the galaxy 
mass $ M_h $ are obtained by solving the Thomas-Fermi eqs.(\ref{pois})-(\ref{den}).
We have also superimposed the
observed values $ \hbar^3\, Q_h/({\rm keV})^4 $ and $ M_h \sqrt{M_\odot / [\rho_0
  \; {\rm pc}^3]} \; \; \left(m/{\rm keV}\right)^4 $ from Table \ref{pgal}.
Notice that the observed values $ Q_h $ from the stars' velocity dispersion are
in fact upper bounds for the DM $ Q_h $. This may explain why the theoretical
Thomas-Fermi curves in the left panel of fig. \ref{halo} appear slightly below
the observational data. Notice also that the error bars of the observational
data are not reported here but they are at least about $ 10-20 \%$.

To conclude, the galaxy magnitudes:  
halo radius, galaxy masses and velocity dispersion
obtained from the Thomas-Fermi quantum treatment for WDM fermion masses in the keV scale are
fully consistent with all the observations for all types of galaxies (see Table \ref{pgal}). 
Namely, fermionic WDM treated quantum mechanically (as it must be) is able to reproduce
the observed sizes of the DM cores of galaxies [3].

It is highly remarkably that in the context of fermionic WDM, the simple stationary
quantum description provided by the Thomas-Fermi approach is able to reproduce such broad variety of galaxies.

\subsubsection{WDM gives the correct abundance of substructures}

It is known since some time through $N$-body simulations
that WDM alleviates the CDM satellite problem [7]
and the CDM voids problem [8].

WDM subhalos turns to be less concentrated than CDM subhalos.
WDM subhalos have the right concentration to host the bright
Milky Way satellites [7].

The ALFALFA survey has measured the velocity widths in galaxies from the 21cm HI line.
This tests substructure formation. The contrast of the ALFALFA survey
with  $N$-body simulations clearly favours WDM over CDM [9].
A particle mass around $ \sim 2 $ keV is favoured by the ALFALFA survey.

In summary, WDM produces the correct substructure abundance at zero redshift.

Data on galaxy substructure for redshift $ z \lesssim 10 $ becomes now available.
In ref. [11] the evolution of the observed AGN luminosity function for $ 3 < z < 6 $
is contrasted with WDM and CDM simulations. WDM is clearly favoured over CDM.
In ref. [12] the number of structures vs. the star formation rate
for  $ z =5, \; 6, \; 7 $ andf $ 8 $ is contrasted  with WDM and CDM simulations.
Again,  WDM is clearly favoured over CDM.

At intermediate scales where WDM and CDM give non-identical results and 
quantum effects are negligible, $N$-body classical simulations are reliable.
Contrasting such $N$-body classical simulations results with astronomical
observations at zero and non-zero redshift {\bf clearly favours} WDM over CDM.

For larger scales $ \gtrsim 100 $ kpc, CDM and WDM $N$-body classical simulations are reliable
and give identical results in good agreement with astronomical observations.

\subsubsection{Detection of keV mass Sterile Neutrinos}

Sterile neutrinos $ \nu_s $ are mainly formed by right-handed neutrinos $ \nu_R $ 
plus a small amount of left--handed neutrinos $ \nu_L $. Conversely, active neutrinos $ \nu_a $
are formed by $ \nu_L $ plus a small amount of $ \nu_R $:
$$ 
\nu_s \simeq \nu_R + \theta \; \nu_L \quad ,  \quad \nu_a = \nu_L + \theta \; \nu_R \; .
$$
Sterile neutrinos were named by Bruno Pontecorvo in 1968.
They are singlets under all symmetries of the Standard Model of particle physics. 
Sterile neutrinos do not interact through weak, electro-magnetic or strong interactions.

WDM $ \nu_s $ are typically produced in the early universe from active neutrinos through mixing,
namely, through a bilinear term $ \theta \; \nu_s \;  \nu_a $ in the Lagrangian.

The appropriate value of the mixing angle $ \theta $
to produce enough sterile neutrinos $ \nu_s $ accounting for the observed total DM
depends on the particle physics model and is typically very small: $ \theta \sim  10^{-3}  -  10^{-4} \; $. 
The smallness of $ \theta $ makes sterile neutrinos difficult to detect in experiments.

Sterile neutrinos can be detected in beta decay and in electron capture (EC) processes
when a $ \nu_s $ with mass in the keV scale is produced {\bf instead} of an active $ \nu_a $:
$$
{}^3H_1 \Longrightarrow {}^3H \! e_{\, 2} + e^- + {\bar \nu}_e \quad ,
\quad {}^{187} Re \Longrightarrow {}^{187} Os  + e^- + {\bar \nu}_e  \; .
$$
In beta decays the electron spectrum is slightly modified at energies around the $ \nu_s $ mass 
($\sim$ keV) when a $ {\bar \nu}_s $ is produced instead of a 
$ {\bar \nu}_e $ in the decay products. Such event can be inferred observing
the electron energy spectrum. A 'kink' should then appear around the energy of the  $ \nu_s $ 
mass.

In electron capture processes like: 
$$
{}^{163}Ho  + e^- \Longrightarrow {}^{163}Dy^* + \nu_e 
$$
when a $ \nu_s $ with mass in the keV scale is produced {\bf instead} of an active $ \nu_e $,
the observed nonradiative de-excitation of the $ Dy^* $ is different
to the case where an active $ \nu_e $ shows up.
 
The available energies for these beta decays and EC are
\be\label{Q}
Q({}^{187}Re) = 2.47 \; {\rm keV} \quad , \quad  Q({}^3H_1) = 18.6 \; {\rm keV} \quad ,\quad  Q({}^{163}Ho) \simeq 2.5 \; {\rm keV}. 
\ee
In order to produce a sterile neutrino with mass $ m , \; Q $ must be larger than $ m $.
However, in order to distinguish the sterile neutrino $ \nu_s $ from a practically massless  
active neutrino $ \nu_a $,
$ Q $ must be as small as possible. This motivates the choice of the nuclei 
with the lowest known $ Q $ in eq.(\ref{Q}).
 
For a theoretical analysis of $ \nu_s $ detection in Rhenium and Tritium beta decay 
see ref. [13] and references therein.

Present experiments searching the small active neutrino mass also look for sterile
neutrinos in the keV scale:

\begin{itemize}
\item{MARE (Milan, Italy), Rhenium 187 beta decay and Holmiun 163 electron capture [18].}
\item{KATRIN (Karlsruhe, Germany), Tritium beta decay [19].}
\item{ECHo (Heidelberg,  Germany), Holmiun 163 EC [20].}
\item{Project 8 (Seattle, USA), Tritium beta decay [21].}
\end{itemize}
The more popular sterile neutrino models nowadays are:
\begin{itemize}
\item{The Dodelson-Widrow (DM) model (1994): sterile neutrinos are produced by
non-resonant mixing from active neutrinos.}
\item{The Shi-Fuller model (SF) (1998): sterile neutrinos are produced by
resonant mixing from active neutrinos.}
\item{$\nu$MSM model (2005): sterile neutrinos are produced by
a Yukawa coupling from the decay of a heavy real scalar field $\chi$.}
\item{Models based on: Froggatt-Nielsen mechanism, flavor symmetries,
$Q_6$, split see-saw, extended see-saw, inverse see-saw, loop mass. Furthermore: scotogenic,
LR symmetric, etc. See for a recent review [14].}
\end{itemize}
WDM particles decoupling ultrarelativistically in the first three WDM particle models behave just as if their
masses were different [2]. The masses of WDM particles in the first three models which give the same
primordial power spectrum can be related according to the formula [2]
(FD = thermal fermions):
$$ \frac{m_{DW}}{\rm keV} \simeq 2.85 \; 
\left(\frac{m_{FD}}{\rm keV}\right)^{\! \frac43} \quad , \quad
m_{SF} \simeq 2.55 \;  m_{FD} \quad , \quad
m_{\nu{\rm MSM}} \simeq 1.9 \; m_{FD} \; .
$$
For the primordial power spectrum, there is a degeneracy between these three models.

Sterile neutrinos $ \nu_s $ decay into active neutrinos $ \nu_a $ plus 
X-rays with an energy $ m/2 $ [15]. The lifetime of $ \nu_s $ is about
$ \sim 10^{11} \times $ age of the universe.
The value of the lifetime depends on the particle physics neutrino model.

These X-rays may be seen in the sky looking to galaxies [16].
See [17] for a recent review.

Some future observations of X-rays from galaxy halos: 
\begin{itemize}
\item{DM bridge between M81 and M82 $ \sim 50$ kpc. Overlap of DM halos.
Satellite projects: Xenia (NASA) [22].}
\item{CMB: WDM decay distorts the blackbody CMB spectrum.
The projected PIXIE satellite mission can measure WDM sterile neutrino mass 
by measuring this distortion [23].}
\end{itemize}
Active neutrinos are very abundant in supernovae explosions and in these explosions
sterile neutrinos are produced too. Hence, bounds on the presence of
sterile neutrinos can be obtained contrasting to  supernovae observations.
The results from supernovae do not constrain $ \theta $ provide
$ 1 < m < 10 $ keV [24].

\subsubsection{Sterile neutrinos and CMB fluctuations}

CMB fluctuations data provide the effective number of neutrinos, N$_{\rm eff}$.
This effective number  N$_{\rm eff}$ is related in a subtle way to the number
of active neutrinos (three) plus the number of sterile neutrinos
with mass much smaller than the electron mass $ m_e $ [25].
As shown in ref. [26]
two Majorana sterile neutrinos with  mass $ m \ll m_e $
are compatible (see fig. 8 in pag. 13 of ref.  [26])
with the Planck results:  N$_{\rm eff} = 3.5 \pm 0.5 \; 
(95\%$; Planck+WP+highL+H$_0$+BAO) [27].

Therefore,  Planck results are {\bf compatible} with one Majorana sterile neutrino
with an eV mass and one Majorana sterile neutrino with a keV mass.
Indeed, other combinations of steriles are compatible with solely the value of N$_{\rm eff}$ [26].
All these analysis are within the standard cosmological model.
\subsubsection{Future Perspectives and Sterile Neutrino Detection}

WDM particle models must explain the baryon asymmetry of the universe.
This is a strong constraint on sterile neutrino models which must be
worked out for each model.

Combining particle, cosmological and galaxy results for sterile neutrinos 
at different mass scales [3,6,28,29]
an appealing {\bf mass} neutrino hierarchy appears:
\begin{itemize}
\item{Active neutrino: $ \sim $ mili eV}
\item{Light sterile neutrino: $ \sim $ eV}
\item{Dark Matter sterile neutrino: $ \sim $ keV}
\item{Unstable  sterile neutrino: $ \sim $ MeV.... }
\end{itemize}
This scheme may represent the future extension of the standard model of particle physics.

In order to falsify WDM, comprehensive theoretical calculations 
showing substructures, galaxy formation and evolution including 
the quantum WDM effects in the dynamical evolution are needed to contrast with
the astronomical observations.
In such WDM theoretical calculations the quantum pressure must be necesarily included.
These calculations should be performed matching the semiclassical
Hartree-Fock (Thomas-Fermi) dynamics in regions where the dimensionless phase-space
density $ \hbar^3 \; Q/m^4 \gtrsim 0.1 $ with the
classical evolution in regions where $  \hbar^3 \; Q/m^4 \ll 1 $. 
These are certainly not easy numerical calculations but they are unavoidable!

Richard P. Feynman foresaw the necessity to include quantum physics in simulations in 1981 [30]
\begin{center}
{\bf ``I'm not happy with all the analyses that go with just the classical theory, because nature isn't classical, dammit, and if you want to make a simulation of nature, you'd better make it quantum mechanical, and by golly it's a wonderful problem, because it doesn't look so easy.'' }
\end{center}
Sterile neutrino detection depends upon the particle 
physics model. There are sterile neutrino models where the
keV sterile is stable and thus hard to detect (see for example [14]).

Detection may proceed through 
astronomical observations of X-ray keV sterile decay
from galaxy halos and by direct detection of steriles in 
laboratory experiments.

Mare [18], Katrin [19], ECHo [20] and Project 8  [21] are expected to provide
bounds on the mixing angles. However, for a particle detection, a 
dedicated beta decay experiment and/or electron capture experiment seem necessary
to find sterile neutrinos with mass around 2 keV.
In this respect, calorimetric techniques seem well suited.

The best nuclei for study are: ${}^{187}$Re and Tritium for beta decay and ${}^{163}$Ho for electron capture

The search of DM particles with mass around 2 keV is a promisory
avenue for future trascendental discoveries.
\begin{center}
{\it References}
\end{center}
\begin{description}
\footnotesize
\item[1] D. Boyanovsky,  H J de Vega, N. G. Sanchez, 	
Phys. Rev.  {\bf D 78}, 063546 (2008).
\item[2] H. J. de Vega, N. G. S\'anchez, Phys. Rev. D85, 043516  (2012)
and  D85, 043517 (2012).
\item[3] C. Destri, H. J. de Vega, N. G. Sanchez, arXiv:1204.3090,
New Astronomy {\bf 22}, 39 (2013) and arXiv:1301.1864.
\item[4] G. Gilmore et al., Ap J, 663, 948 (2007).
M. Walker, J. Pe\~narrubia, Ap. J. 742, 20 (2011).
P. Salucci et al., MNRAS, 378, 41 (2007).
J. D. Simon, M. Geha,  Ap J, 670, 313 (2007) and references therein.
J. P. Brodie  et al., 
AJ, 142, 199 (2011). B. Willman and J. Strader, AJ, 144, 76 (2012).
J. D. Simon et al., Ap. J. 733, 46 (2011) and references therein.
J. Wolf  et al., MNRAS, 406, 1220 (2010) and references therein.
G. D. Martinez et al., Ap J, 738, 55 (2011). 
\item[5] H. J. de Vega,  P. Salucci, N. G. Sanchez, 
New Astronomy {\bf 17}, 653 (2012) and references therein.
\item[6] H. J. de Vega, N. G. S\'anchez, 
Mon. Not. R. Astron. Soc. {\bf 404}, 885 (2010) and
Int. J. Mod. Phys. {\bf A 26}, 1057 (2011).
\item[7] P. Col\'{\i}n, O. Valenzuela, V.  Avila-Reese, Ap J, 542, 622 (2000).
J. Sommer-Larsen, A. Dolgov, Ap J, 551, 608 (2001).
L. Gao and T. Theuns,  Science, 317, 1527 (2007).
M. R. Lovell et al., MNRAS, 420, 2318 (2012).
\item[8] A. V. Tikhonov et al., MNRAS, 399, 1611 (2009).	
\item[9] E. Papastergis et al., Ap J, 739, 38 (2011),
J. Zavala et al., Ap J,	700, 1779 (2009).
\item[10] A. V. Macci\`o, S. Paduroiu, D. Anderhalden, A. Schneider, B. Moore,
MNRAS, 424, 1105 (2012). S. Shao et al. arXiv:1209.5563, MNRAS in press.
\item[11] N. Menci, F. Fiore, A. Lamastra, arXiv:1302.2000.
\item[12] L. Danese, H. J. de Vega, A. Lapi, P. Salucci, N. Sanchez (in preparation).
\item[13] H J de Vega, O. Moreno, E. Moya, M. Ram\'on Medrano, N. S\'anchez, Nucl. Phys. {\bf B866}, 177 (2013).
\item[14] A. Merle, arXiv:1302.2625.
\item[15] R N Mohapatra, P B Pal, `Massive neutrinos in physics and astrophysics', 
World Scientific, Singapore, 2004.
\item[16] M. Loewenstein, A. Kusenko, P. L. Biermann, Astrophys.J. 700 (2009) 426-435
 M. Loewenstein, A. Kusenko, Astrophys.J. 714 (2010) 652 and 751 (2012) 82.
\item[17] C. R. Watson et al. JCAP, 03, 018 (2012). 
\item[18] http://mare.dfm.uninsubria.it/frontend/exec.php
\item[19] http://www.katrin.kit.edu/
\item[20] L. Gastaldo, lecture at the 3rd. NuMass Workshop, Milano Bicocca, Italy, February 2013.
\item[21] http://www.npl.washington.edu/project8/
\item[22] http://xenia.msfc.nasa.gov/
\item[23] A. Kogut et al., JCAP 07, 025 (2011).
\item[24] G. Raffelt, S. Zhou, PRD 83, 093014 (2011).
\item[25] G. Steigman, Adv. in High Energy Phys. 268321 (2012).
\item[26] G. Steigman, arXiv:1303.0049.
\item[27] P. A. R. Ade et al., Planck 2013 results. XVI. arXiv:1303.5076.
\item[28] J. Kopp et al. arXiv:1303.3011 and references therein.
\item[29] S. N. Gninenko, Phys. Rev. D85, 051702(R) (2012) and references therein.
\item[30] R. P. Feynman, Lecture at the 1st. Conference on Physics and Computation, 
MIT 1981, Int. J. Theor. Phys. {\bf 21},467(1982).
\end{description}

 \newpage\subsection{G.J. Barker: ``Neutrino Mass Hierarchy From Long Baseline Accelerator Experiments''}
\begin{description}
\it\small 
\setlength{\parskip}{-1mm}
\item[]Dept. of Physics, University of Warwick, Coventry, UK
\end{description}
%
%
\subsubsection{Introduction}
Long baseline neutrino oscillation projects based on accelerator-derived
neutrino beams have the ultimate goal of observing CP-violation in
neutrinos. To achieve this, an unambiguous determination of the mass hierarchy
must first be made since matter effects also introduce a different oscillatory
behaviour between neutrinos and anti-neutrinos which must be unfolded from any
CP-violating effects measured. 
A relatively large value for $\theta_{13}$~has meant that `conventional' LBL
experiments studying oscillations of a beam of $\nu_{\mu}$'s with baselines of $1000\,$km or more,
can measure the MH independently of the value of $\delta_{CP}$.   
In what follows, we briefly report on projects, either running or planned, which could
conceivably provide results on the MH within a decade or so from now: NO$\nu$A+T2K,
LBNO and LBNE.  

\subsubsection{The NO$\nu$A and T2K Experiments}
Both the  NO$\nu$A and T2K experiments are conventional LBL projects
consisting of a near and far detector measuring a $\nu_{\mu}$~beam at a small
off-axis angle which provides a quasi mono-energetic beam optimised to measure
small $\theta_{13}$. 
NO$\nu$A[1] works with the Fermilab NUMI beam which is planned to begin operation at a
power of $700\,$kW in May 2013. The near and far detectors are identical
liquid scintillator devices with a mass of $0.3\,$kt and $14\,$kt
respectively, separated by a baseline of $810\,$km. The T2K project[2]
utilises Super-Kamiokande as a far detector while the near detector is a
multi-purpose tracking and calorimetry device optimised to measure the beam
flux and flavour content as well as make neutrino interaction cross-section
measurements. T2K is based on the JPARC beam (currently $200 \,$kW ramping up
to $700 \,$kW by 2019) and the baseline is $295 \,$km. The results show that
NO$\nu$A can determine the MH if Nature's choice of $\delta_{CP}$~is in the
favourable half-plane. T2K brings some extra sensitivity but cannot raise it
above the $90 \% \,$C.L. over the full 
  $\delta_{CP}$~range.    

\begin{figure}[b]
        \centering
        \includegraphics[scale=0.8]{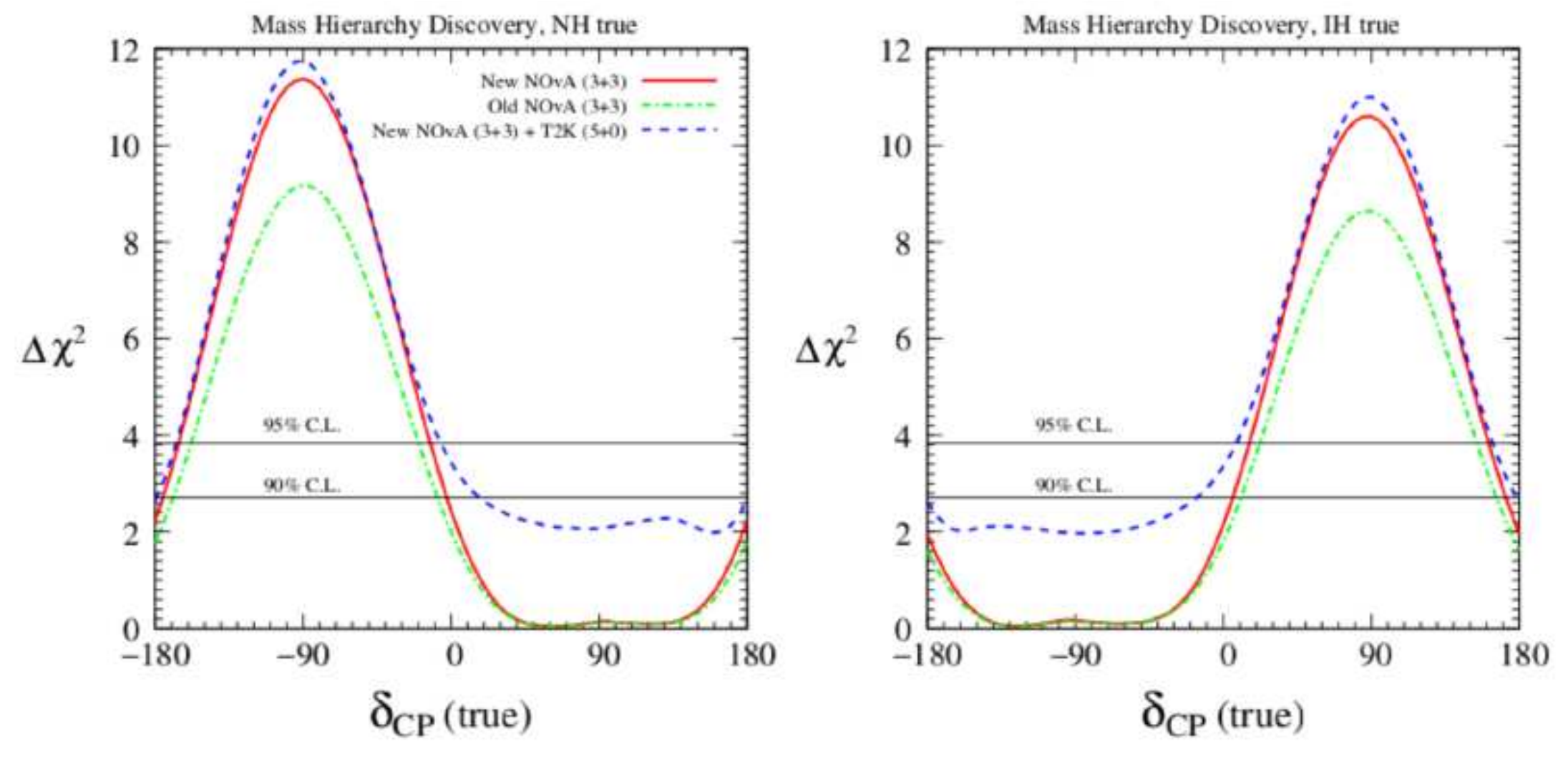}
        \caption{\label{NovaT2K}The MH discovery sensitivity of NO$\nu$A alone
          and combined with T2K (Left) for normal hierarchy and (Right) for
          inverted hierarchy. `New' NO$\nu$A refers to a new analysis
          event selection optimised for the measured (large) value of
          $\theta_{13}$. From [3].}
\end{figure}

Figure~\ref{NovaT2K} illustrates the sensitivity [6]
 of NO$\nu$A and
NO$\nu$A+T2K from the observation of $\nu_{\mu} \rightarrow \nu_{e}$
oscillations assuming three years $\nu_{\mu}$ and three years
$\overline{\nu}_{\mu}$ running for  NO$\nu$A and five years  $\nu_{\mu}$
operation for T2K.

\subsubsection{Long Baseline Neutrino Oscillation experiment (LBNO)}
LBNO[4] is a proposed next-generation LBL experiment with considerably
better physics sensitivity to MH (and CP-violation) than NO$\nu$A/T2K. The
project consists of a new
conventional beam-line facility at CERN (CN2PY) pointed at a deep
underground neutrino observatory located at the Pyh\"{a}salmi mine in Finland,
$2300 \,$km away. The physics gains of the LBNO proposal over existing
experiments comes mainly from:
\begin{itemize}   
\item[(1)] The beam is on axis, providing a wide spread of neutrino
  energies. Measurements of $\nu_{\mu}$~disappearance and of
  $\nu_{e},\nu_{\tau}$~appearance, as a function of energy and for
  neutrino/antineutrino separately, provide a way to unfold the effects of MH
  and CP-violation for definitive measurements. The long baseline also
  guarantees superb sensitivity to the MH through large matter effects.   
\item[(2)] A far detector based on a double phase Liquid Argon Time Projection
  Chamber (LAr TPC) technology provides a detector with superior signal efficiency and
  background rejection but with a similar mass ($20 \,$kt) as Super-Kamiokande
  and NO$\nu$A.A staged deployment approach could see the detector mass
  eventually rising to $100 \,$kt.  
\item[(3)] Based on already planned upgrades to the SPS intensity the CN2PY beam
power will start at $\sim 770 \,$kW but with scope to rise to $2 \,$MW via a future
upgrade of the LHC injection chain with a new HP-PS synchrotron. 
\item[(4)] The  Pyh\"{a}salmi location is the deepest mine in Europe at $\sim
  1440 \,$m and this fact, in combination with the large target mass available,
  provides an excellent opportunity for a forefront programme in nucleon decay
  searches and particle-astrophysics topics. 
\end{itemize}

\begin{figure}
        \centering
        \includegraphics[scale=0.85]{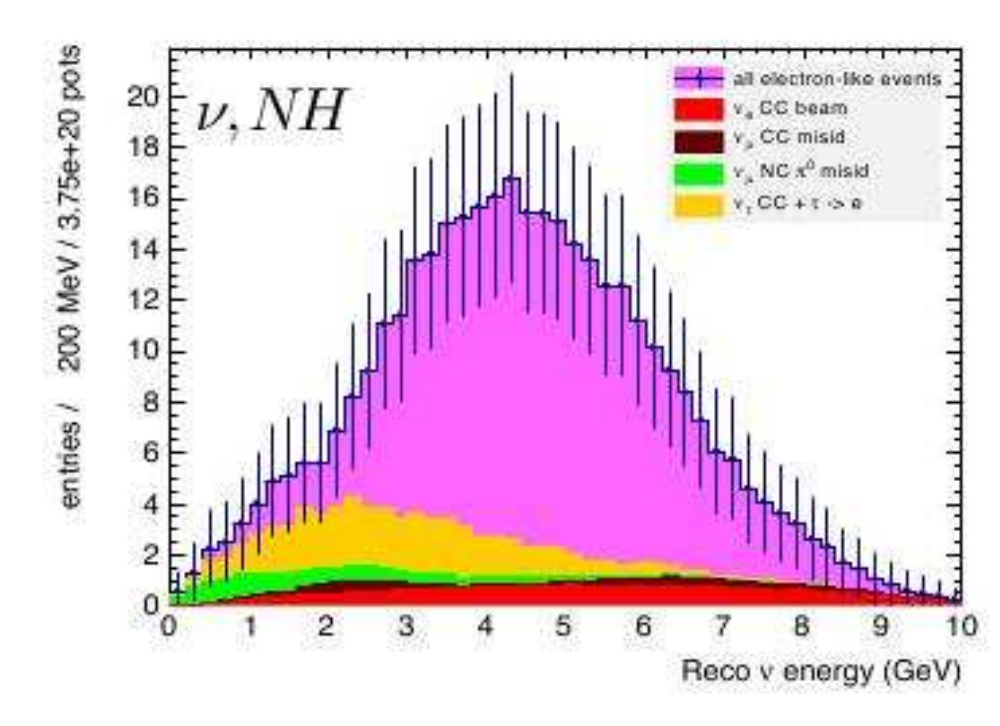}
\includegraphics[scale=0.85]{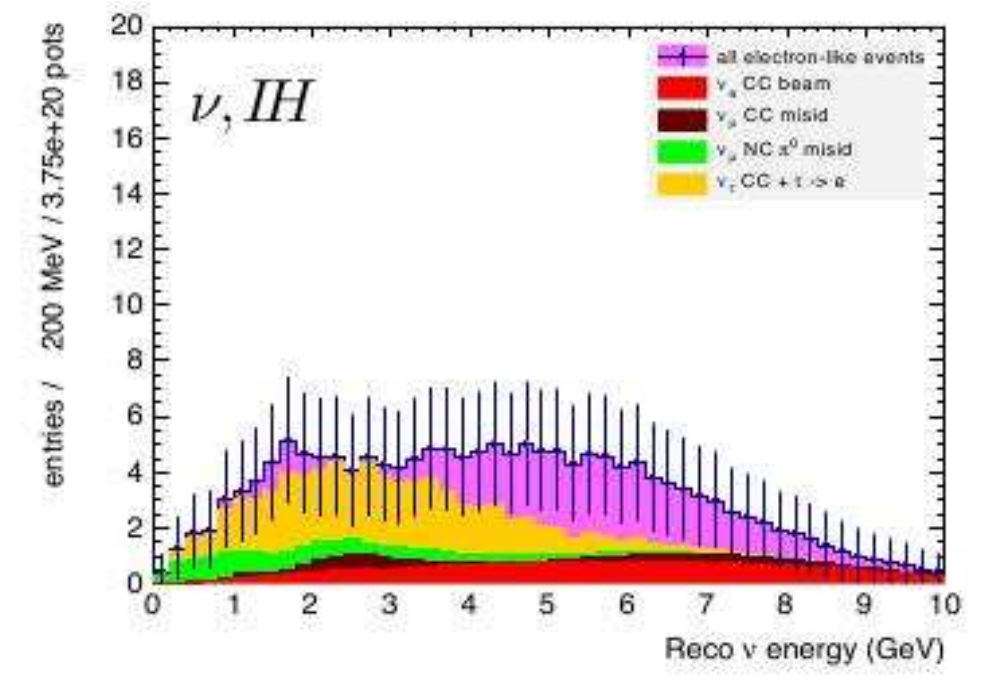}
        \caption{\label{EAPPEAR} LBNO reconstructed neutrino energy for
          electron-like final state signal events and all background
          contributions listed. (Left) normal hierarchy, (Right) inverted
          hierarchy. Assuming $\delta_{CP}=0$~and $3.75 \times
          10^{2}$ protons on target. From [4].}
\end{figure}

\begin{figure}
        \centering
        \includegraphics[scale=1.]{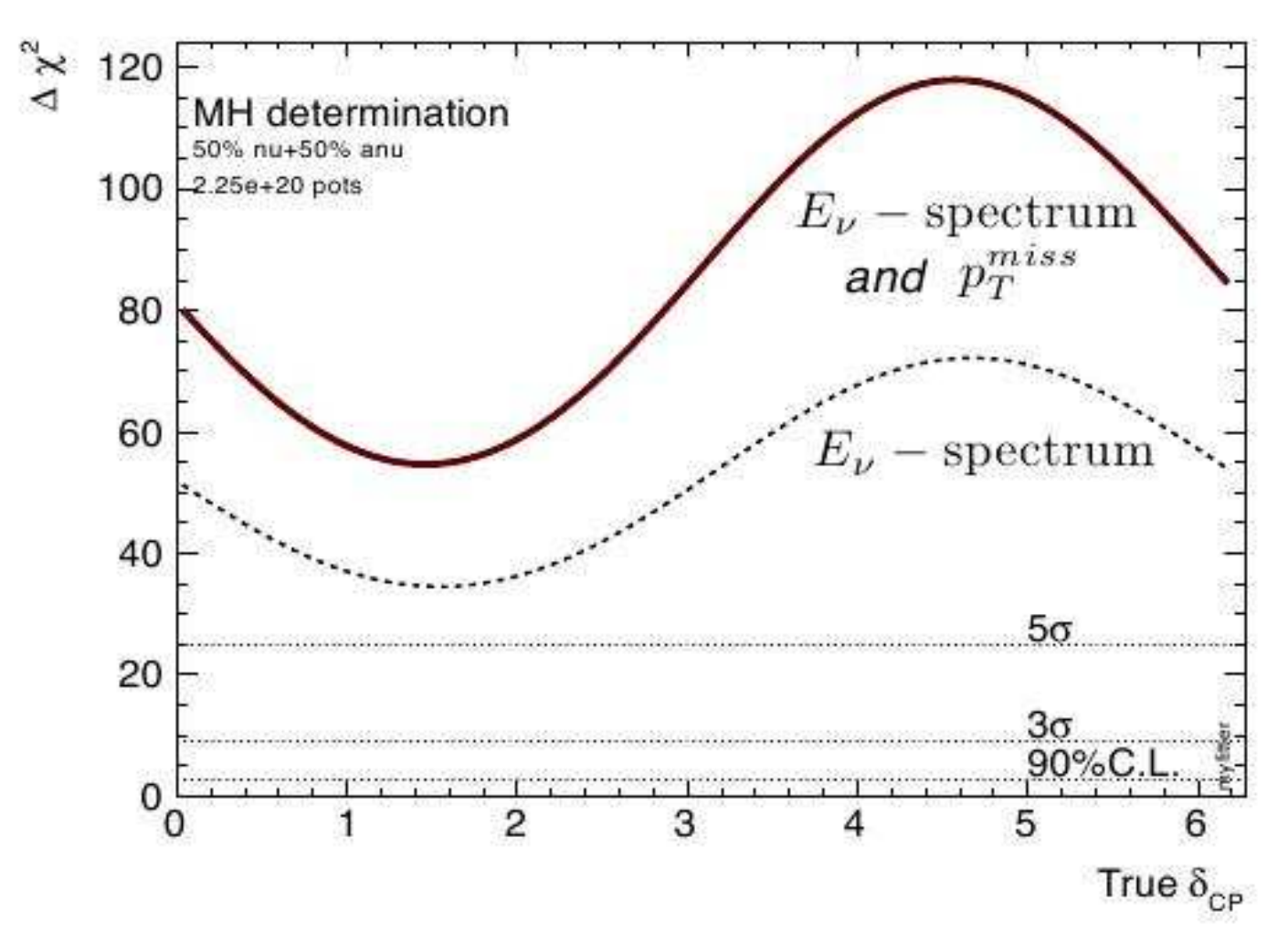}
        \caption{\label{LBNOMH} $\Delta  \chi^2$~for MH discrimination with LBNO as a
          function of the true $\delta_{CP}$~value based on $2.25 \times
          10^{20}$~protons on target. The two curves represent the standard
          analysis and an additional technique to reduce the tau background
          based on missing $p_T$. From [4].}
\end{figure}

Figure~\ref{EAPPEAR} shows the event energy spectrum expected in the
electron-appearance channel after around three years of data taking. The large
impact of the MH is striking and the largest background component is seen to
be tau decays to electrons.   
Figure~\ref{LBNOMH} presents the sensitivity to a MH determination based on
$\Delta \chi^2$~which shows that a, greater than, $5 \sigma$ discrimination
(independent of $\delta_{CP}$) is
possible from what amounts to only the first few years of LBNO  in
it's initial configuration. 

An Expression of Interest for LBNO[4] was submitted to CERN in 2012. If the
project were to receive approval by 2015, excavation and construction would
take place up to 2021 with the first physics run beginning in 2023. 

\subsubsection{Long Baseline Neutrino Experiment (LBNE)}
\begin{figure}
        \centering
        \includegraphics[scale=1.]{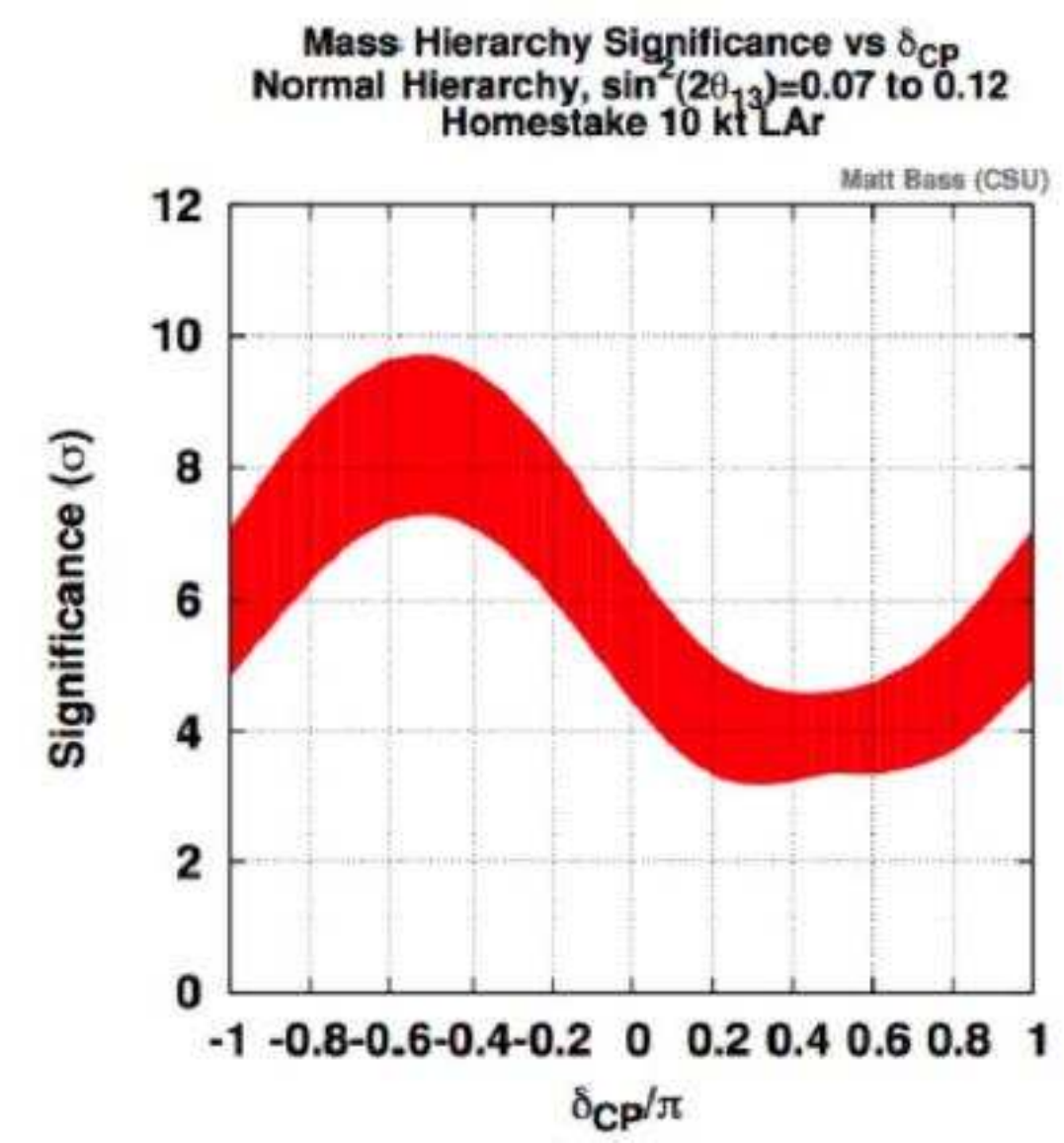}
        \caption{\label{LBNEMH} Significance for determining MH from LBNE as a function
          of  true $\delta_{CP}$ including the results expected from  NO$\nu$A
          and T2K. Projections are for $5+5$ years of $700 \,$kW $\nu
          \bar{\nu}$ with a $10 \,$kt detector at Homestake combined with $3+3$
          years of  NO$\nu$A running at $700 \,$kW and 6 years of nominal T2K
          neutrino data. The width of the band corresponds to varying $\sin^2
          \theta_{13}$ between $[0.07, 0.12]$.  From [5].}
\end{figure}

The LBNE project[5] is a next generation LBL oscillation proposal consisting
of a conventional neutrino beam generated at
Fermilab and fired $1300 \,$km to the Homestake mine. In common with LBNO, the
far detector is a LAr TPC, but an important difference is that in the first phase
of LBNE the far detector will have a mass of only $10 \,$kt and will be
located on the surface. This has important impact on the physics sensitivity
achievable initially but the project has plans for the mass to increase
eventually to $35 \,$kt and to be located deep underground at Homestake. The
initial beam power will be $700 \,$kw but with plans (Project X) for this to
increase to $2.3 \,$MW. Figure~\ref{LBNEMH} shows the sensitivity to the MH, as
a function of true $\delta_{CP}$, as expected from phase one of the project
and in combination with the data from  NO$\nu$A and T2K.
LBNE has attained CD1 approval in the USA and the first phase is planned to
start in 2022. 

\subsubsection{Conclusion}
If the value of $\delta_{CP}$~is favourable, NO$\nu$A and T2K data will be
able to determine the mass hierarchy before the turn-on of the next-generation
of conventional LBL project. Projects starting early in the next decade, such as LBNO and upgrades to the
first phase of LBNE, will be optimised to search for CP-violation and will
have the capability to pin-down the mass hierarchy with high significance over
the full range of $\delta_{CP}$.
     
\subsubsection{Acknowledgements}
The author would like to thank the organisers of NuMass2013 for an excellent
workshop and acknowledges support for some of the work described here from the UK Science and
Technology Facilities Council and
from the FP7 Research Infrastructure Design Study LAGUNA-LBNO (Grant Agreement
No. 284518 FP7-INFRA-2011-1).

\begin{center}
{\it References}
\end{center}
\begin{description}
\footnotesize	
\item[1] Nova Collab.,The NO$\nu$A TDR, FERMILAB-DESIGN-2007-01 (2007)
\item[2] T2K Collab., Nucl. Instru. Meth. {\bf 659} (2011) 106-135 
\item[3] S.K. Agarwalla {\it et al.}, arXIV:1208.3644 [hep-ph] (2012)
\item[4] A. Stahl {\it et al.}, LBNO Expression of Interest,
  CERN-SPSC-2012-021/SPSC-EOI-007 (2012)
\item[5] The Long-Baseline Neutrino Experiment submission to the European
  Strategy Preparatory Group, (2012)
\item[6] Estimated by minimising  $\Delta \chi^2$ where  $\Delta \chi^2_{min} \geq 2.71 (3.84)$
corresponds to $90\%(95\%)\,$C.L.
%
\end{description}

\clearpage
 \newpage\subsection{A. De R\'ujula and M. Lusignoli: ``The benefits and drawbacks of calorimetry''}
\begin{description}
\it\small 
\setlength{\parskip}{-1mm}
\item[\small\it ADR.:]CERN, IFT(UAM), Madrid, Spain
\item[\small\it ML.:]Sapienza, Universit\`a di Roma, and INFN, Sezione di Roma, Italy
\end{description}
%

In spite of the time elapsed since they were first discussed [1] and theoretically
analyzed in detail [2], the fundamentals of a calorimetric attempt to constrain
or measure the electron-neutrino mass in the decay of $\rm ^{163}Ho$ are
still a subject of puzzlement. New hopes in this direction begin
to materialize and it may be adequate to review the subject.

Consider an ``object": an atom, a molecule --or a full detector-- that $\beta$-decays 
(or inverse $\beta$-decays, in the electron-capture case) with emission of a 
right-handed (or left-handed) electron antineutrino (neutrino). 
Let the $Q$-value be defined, as usual, as the mass difference
between the parent object and the final object in its lowest-energy ``ground" state.
In the cases of interest to the measurement of the neutrino mass, the recoil energy
of the daughter object is totally negligible.

The final object may be left in excited states of energy $E_n$ relative to the ground 
state. In a $\beta$-decay experiment in which the electron energy is measured, energy
conservation implies $Q=E_e+E_\nu+E_n$, so that the electron spectrum is
a superposition of contributions whose end-points are at $E_e=Q-E_n-m_\nu$.
The spectral shape from which $m_\nu$ is to be inferred is complicated.
This is the known and increasingly well understood ``atomic or molecular" problem
[3].

In the electron-capture (EC) case, the elementary processes are more complex.
The daughter object may be left with a hole in the atomic orbital from which
the electron was captured and ``subsequently" suffer a first de-excitation by
the emission of an X-ray, with the result of a hole in a higher orbital. The same hole
may be left over by ``direct" nuclear capture from the higher orbital,
accompanied by an ``internal bremsstrahlung" photon (IBEC). It was Feynman who
first realized that these processes, having the same initial and final states,
are quantum-mechanically indistinguishable.

Extra complications of EC process are that, but for the most bound
atomic orbitals, the dominant hole de-excitations are emissions of other electrons,
called Auger, Koster-Kronig or super-Koster-Kronig transitions, depending
on the orbitals vacated by the process. It may even happen that the initial
capture ``instantaneously" leaves two holes in the daughter object, because
of the mismatch of the atomic orbitals of parent and daughter atoms.
Close to the energies corresponding to orbital energy differences, interferences
between the corresponding ``Breit-Wigner" shapes may be more relevant
than estimated in [2]. The detailed theory of EC is fiendishly complicated [1,4].

The thesis, which we shall substantiate and qualify, is: {\bf In a ``calorimetric" 
measurement of EC, none of the above complications
are relevant to the neutrino-mass-sensitive shape of the endpoint of calorimetric 
energies} [1,2].

\begin{figure}[htbp]
\begin{center}
\includegraphics[width=0.74\textwidth]{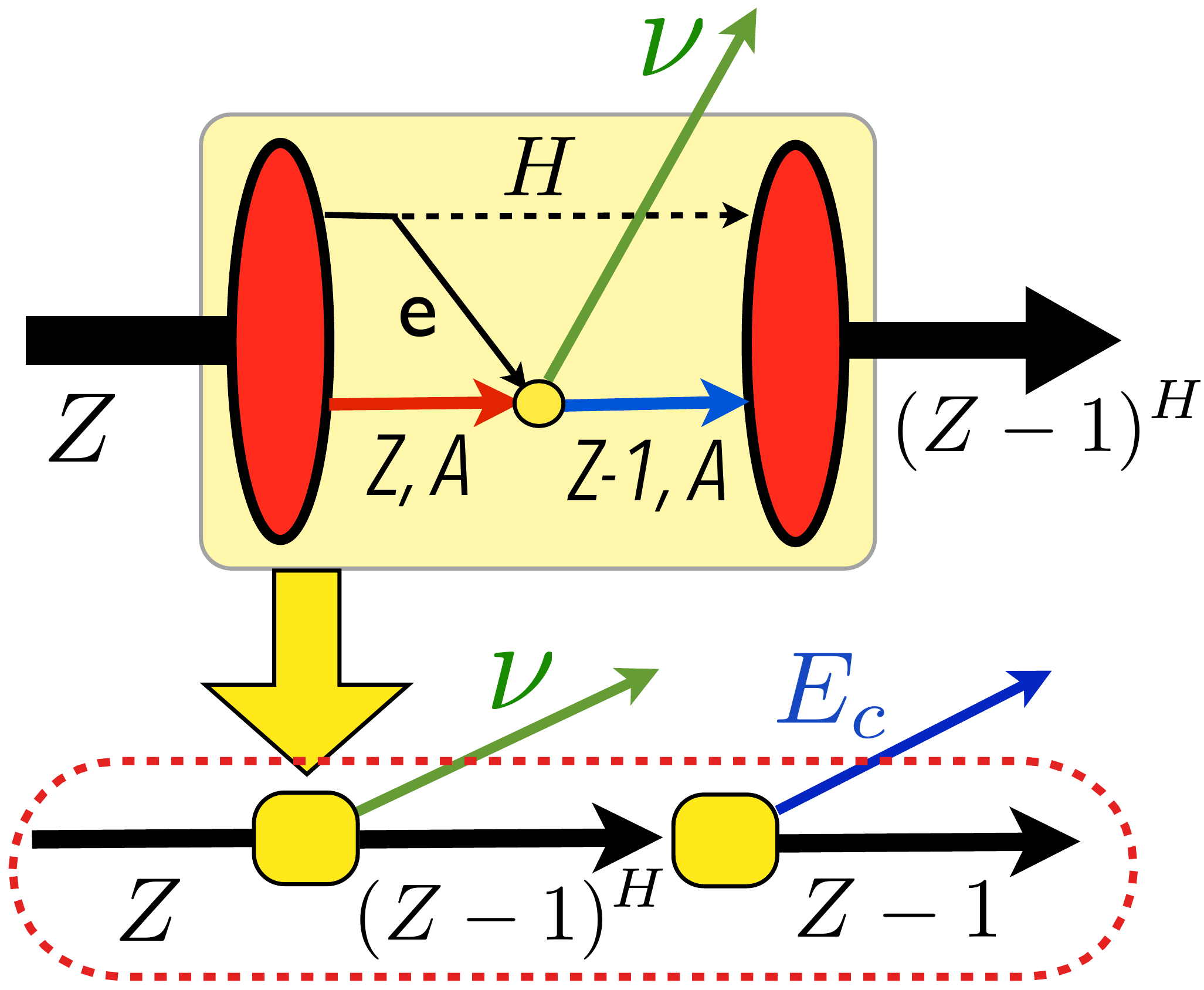}
\caption{Effective theory of electron capture [2]. The upper capsule embodies the
details for the decay into a daughter atom with an electron ``hole" $H$. The lower
(dashed) capsule embodies also the transition to the detector's final ground state. The
calorimetric energy $E_c$ is not meant to escape, but to be converted into a signal.
\label{fig:EffectiveEC}}
\end{center}
\end{figure}

Consider for definiteness an $\rm ^{163}Ho$ atom in a detector, decaying into an 
unstable state, $\rm ^{163}Dy^n$, by the emission of a neutrino, as in the colored
boxed part of 
Fig.~\ref{fig:EffectiveEC}. The subsequent transitions end up in Dy in its ground state: 
$\rm ^{163}Dy^n  \rightarrow {\rm ^{163}Dy}+\it E_c$, where $E_c$ is the
measurable calorimetric energy. Energy conservation implies $Q=E_c+E_\nu$,
{\bf irrespective} of the channels via which Dy de-excited. Here
$Q$ is the mass difference between the detector before ($D_b$) and after ($D_a$) 
the neutrino
(and nothing else) escaped from it. The overall process, $D_b\to D_a+E_c+E_\nu$, is
a three-body decay with kinematics --and the consequent neutrino-mass sensitivity-- 
as simple as the ones of neutron decay in flight.
In this case, it would be useless to worry that it is actually a down quark within the 
neutron that decays into an up quark within the proton. One would simply draw
a sphere around the bound quarks, call $Q=m_n-m_p$, and forget all
nucleons' inner details. In a calorimetric experiment, 
this ``sphere" is the detector, represented by the dashed capsule in Fig.~\ref{fig:EffectiveEC}, 
from
which $E_c$ is not meant to escape, but to be converted into an observable signal.

Implicit in the previous paragraph is the hypothesis that the de-excitation time
of $\rm ^{163}Dy^n$ to its ground state is faster than $\tau_c={\cal O}(10^{-3})$ s, 
a typical duration of a complete single-event energy-collection in current
calorimetric measurements. Atomic excited states having inverse
widths of ${\cal O}(1)$ eV$^{-1}\sim 10^{-15}$ s, this seems to be a safe
expectation, barring the existence of unforeseen metastable final states.

A more serious consideration is the possibility that Ho atoms in the detector
be bound, not to one type of chemical neighbourhood, but to more than one [5].
That would mean that the calorimeter is a sum of detectors with
$Q$-values that may differ by an eventually significant amount.

We argued in [2] that the matrix element for electron capture in $\rm ^{163}Ho$
may be very well approximated in an ``effective" theory extraordinarily simpler
than a first-principle QED approach [1]. The trick consists, as in the Fig.~\ref{fig:EffectiveEC}, in 
approximating the process as a two-step one. First a two-body decay
$\rm ^{163}Ho  \rightarrow {\rm ^{163}Dy^n}+\nu_e$, with $\rm Dy^n$ any
of the relevant daughter states, to be summed over. Second, the
de-excitation $\rm ^{163}Dy^n  \rightarrow {\rm ^{163}Dy}+\it E_c$,
the details of which need not be specified. The differential decay rate is of the form:
\begin{eqnarray}
{d\Gamma\over dE_c}&\propto& (Q-E_c)\sqrt{(Q-E_c)^2-m_\nu^2}\;\sum_n
\varphi_n^2(0)\,{\Gamma_n\over 2\pi}\,{1\over (E_c-E_n)^2+\Gamma_n^2/4}
\nonumber\\
&\rightarrow& {\cal K}\;(Q-E_c)\sqrt{(Q-E_c)^2-m_\nu^2}\;\;{\rm close\,to\,the\,endpoint.}
\label{eq:resonances}
\end{eqnarray}
with a common endpoint at $E_c=Q-m_\nu$ for all $n$ and
${\cal K}$ constant, to a level of precision to be discussed below.
The explicit form in terms
of wave functions at the origin $\varphi_n(0)$ and widths $\Gamma_n$ is indicative,
in practice they ought to be substituted by the observed widths and spectral peak ratios,
for a precise description of the calorimetric spectrum.

At first sight the expression \ref{eq:resonances} does not cover processes such
as the ``instantaneous" production of a final state with two vacancies ($l$ and $m$)
in the daughter Dy atom:
\begin{equation}
{\rm Ho}\rightarrow {\rm Dy}^{l,m}+e+\nu,
\label{eq:2holes}
\end{equation}
a three-body decay with the customary extended phase space for the distribution
of electron energies. But the process is quantum-mechanically identical to
another ``classical" view of its interpretation.
Namely electron capture leaving a hole in an orbital $n$, followed by an Auger or Koster-Kronig transition in which the hole migrates to $m$ and an $l$ electron is ejected.
As an example, consider $n=\rm MI$, $m=\rm MII$, $l=\rm NI$. This later process is resonant in that
the ejected electron spectrum peaks at the mass difference between Dy[MI] and Dy[MII,NI]
and the total calorimetric energy peaks at the Ho - Dy[MI] mass difference. As for all processes, the calorimetric energy, as merely dictated by energy conservation, extends all the way to its 
endpoint at  $Q-m_\nu$.

There is only one relevant process not subject to the two-fold ``classical" interpretation we just
discussed: the ``instantaneous" decay 
\begin{equation}
{\rm Ho}\rightarrow {\rm Dy^{MI,NI}}+e+\nu.
\label{eq:2darnedholes}
\end{equation}
The process is possible thanks to the slightly incomplete overlap between the
wave function of the NI electron in Ho and in Dy with an MI vacancy. The
charge that the corresponding electrons feel is the same, but the charge
distribution is slightly different, since, as ``seen" by an NI electron,
 the MI electron in Ho does not
completely screen the extra proton that Ho has, relative to Dy.
The decay channel of Eq. \ref{eq:2darnedholes} is non-resonant and has
a negligible rate relative to the resonant processes reflected in Eq.\ref{eq:resonances}.
More importantly, the shape of its calorimetric endpoint is, once again, that of the second
line of Eq.~\ref{eq:resonances}.

Yet another sense in which Eq.~\ref{eq:resonances} is ``classical" is that it is a sum
of squared amplitudes and not an amplitude-sum squared. The neglected interferences
must be small, as discussed in detail in [2], because the dominant decay channels for Dy 
with different electronic holes are different and consequently non-interfering. The 
interferences, explicitly calculated in [2], are at most at the
percent level close to the spectral endpoint as described by Eq.~\ref{eq:resonances}.
They are made even less relevant by the argument in the next paragraph.

\begin{figure}[htbp]
\begin{center}
\includegraphics[width=0.50\textwidth]{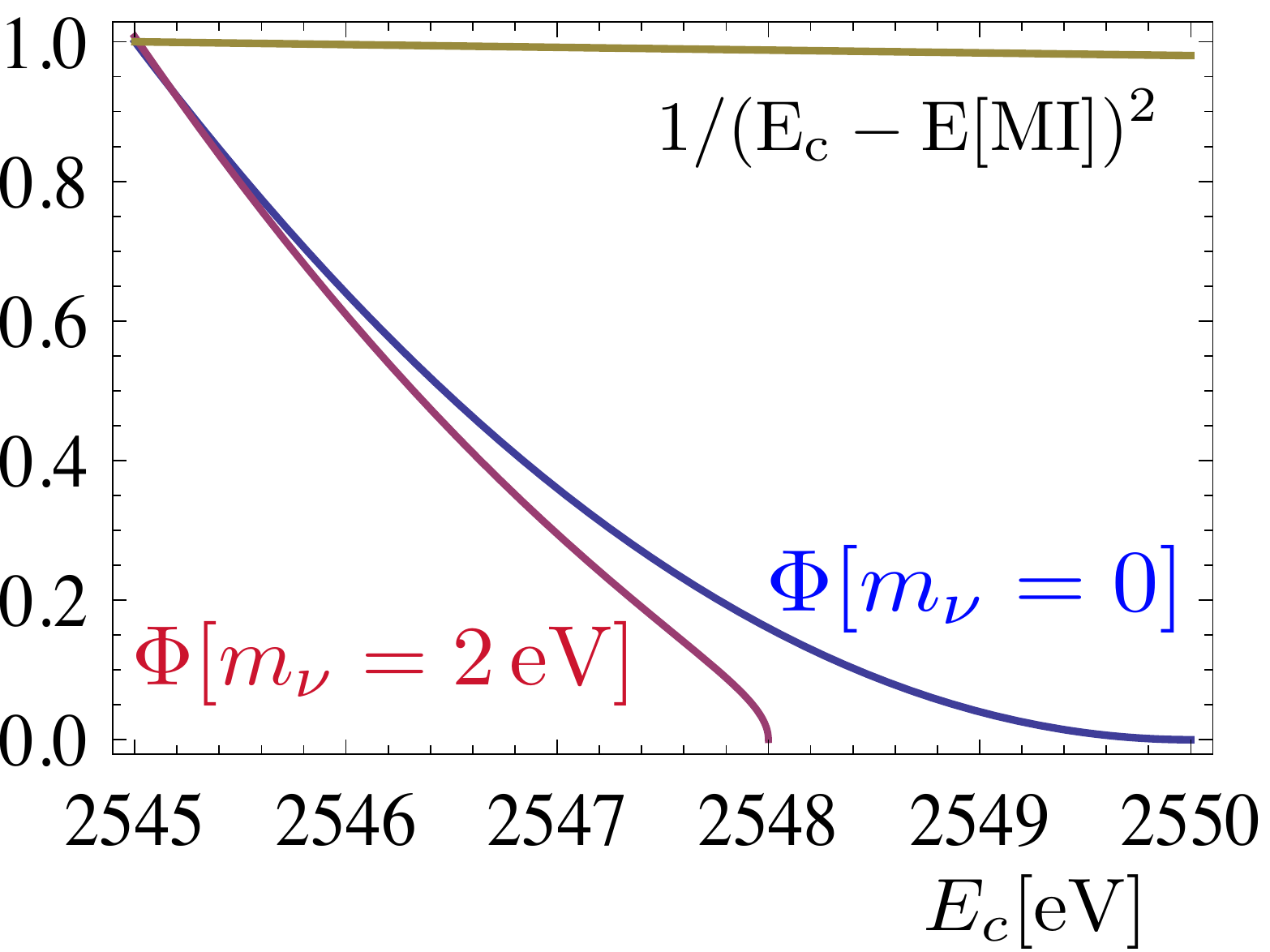}
\caption{Shapes at the endpoint of the $E_c$ spectrum in Ho decay, with an assumed
$Q=2.55$ keV. The $\Phi[m_\nu]$ lines are the phase-space function for two choices of
$m_\nu$. The line above them reflects the amount of energy-dependence expected
for the squared matrix element. All curves are normalized to unity at the lowest $E_c$
in the figure.
\label{fig:Endpoints}}
\end{center}
\end{figure}

An important question is the range of the largest $E_c$ values for which $\cal K$ in 
Eq. \ref{eq:resonances} may in practice be taken to be a constant. The answer
depends on the $^{163}$Ho decay $Q$-value, still insufficiently well measured. 
Consider the example $Q=2.55$ keV and recall that $E_n\approx 2.05$ keV for
$n=\rm MI$ Dysprosium, the state of closest energy to the endpoint. In Fig.~\ref{fig:Endpoints}
we have plotted the phase-space factor of Eq.~\ref{eq:resonances} 
for $m_\nu=0$ and  $m_\nu=2$ eV,
as well as the squared matrix element, whose variation near the endpoint is
essentially that of the function $1/(E_c-E\rm [MI])^2$, with all curves normalized
at the lowest $E_c$ in the plot. The point that this figure conveys is that the variation of
the matrix element, thus simplified or not, is governed by atomic singularities located
at the electron binding energies in Dy, as dictated by arguments as general as causality
and analyticity. The precise absolute value of the matrix element may be hard to compute,
but its variation cannot be large enough to be relevant in practice, unless the Q value
happened to fall within a few widths of a resonance. Otherwise, the figure speaks for itself.

Relative to the measurements of Tritium $\beta$-decay, $^{163}$Ho
calorimetry has its drawbacks. One of them is that
the spectrum is unavoidably measured over its entire range, very significantly reducing 
the fraction of neutrino-mass sensitive events and calling for ``farms" of calorimeters.
Another one is that the effort and advanced technical tools required
to be competitive are being developed only recently, with decades of delay
relative to $\beta$ decay. But micro-calorimeters have the irresistible aesthetic advantage
of being tiny contraptions to measure a tiny mass!

%
\begin{center}
{\it References}
\end{center}
\begin{description}
\footnotesize	
\item[1] A. De R\'ujula, Nucl.Phys. {\bf B188} (1981) 414
\item[2] A. De R\'ujula and M. Lusignoli, Phys.Lett. {\bf B118} (1982) 429
\item[3] See, for instance, C. Weinheimer, these proceedings.
\item[4] B.A. Zon and L.P. Rapoport, Sov. J. Nucl. Phys. {\bf 7} (1968) 330;
B.A. Zon, Sov. J. Nucl. Phys. {\bf 13} (1971) 554
\item[5] We are indebted to M. W. Rabin for this warning.
\end{description}

 \newpage
\subsection{N. Wandkowsky: ``Commissioning of the KATRIN main spectrometer''}
\begin{description}
\it\small 
\setlength{\parskip}{-1mm}
\item[\small\it N.W.:]Institute for Nuclear Physics, Karlsruhe Institute of Technology, Germany
\end{description}
\subsubsection{Introduction}

The Karlsruhe Tritium Neutrino (KATRIN) experiment~[1] is designed to determine the effective electron antineutrino mass $m_{\bar{\nu}_{e}}$ with a sensitivity of 0.2~eV/c$^{2}$ in a model-independent way.
Investigating the kinematics of tritium $\beta$-decay close to the endpoint $E_{0} = 18.6$~keV gives access to $m_{\bar{\nu}_{e}}$.
However, only a fraction of $10^{-13}$ of all decay electrons is emitted with energies in the interesting region.
Therefore, the background rate has to be kept sufficiently low ($<10^{-2}$ counts per second (cps)).

\begin{figure}[ht!]
 \centering
    \includegraphics[width=0.8\textwidth]{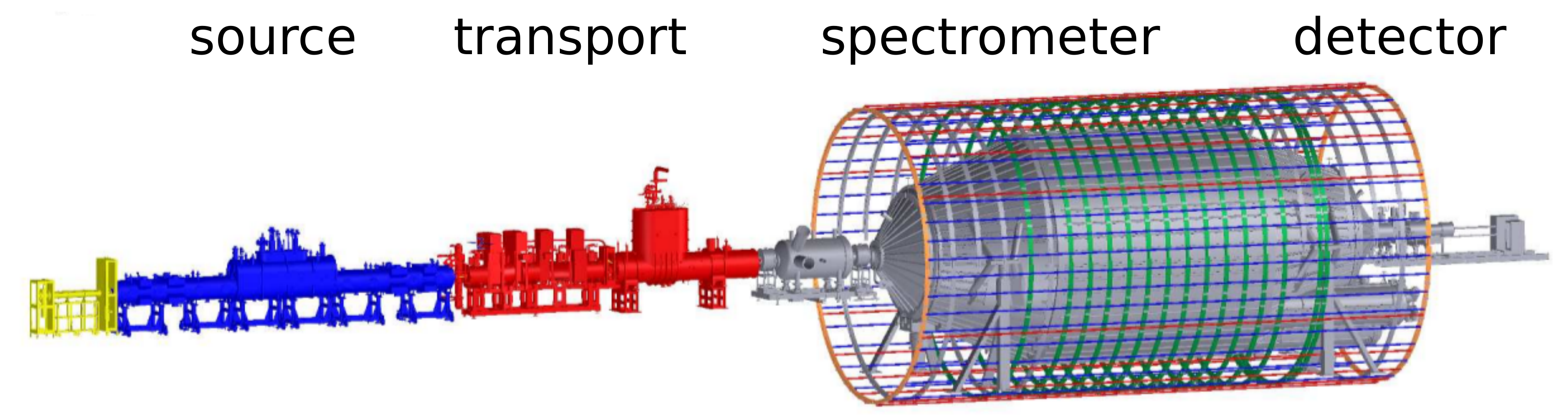}
 \caption{Overview of the KATRIN exprimental setup: decay electrons are guided from the source through the transport section (responsible for tritium removal) to the spectrometer (responsible for electron energy analysis). Electrons with sufficient kinetic energy can pass the potential barrier and are counted at the detector.}
 \label{fig:KATRINSetup}
\end{figure}

The decay electrons are guided by strong magnetic fields (up to 6~T) from the source, through the transport section towards the spectrometer section of the experiment, which is responsible for the energy analysis (see fig.~\ref{fig:KATRINSetup}).
The main spectrometer working principle is based on the MAC-E (Magnetic Adiabatic Collimation and Electrostatic) filter~[2].
Superconducting magnets at the entrance and exit of the spectrometer provide a magnetic guiding field which decreases rapidly towards the spectrometer center ($B_{\text{max}}=6$~T, $B_{\text{min}}=3\cdot 10^{-4}~$T).
Electrons with a kinetic energy $E_ {\text{kin}}$ perform a cyclotron motion around the magnetic field lines.
When propagating to the low field region, the magnetic gradient transforms the signal electrons' transversal energy component $E_{\perp}$ into longitudinal energy $E_{||}$ ($E_ {\text{kin}} = E_{\perp}+E_{||}$).
This is essential because the electrostatic potential $U_{0}$ which is applied to the spectrometer vessel only affects $E_{||}$.
As the transformation $E_{\perp}\rightarrow E_{||}$ cannot be complete for an isotropic source, the corresponding energy resolution is
$$
\Delta E = E_{0}\cdot\frac{B_{\text{min}}}{B_{\text{max}}} = 0.93~\text{eV}.
$$
Only those electrons with $E_{||}>qU_{0}$ (q being the signed electron charge) are able to pass the electrostatic potential barrier and will be counted at the focal plane detector.

In general, there are several sources of background in a MAC-E filter, which will be described in section~\ref{sec:sources}.
Section~\ref{sec:mechanism} will describe the background mechanisms and will show that, without appropriate countermeasures (section~\ref{sec:suppression}), the background would exceed the design limit of $<10^{-2}$~cps.
The focus will be the current status of the KATRIN main spectrometer with regard to the above mentioned topics.

\subsubsection{Background sources}
\label{sec:sources}

\begin{figure}[ht!]
 \centering
    \includegraphics[width=0.8\textwidth]{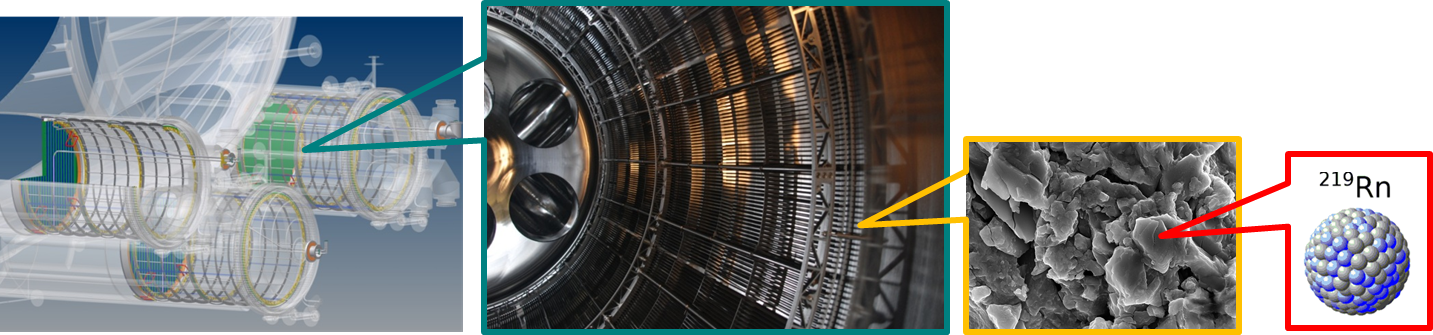}
 \caption{The porous medium of the getter pump, which is installed in the three pump ports, is a source for radon atoms, which can enter the spectrometer volume and produce background when undergoing nuclear $\alpha$-decay.}
 \label{fig:RadonSource}
\end{figure}

There are three main sources of background in a MAC-E filter
\begin{enumerate}
  \item electrons from Penning traps,
  \item secondary electrons from the vessel surface, and
  \item stored electrons in the spectrometer volume.
\end{enumerate}
\textbf{Penning traps}: Careful design considerations concerning the electro-magnetic layout of the main spectrometer can avoid the occurrence of harmful Penning traps.
The corresponding design of electrode and magnet components was carried out taking into account all the experiences from the Mainz experiment~[3] and from the pre-spectrometer test experiment~[4].
Therefore, we are confident to have minimized Penning traps so that they are of no concern for background within KATRIN.\\
\textbf{Secondary electrons}: Secondary electrons are created if muons (present in cosmic radiation) or $\gamma$'s (produced by radioactive decays) interact with the stainless steel surface of the spectrometer.
If these electrons penetrate into the sensitive volume of the spectrometer -- the flux tube -- they can contribute to the background.\\
\textbf{Stored electrons}: Electrons can also be created directly within the flux tube, i.e.~as a result of decay processes of radioactive atoms.
Measurements at the pre-spectrometer~[5] revealed a radon-induced background component, while measurements in Mainz~[3] and Troitsk~[6] suffered to some extent from tritium-induced background.
\begin{itemize}
  \item Tritium-induced: If tritium from the source would penetrate to the spectrometer, it could decay there, emitting electrons with a broad energy spectrum (0-18.6~keV).
  \item Radon-induced: It was shown that ${}^{219}$Rn from the non-evaporable getter (NEG) material, which is used as a chemical pump for hydrogen, contributed significantly to the background rate when undergoing $\alpha$-decay in the spectrometer volume (see fig.~\ref{fig:RadonSource}).
  Furthermore, the vessel itself and auxiliary equipment attached to it was identified as a source for ${}^{219,220}$Rn due to the natural abundance of radioactive isotopes in structural materials.
  The isotope ${}^{222}$Rn is rather long-lived ($\tau=5.5$~d), and thus does not contribute significantly to the background due to the short pump-out times on the order of minutes.
  Although radon is an $\alpha$-emitter, there are a variety of processes generating electrons during the decay: conversion, inner shell shake-off, relaxation (Auger process) or shell reorganization.
  Details about these processes can be found in~[7].
  The electron energy spectrum ranges from a few~eV up to several hundred~keV.
\end{itemize}

\subsubsection{Background processes in MAC-E filters}
\label{sec:mechanism}

\begin{figure}[ht!]
 \centering
    \includegraphics[width=0.6\textwidth]{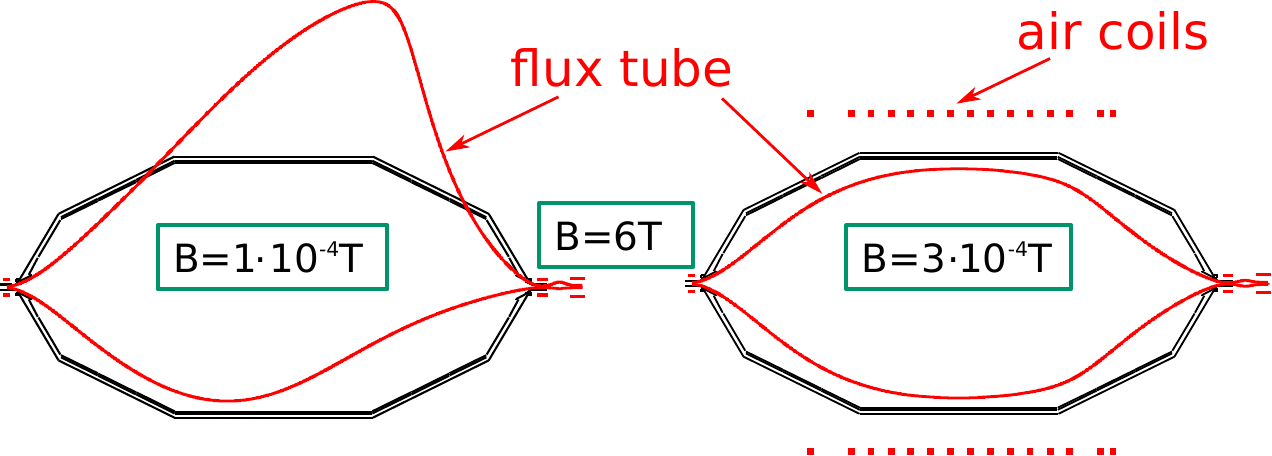}
 \caption{Background suppression by magnetic shielding. Left: The flux tube without air coil system touches the vessel wall, which reduces the signal and increases the background rate. Right: The symmetric flux tube, which is created by the large volume air coil system, suppresses background from the vessel walls.}
 \label{fig:LFCSBackground}
\end{figure}

\begin{figure}[ht!]
 \centering
    \includegraphics[width=0.3\textwidth]{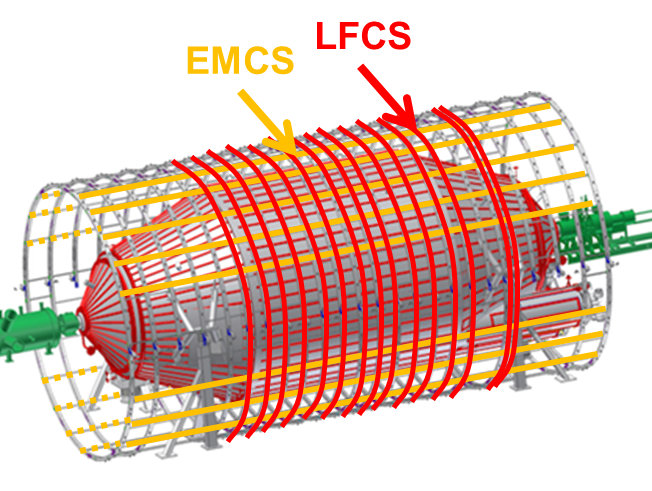}
    \includegraphics[width=0.3\textwidth]{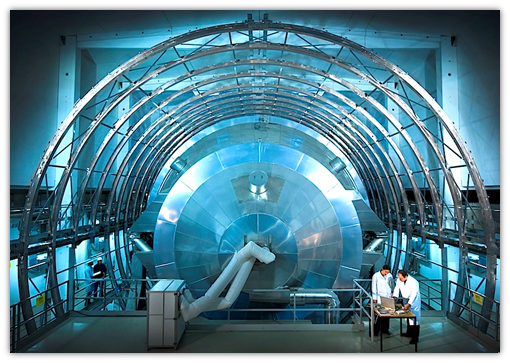}
 \caption{Air coil system. Left: Schematic drawing of EMCS (earth magnetic field compensation system) and LFCS (low field correction system), surrounding the main spectrometer. Right: Photograph of the installed systems.}
 \label{fig:AirCoil}
\end{figure}

Having identified several sources of electrons, it is important to understand the mechanism of background production in the complex electromagnetic configuration of the main spectrometer.\\
\newline
\textbf{Secondary electrons}: Electrons, which are created at the vessel surface, only contribute to the background if they enter the flux tube, as the magnetic field restricts the electron motion to cyclotron paths along the field lines.
Fig.~\ref{fig:LFCSBackground} (left) shows the flux tube, which is created by the superconducting solenoids and the earth magnetic field.
The fact that field lines directly touch the vessel wall has two negative effects:
\begin{itemize}
  \item signal electrons, which follow these field lines are guided to the vessel wall, which results in reduced statistics,
  \item background electrons, which are created at the positions where these field lines connect to the walls, can be guided directly to the detector.
\end{itemize}
Therefore, the flux tube is shaped and adjusted by two large volume air coil systems~[8] (see fig.~\ref{fig:AirCoil}):
\begin{itemize}
  \item EMCS (earth magnetic field compensation system): Two systems, consisting of vertical and horizontal current loops which span the length of the main spectrometer, are responsible for the compensation of the vertical and horizontal (perpendicular to beam direction) earth magnetic field. The resulting magnetic field configuration shows symmetry around the beam axis. However, the flux tube still extends beyond the spectrometer boundaries.
  \item LFCS (low field correction system): A system of vertical coils around the beam axis was installed to fit the flux tube into the spectrometer (see fig.~\ref{fig:LFCSBackground} (right)). In addition to delivering a certain necessary field strength, the flexibility of the system (14 independent coils) allows to fine tune the field shape to further suppress background and to improve the signal electron guiding through the MAC-E filter.
\end{itemize}
A photograph of the successfully installed air coil system is shown in fig.~\ref{fig:AirCoil} (right).
This system will allow to reduce background from secondary electrons by a factor of $\sim10^{4}-10^{5}$, depending on the strength of magnetic field disturbances, which break the axial symmetry and enhance the motion of background electrons into the flux tube.

\begin{figure}[ht!]
 \centering
    \includegraphics[width=0.30\textwidth]{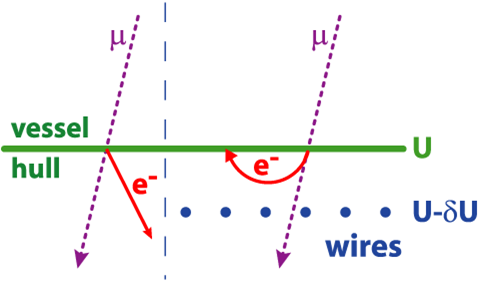}
    \includegraphics[width=0.25\textwidth]{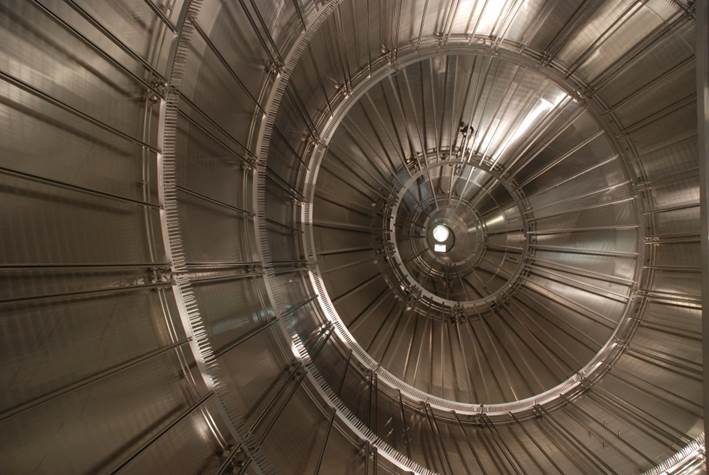}
 \caption{Background suppression by electrostatic shielding. Left: Working principle: The more negative potential of the wire electrode rejects electrons with energies smaller than the potential difference between vessel and wire layer. Right: Photograph of the installed wire modules.}
 \label{fig:IEBackground}
\end{figure}

\begin{figure}[ht!]
 \centering
    \includegraphics[width=0.5\textwidth]{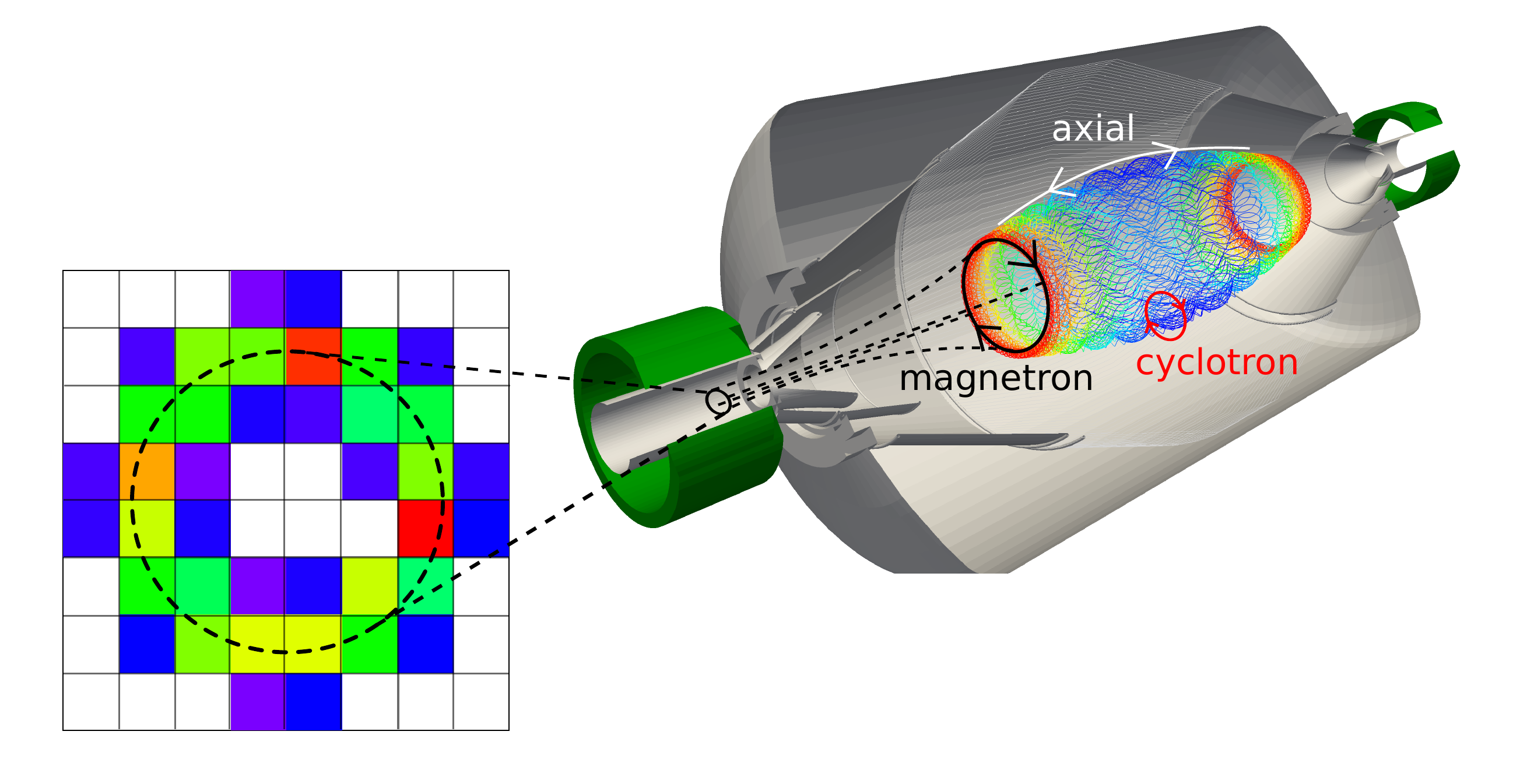}
 \caption{Simulation of a stored electron~[10]. On its path (composed of cyclotron, axial and magnetron motion) a stored electron can ionize residual gas molecules, creating low-energy secondary electrons, which leave the magnetic mirror trap. A stored primary produces up to several thousands of secondaries which leads to a characteristic ring pattern on the detector.}
 \label{fig:TrappedElectrons}
\end{figure}

In addition to the dominant magnetic shielding of secondary electrons from the wall, also electrostatic shielding is applied.
The whole $690~\text{m}^{2}$ inner surface of the spectrometer vessel is covered by a two-layer wire electrode system (see fig.~\ref{fig:IEBackground}).
The working principle is shown in fig.~\ref{fig:IEBackground} (left): A more negative potential, which is applied to the wire electrode, repels electrons with energies smaller than the potential difference between vessel and wire layer.
A two-layer system is even more advantageous since the inner layer (using even thinner wires than the outer layer) will shield the outer layer and the wire module holding structures.
The installation of the system, shown in fig.~\ref{fig:IEBackground} (right), was completed mid 2012.
By this method, the background is expected to be reduced by an additional factor $\sim$100.\\
\newline
\textbf{Stored electrons}: Electrons can also be created inside the transported flux tube.
When flying towards the increasing magnetic field at the entrance and exit region of the spectrometer, the longitudinal momentum of electrons gets transformed into transversal momentum.
Electrons are reflected by the magnetic mirror if their longitudinal momentum vanishes.
Generally, all electrons generated in the spectrometer volume with a starting transversal energy component $E_{\perp}>1$~eV are stored, which results in a large storage probability ($>90\%$) for electrons accompanying tritium and radon decays.\\
The motion of stored electrons is composed of a fast cyclotron motion around the magnetic field line, an axial motion between the magnetic mirrors and a slow magnetron motion due to the $\textbf{E}\times\textbf{B}$ and $\nabla\textbf{B}\times\textbf{B}$ drifts, as shown in fig.~\ref{fig:TrappedElectrons}.
During this motion, they can ionize residual gas molecules and create secondary electrons of low energy which can thus leave the magnetic mirror trap.
Depending on the starting kinetic energy, a single primary electron can produce up to several thousands of secondary electrons.
Due to the distinct motion of the primary electron, the corresponding secondaries produce a characteristic ring structure at the detector.
Although this feature in principle would allow to remove this background component by appropriate cuts, this is not feasible due to the extremely long storage times of up to several hours for a single primary electron (a result of the low interaction probability in an ultra high vacuum of $10^{-11}$~mbar).
Extrapolations from the pre-spectrometer to the much larger main spectrometer~[9] imply an expected background rate of $\sim6\cdot10^{-2}$~cps, which exceeds the design limit of $1\cdot10^{-2}$~cps, and which would reduce the statistical neutrino mass sensitivity from 0.15~eV to 0.38~eV, if no countermeasures are taken.

\subsubsection{Background suppression techniques}
\label{sec:suppression}

Although the MAC-E filter offers intrinsic background reduction techniques (magnetic and electrostatic shielding), these only affect electrons which are created outside of the flux tube.
As pointed out above, the background due to stored electrons could severely influence the neutrino mass sensitivity.
Therefore, additional counter measures have to be applied.
The most effective way to remove this background contribution is to prevent the radioactive atoms from entering the main spectrometer in the first place.
This \emph{passive} background reduction technique is realized by a LN$_{2}$ cooled baffle, which blocks the line-of-sight from the getter pump (where radon is being emitted) to the spectrometer volume.
Radon will stick to the cryo-surface and decay there (sticking probability $>90\%$).
\begin{figure}[ht!]
 \centering
    \includegraphics[width=0.5\textwidth]{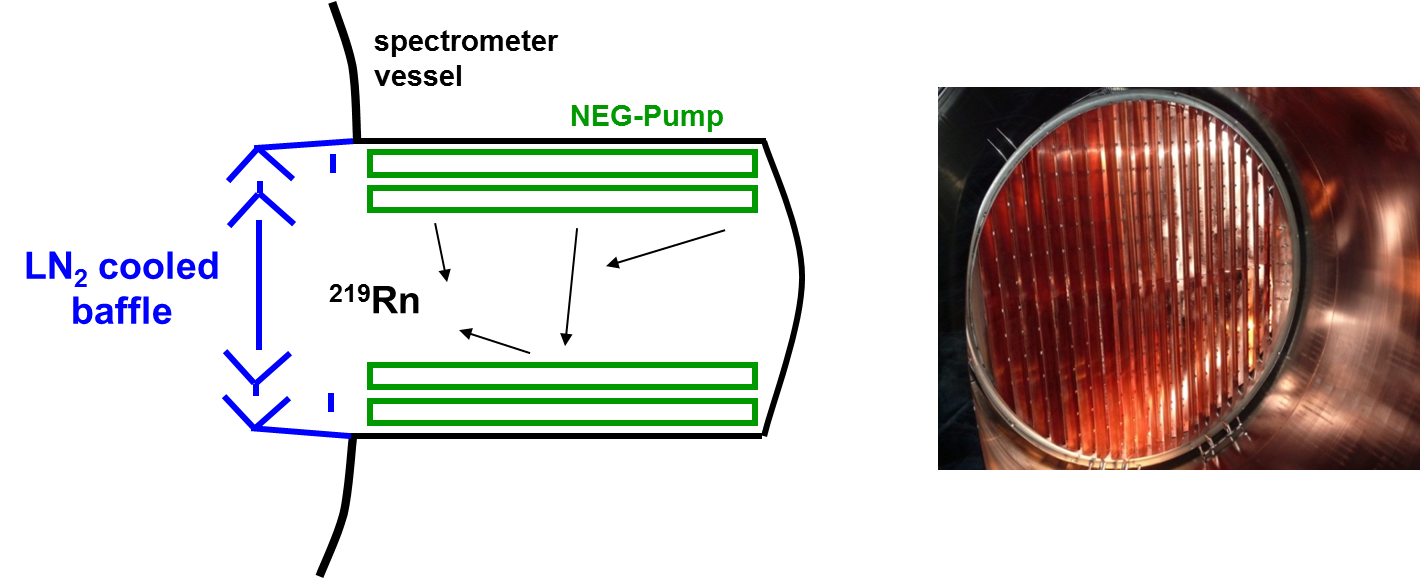}
 \caption{Passive background suppression. Left: Schematic drawing of the working principle of a LN$_{2}$ cooled baffle. Radon atoms stick to the cryo-surface and decay before they can enter the spectrometer. Right: Photograph of the installed baffle.}
 \label{fig:PassiveBackground}
\end{figure}
Tritium atoms, however, will barely be affected by the baffle because of its LN temperature range.
Consequently, additional \emph{active} background reduction techniques have to be applied.
Three promising methods are under investigation (see fig.~\ref{fig:ActiveBackground}):

\textbf{Electric dipole}: The inner electrode system is subdivided electrically into two half-shells which can be put on different potentials. The resulting electric field acts on the stored electrons ($\textbf{E}\times\textbf{B}$) and distorts their path in radial direction. If the dipole field is strong enough, electrons are drifted to the walls where they are absorbed. In the main spectrometer, a dipole field of 100~V/m can be applied which is sufficient to remove electrons with energies below $\sim$1~keV.

\textbf{Electron Cyclotron Resonance}: Instead of operating the inner electrode system in a static dipole mode, an applied high-frequency (HF) will stochastically heat electrons if in resonance with the HF field (a phenomenon well known in plasma physics) until they eventually hit the vessel surface~[11]. The advantage of this method is that it can efficiently remove electrons of all energies. However, the technical realization is more complicated and has to ensure the integrity of the inner electrode system.

\textbf{Magnetic pulse}: As stored electrons are guided magnetically, bending the magnetic field lines towards the spectrometer walls turns out to efficiently remove electrons of all energies from the volume. The appropriate manipulation of the magnetic field is easily possible with the air coil system described above. However, as pointed out, secondary electrons from the walls can in turn enter the flux tube and contribute to the background. The efficiency of this method therefore strongly depends on the relative contribution of stored electrons and muon-induced electrons to the overall spectrometer background.
\begin{figure}[ht!]
 \centering
    \includegraphics[width=0.7\textwidth]{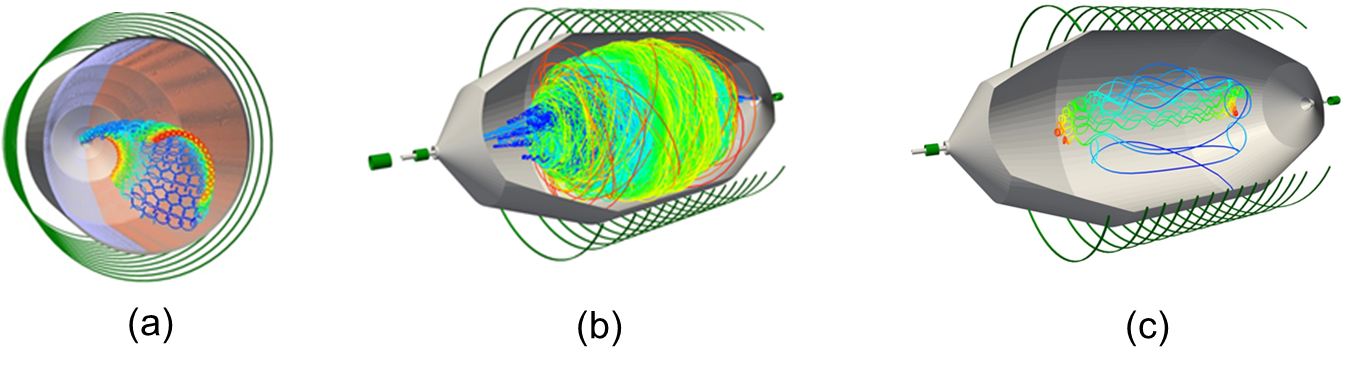}
 \caption{Active background suppression techniques. (a) Electric dipole, (b) Electron Cyclotron Resonance, (c) Magnetic Pulse. Details on the different methods can be found in the main text.}
 \label{fig:ActiveBackground}
\end{figure}
It has to be noted that these methods only affect already stored electrons.
Because these electrons start to produce background (via ionization) immediately after their creation, the methods discussed above are ideally applied with a frequency not exceeding the duration between two succeeding ionization events.
Since the methods also affect the signal electrons, data taken during the active background removal period have to be rejected.
To keep this ``dead time'' as low as possible, the interaction probability of the primary stored electrons has to be reduced.
This is achieved by improving the vacuum in the spectrometer to a level of $\mathcal{O}(10^{-11})$~mbar.

\begin{figure}[ht!]
    \includegraphics[width=0.4\textwidth]{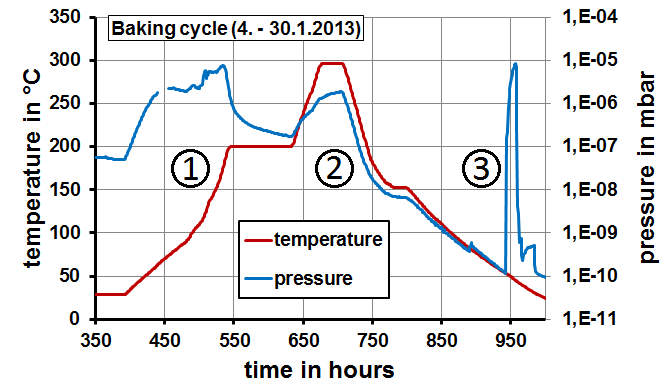}
    \includegraphics[width=0.4\textwidth]{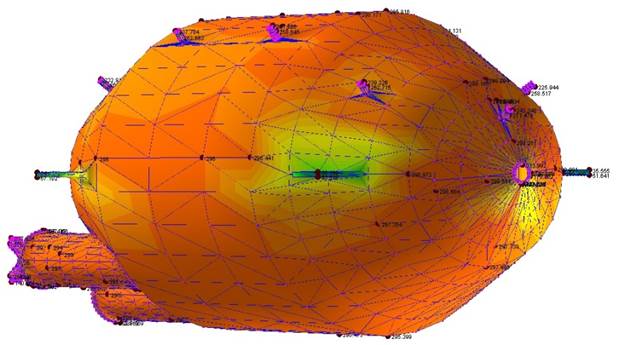}
 \caption{Bake-out of the main spectrometer. Left: Bake-out procedure. (1): Heating to $200^{\circ}$C to remove water from the vessel surface; (2): Heating to $300^{\circ}$C to activate the getter pump; (3) Cool-down and leak development. Right: Temperature profile at $200^{\circ}$C with inhomogeneities $<1^{\circ}$C on the main surface.}
 \label{fig:BakeOut}
\end{figure}

Due to an exposure to ambient air during the installation work within the main spectrometer, outgassing of the vessel surface is dominated by water.
Heating up to temperatures $\sim200^{\circ}$C is most efficient in cleaning the vessel surface.
This so-called ``bake-out'' was performed early 2013.
The procedure is shown in fig.~\ref{fig:BakeOut} (left).
Period (1) aimed to remove the majority of the adsorbed water, which is confirmed by the corresponding pressure rise. 
Period (2) was needed to ``activate'' the chemical getter pump (i.e. clean the getter surface).
During cool-down (period (3)), the pressure curve follows the temperature curve as expected.
At a temperature of about 50$^{\circ}$, a leak developed at one of the $\sim120$ flanges.
However, after fixing the leak, the pressure nearly reached back to the former value.
Fig.~\ref{fig:BakeOut} (right) shows the temperature distribution on the vessel surface at a temperature of $\sim200^{\circ}$C.
The maximum deviation on the main surface is $<1^{\circ}$C.
This is essential to avoid any mechanical stress, especially on the fragile inner electrode system.\\
The current pressure of $\sim10^{-10}$~mbar is about an order of magnitude larger than the design value, which will however allow a detailed investigation of background and transmission characteristics of the main spectrometer.

\subsubsection{Conclusion}
\label{sec:conclusion}

KATRIN is targeted to determine the effective electron antineutrino mass $m_{\bar{\nu}_{e}}$ with a sensitivity of 0.2~eV/c$^{2}$.
A central design requirement is a background rate $<10^{-2}$~cps.
The main background sources are muon- or $\gamma$-induced secondary electrons from the wall and radon- or tritium-induced stored electrons in the volume of the spectrometer.
Secondary electrons are largely suppressed (reduction factor $10^{5}$) by the successfully installed external air coil and inner electrode systems.
A reduction of stored electron background is achieved by \emph{passive} and \emph{active} techniques.
Cryo-cooled baffles prevent the radon from entering the spectrometer in the first place, while application of the electric dipole, of the electron cyclotron resonance or of the magnetic pulse will efficiently remove stored primary electrons.
In combination with the ultra high vacuum conditions ($\leq 10^{-10}$~mbar), a background reduction below the design limit seems possible.

\begin{center}
{\it References}
\end{center}
\begin{description}
\footnotesize	
\item[1] J.~Angrik {\it et al.} (KATRIN Collaboration), KATRIN Design Report, Wissenschaftliche Berichte FZKA 7090 (2004)
\item[2] A.~Picard {\it et~al.}, A solenoid retarding spectrometer with high resolution and transmission for keV electrons, Nucl. Instrum. Meth. B \textbf{63} (1992) 3.
\item[3] J. Bonn {\it et~al.}, The mainz neutrino mass experiment, Nucl. Phys. B  \textbf{91} (2001) 1-3.
\item[4] F. Fr{\"a}nkle, Background Investigations of the KATRIN Pre-Spectrometer, PhD thesis, KIT, 2010.
\item[5] F. Fr{\"a}nkle {\it et~al.}, Radon induced background processes in the KATRIN pre-spectrometer, Astropart. Phys. \textbf{35} (2011) 128.
\item[6] V.~M.~Lobashev {\it et~al.}, Direct search for neutrino mass and anomaly in the tritium beta-spectrum: Status of Troitsk neutrino mass experiment, Nucl. Phys. B Proc. Suppl. \textbf{91} (2001) 280.
\item[7] N.~Wandkowsky {\it et~al.}, Modeling of electron emission processes accompanying Radon-$\alpha$-decays within electrostatic spectrometers, \emph{to be published}.
\item[8] F. Gl{\"u}ck {\it et~al.}, Electromagnetic design of the KATRIN large-volume air coil system, \emph{to be published}.
\item[9] S.~Mertens {\it et~al.}, Background due to stored electrons following nuclear decays at the KATRIN experiment, Astropart. Phys. \textbf{41} (2013) 52
\item[10] N. Wandkowsky {\it et~al.}, Validation of a model for Radon-induced background processes in electrostatic spectrometers, \emph{to be published}.
\item[11] S.~Mertens {\it et~al.}, Stochastic Heating by ECR as a Novel Means of Background Reduction in the KATRIN spectrometers, JINST \textbf{7} (2012) P08025.
\end{description}

 \newpage\subsection{K. Blaum, A. Doerr, C. E. Duellmann, K. Eberhardt, S. Eliseev, C. Enss, A. Faessler, A. Fleischmann, L. Gastaldo$^*$, S. Kempf, M. Krivoruchenko, S. Lahiri, M. Maiti, Yu.N. Novikov, P. C.-O. Ranitzsch, F. Simkovic, Z. Szusc, M. Wegner  : ``The Electron Capture $^{163}$Ho Experiment ECHo''}
\begin{description}
\it\small
\setlength{\parskip}{-1mm}
\item[\small\it K.B., A.D., S.E.:]Max-Planck Institute for Nuclear Physics Heidelberg, Germany
\item[\small\it C.E.D.:]1) Institute for Nuclear Chemistry, Johannes Gutenberg University Mainz, Germany\\
2) GSI Helmholtzzentrum für Schwerionenforschung, 64291 Darmstadt, Germany\\
3) Helmholtz Institute Mainz, 55099 Mainz, Germany 
\item[\small\it K.E.:]1) Institute for Nuclear Chemistry, Johannes Gutenberg University Mainz, Germany\\
2) Helmholtz Institute Mainz, 55099 Mainz, Germany
\item[\small\it C.E., A.F., L.G., S.K., P.C.O.R., M.W.:]Kirchhoff Institute for Physics, Heidelberg University, Germany
\item[\small\it A.F.:] Institute for Theoretical Physics, University of Tuebingen, Germany
\item[\small\it M.K.:]Institute for Theoretical and Experimental Physics Moscow, Russia
\item[\small\it S.L.:]Saha Institute of Nuclear Physics, Kolkata, India
\item[\small\it M.M.:]Department of Physics, Indian Institute of Technology Roorkee, India
\item[\small\it Y.N.:]Petersburg Nuclear Physics Institute, Russia
\item[\small\it F.S.:]Department of Nuclear Physics, Comenius University, Bratislava, Slovakia
\item[\small\it Z.S.:]Institute of Nuclear Research of the Hungarian Academy of Sciences
\item[$^*$] Corresponding author, Loredana.Gastaldo@kip.uni-heidelberg.de
\end{description}
%

The Electron Capture $^{163}$Ho Experiment, ECHo, has the aim to investigate the electron neutrino mass in the energy range below $1\,$eV by a high precision and high statistics calorimetric measurement of the $^{163}$Ho electron capture (EC) spectrum.  $^{163}$Ho decays by capturing an electron from the inner atomic shells to an excited state of the $^{163}$Dy atom with a half-life of $\tau_{1/2}\approx 4570$ years and a recommended value for the energy available to the decay $Q_{\mathrm{EC}}\approx \, 2.5\,$keV which still has a large uncertainty since different experiments have measured values between 2.3 and 2.8 keV, as discussed in Section \ref{QEC}. The atomic de-excitation is a complex process which includes cascades of both x-rays and electron emissions (Auger electrons and Coster-Kronig transitions). By performing a calorimetric measurement of the de-excitation spectrum, i.e. by measuring for each event the sum of the energies of all emitted photons and electrons wit
 h the same detector, the sensitivity to the neutrino mass is increased. In order to perform a calorimetric measurement of the $^{163}$Ho EC spectrum, the $^{163}$Ho source has to be: i) part of the sensitive volume of detector in order not to have energy losses in the source that are not measured, ii) homogeneously distributed so that the detector response has not position dependent effects and iii) completely contained in the detector in order to ensure a quantum efficiency for the emitted particles of $100\%$.

In [1], several scenarios to reach the sub-eV sensitivity for the neutrino mass by the analysis of the calorimetrically measured $^{163}$Ho EC spectrum have been described. According to this work, the performance required for the detectors to measure the $^{163}$Ho EC spectrum are really demanding: first of all it should be possible to prepare the detector with the $^{163}$Ho source fully contained in the sensitive volume, then all the particles emitted in the $^{163}$Ho EC need to be detected with the same efficiency, an energy resolution $\Delta E_{\mathrm{FWHM}}$ better then 10 eV is asked and the signal rise-time $\tau_{\mathrm{r}}$ has to be short, possibly below $1\,\mu$s. Moreover in order to reach the aimed sensitivity, a statistics of $10^{14}-10^{16}$ counts in the full spectrum need to be acquired, depending on different combinations of detector parameters and on the defined value of $Q_{\mathrm{EC}}$. The measurement of the full statistics within a reasonable time
  of few years requires a total $^{163}$Ho activity of more than 1 MBq. The production of the $^{163}$Ho source with the needed activity and purity is a very important aspect of the ECHo project. In fact it is required that the presence of long living radioactive contaminants is negligible so that their contribution to the background of the $^{163}$Ho EC spectrum is also negligible and the presence of material from the target that has been used to produce the $^{163}$Ho atoms has to be about $10^{-2}\,-\,10^{-1}$ times the amount of $^{163}$Ho.

Once the high purity $^{163}$Ho source will be available and it will be embedded in the energy absorbers of detectors able to measure with high precision the $^{163}$Ho EC spectrum, it will be possible to perform the high statistics calorimetric measurement of the $^{163}$Ho EC spectrum. In order to extract a limit on the neutrino mass in the sub-eV range from this measurement, it is important to improve the knowledge of the expected $^{163}$Ho EC spectrum. A large part of the ECHo collaboration is working to determine theoretically and experimentally the parameters describing the $^{163}$Ho EC spectrum. In particular one of the important goals of the ECHo experiment is to define the energy $Q_{\mathrm{EC}}$ available to the $^{163}$Ho decay as the mass difference between $^{163}$Ho and $^{163}$Dy within an uncertainty of 1 eV. This value will be used as a reference point to investigate the high energy part of the $^{163}$Ho EC spectrum. Moreover it is important to quantify t
 he modifications to the $^{163}$Ho EC spectrum due to solid state effects generated by the fact that the $^{163}$Ho ions are embedded in a solid.

\begin{figure*}
  \includegraphics[angle=0, width=.50\textwidth]{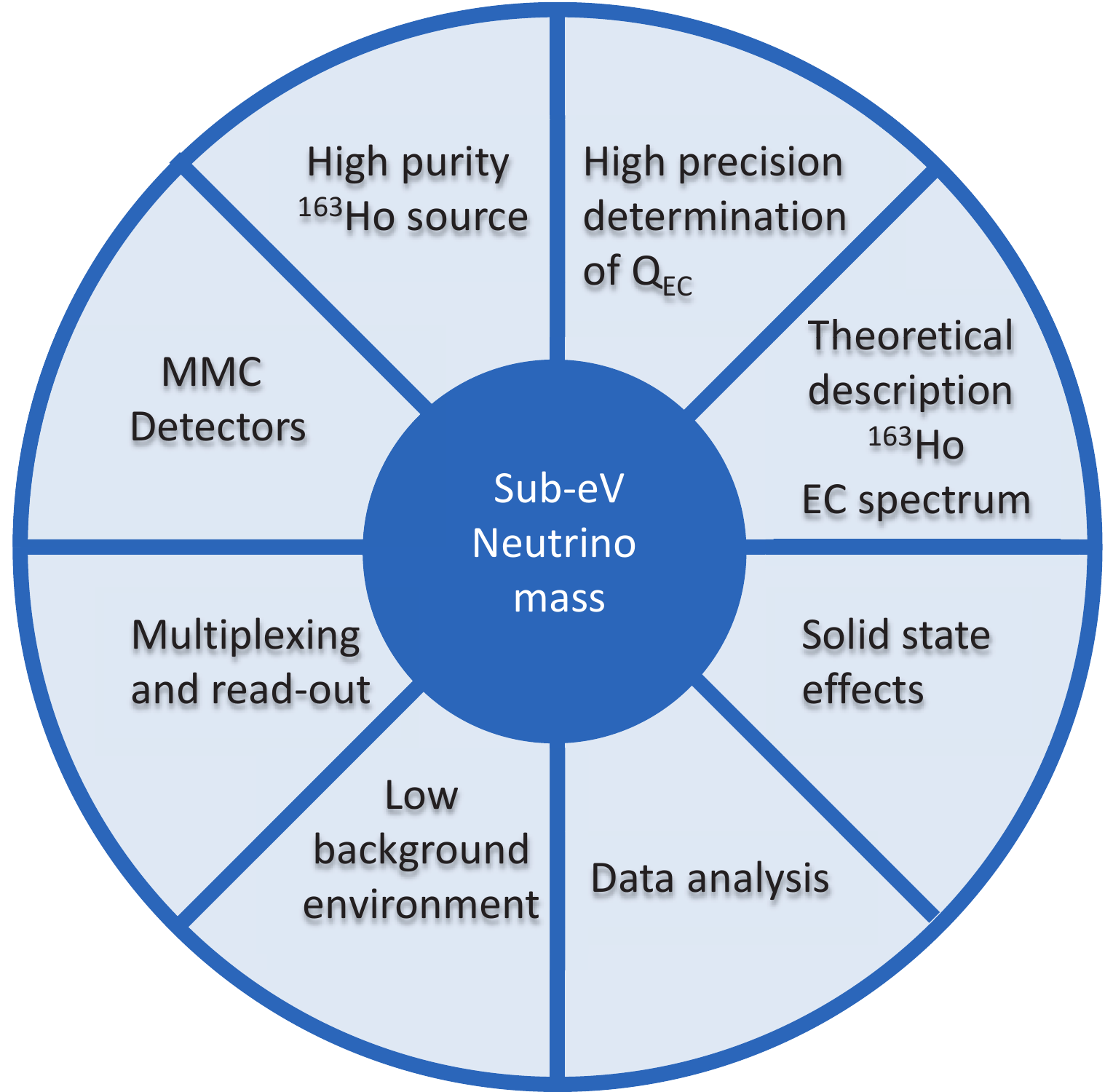}
  \caption{ Diagram showing the structure of the ECHo experiment.  \vspace{-5mm}}
  \label{ECHo_structure}
\end{figure*}

Fig. \ref{ECHo_structure} shows a diagram representing all the investigation routes which compose the ECHo experiment. In the following a few aspects of the ECHo experiment will be described. First of all the detector technology that will be used in ECHo will be introduced as well as the proposed read-out scheme. The results achieved by the first detector prototype will be also discussed. The challenges of producing and purifying the high activity and high purity $^{163}$Ho source will be presented and finally the high precision measurement of the $Q_{\mathrm{EC}}$ value will be described.

\subsubsection{Low temperature metallic magnetic calorimeters and microwave multiplexing}
Presently the only detectors that can measure energy below 3 keV with high precision are low temperature micro-calorimeters [2]. They make use of the calorimetric
principle where the absorption of energy produces an increase
of the detector temperature $\Delta T$ proportional to the deposited energy $\Delta E$ and to the
inverse of the detector heat capacity $C_{\mathrm{tot}}$. These detectors work at temperatures below 100 mK where the phononic and electronic contributions to the heat capacity is smallest. A small heat capacity is one of the key parameters to have a good signal to noise ratio. For detectors developed to measure soft x-rays, the heat capacity is of the order of $1\,$pJ/K or even smaller. It is then clear that only a small fraction of the activity of $^{163}$Ho, required to reach the sub-eV sensitivity to the neutrino mass, can be embedded in a single detector without degrading its performance by increasing its heat capacity. Moreover the reduced activity per detector is also important to reduce the so-called un-resolved pile-up problem. The un-resolved pile-up consists in the impossibility to resolve two or more events happening in the same detector within a time interval of less than the signal rise-time. In this case the detector would show a single event, with an energy gi
 ven approximately by the sum of the energy of the single events, which contributes to the background. In first approximation the fraction of un-resolved pile-up events is given by the product of the total activity in the detector and the pulse rise-time. Therefore in order to reduce these un-wanted events a fast detector response and a reduced activity per pixel are important features of the detector. On the other hand the possibility to increase the activity per pixel will reduce the number of detectors needed to perform the experiment. It is therefore important to optimize the activity per pixel in order on one hand  to have high energy resolution and low intrinsic background and on the other hand to reduce the number of pixels.

As it has been discussed in the previous section, the performance required to the detectors which allows for the achievement of the sub-eV sensitivity to the electron neutrino mass are: high precision detection of the energy of electrons and photons below 3 keV which corresponds to an energy resolution $\Delta E_{\mathrm{FWHM}}\,\leq\,10\,$eV, a fast signal rise-time $\tau_{\mathrm{r}}\,\leq\,1\, \mu$s and a read-out scheme that makes possible a fast and high pixels density multiplexing system.

Within the ECHo experiment low temperature metallic magnetic calorimeters (MMCs) [3] will be used. MMCs are energy dispersive
detectors typically operated at temperatures below $50\,$mK. In first approximation these detectors consist of a particle absorber, where the energy is deposited, tightly connected to a temperature sensor which is then weakly connected to a thermal bath. As for all the other micro-calorimeters, also for MMCs the deposition of energy in the absorber leads to an increase of the detector temperature. The temperature sensor of the MMCs is a paramagnetic alloy which resides in a small magnetic field. The change of temperature leads to a
change of magnetization of the sensor which is read-out as
a change of flux by a low-noise SQUID magnetometer. The
sensor material, presently used for MMCs, is a dilute alloy of erbium in gold,
Au:Er. The concentration of erbium ions in the sensor can be chosen to
optimize the detector performance and usually varies between 200
ppm and 800 ppm.
The spectral resolving power of a state of the art MMCs
for soft x-rays is above 2000. For completely micro-structured detectors, an energy resolution of $\Delta E_{\mathrm{FWHM}}\,=\,2\,$eV at $6\,$keV and a signal rise-time $\tau_{\mathrm{r}}\,=\,0.09\, \mu$s [4] have been achieved. Moreover the read-out scheme for MMCs is compatible with several multiplexing techniques developed for low temperature micro-calorimeters, in particular with the microwave multiplexing as will be discussed in the following. The achieved performance suggests that MMCs are suitable detectors for measuring the high precision and high statistics EC spectrum of $^{163}$Ho.

A first test experiment to investigate the behavior of a MMC detector with $^{163}$Ho ion-implanted in the absorber has been successfully performed [5]. The $^{163}$Ho was produced at ISOLDE-CERN [6] by irradiating with a proton beam a Ta-W target. After surface ionization, a mass-selected beam with ions having mass $163\,$u was directed onto the detector chip and collimated on a surface having the diameter of about $2\,$mm. 
\begin{figure*}
  \includegraphics[angle=0, width=.95\textwidth]{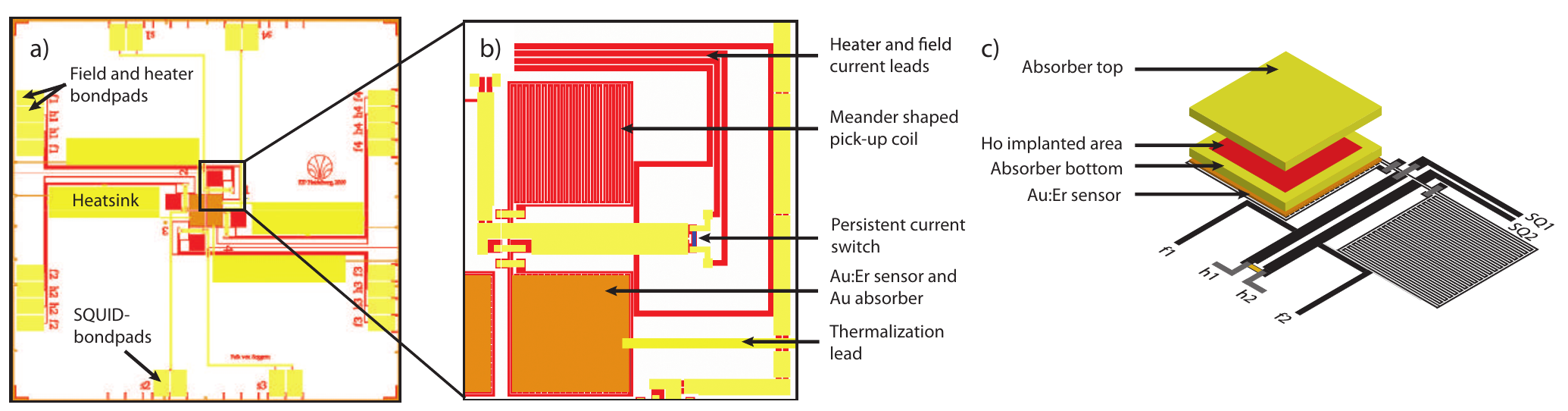}
  \caption{a) Schematic of a MMC detector chip for the $^{163}$Ho implantation experiment. The four double-meander pick up coils are located in the center of the chip. b) The magnified region shows a detailed picture of one detector. Only one side of the two double-meander pick-up coil is equipped with sensor and absorber. c) A simplified three-dimensional picture of one detector is shown. In red is indicated the position where the $^{163}$Ho ions are implanted. Reprinted from [7]  \vspace{-5mm}}
  \label{Det_scheme}
\end{figure*}
Fig. \ref{Det_scheme}a) shows the layout of the first prototype of detector chip for the measurement of the $^{163}$Ho EC spectrum. The chip is equipped with four detectors. The detectors are based on the niobium double-meander pick-up coil geometry [3]. The details of the single detector are shown in the magnification, Fig. \ref{Det_scheme}b). In particular only one side of the double-meander pick-up coil has been equipped with Au:Er sensor and absorber to better characterize the thermo-dynamical properties of sensor and absorber materials. The description of the chip design and fabrication is given in [5]. A schematic cross-section of the detector is shown in Fig. \ref{Det_scheme}c).
The absorber is composed of two gold layers, each of dimensions $190\,\times\,190\,\times\,5\,\mu$m$^3$. On top of the first gold layer, indicated as "absorber bottom", the area where the $^{163}$Ho is implanted is indicated in red and has dimensions $160\times160\,\mu$m$^2$. This area is smaller than the absorber area in order to avoid loss of energy through the side walls of the absorber and therefore to achieve the complete quantum efficiency. During the implantation process, the complete chip besides the four $160\times160\,\mu$m$^2$ squares was protected with photo-resist. 
The $^{163}$Ho activity in each pixel was approximately $10^{-2}$ Bq, corresponding to about $10^{10}$ implanted ions. A few tests have been performed with these detectors showing that the implantation process did not degrade the performance of the MMC [5]. With this first measurements, the presently most precise calorimetric measurement of the $^{163}$Ho EC spectrum was obtained. Fig. \ref{163Hospectrum} shows the measured spectrum. The largest background is due to the EC of $^{144}$Pm that was mass-selected and implanted as PmF$^+$ together with the $^{163}$Ho ions. Presently many efforts are dedicated to the production of high purity $^{163}$Ho sources as will be discussed in the next section. The detailed description of this $^{163}$Ho EC spectrum is discussed in [7]. The measured energy resolution was $\Delta E\,\simeq\,12\,$eV and the rise-time $\tau_{\mathrm{r}}\,\simeq\,100\,$ns. The position of the energy peaks was defined within few eV and the obtained best estimati
 on of the total energy available to the $^{163}$Ho decay was $Q_{\mathrm{EC}}\,=\,(2.80\,\pm\,0.08)\,$keV.
\begin{figure*}
  \includegraphics[angle=0, width=.80\textwidth]{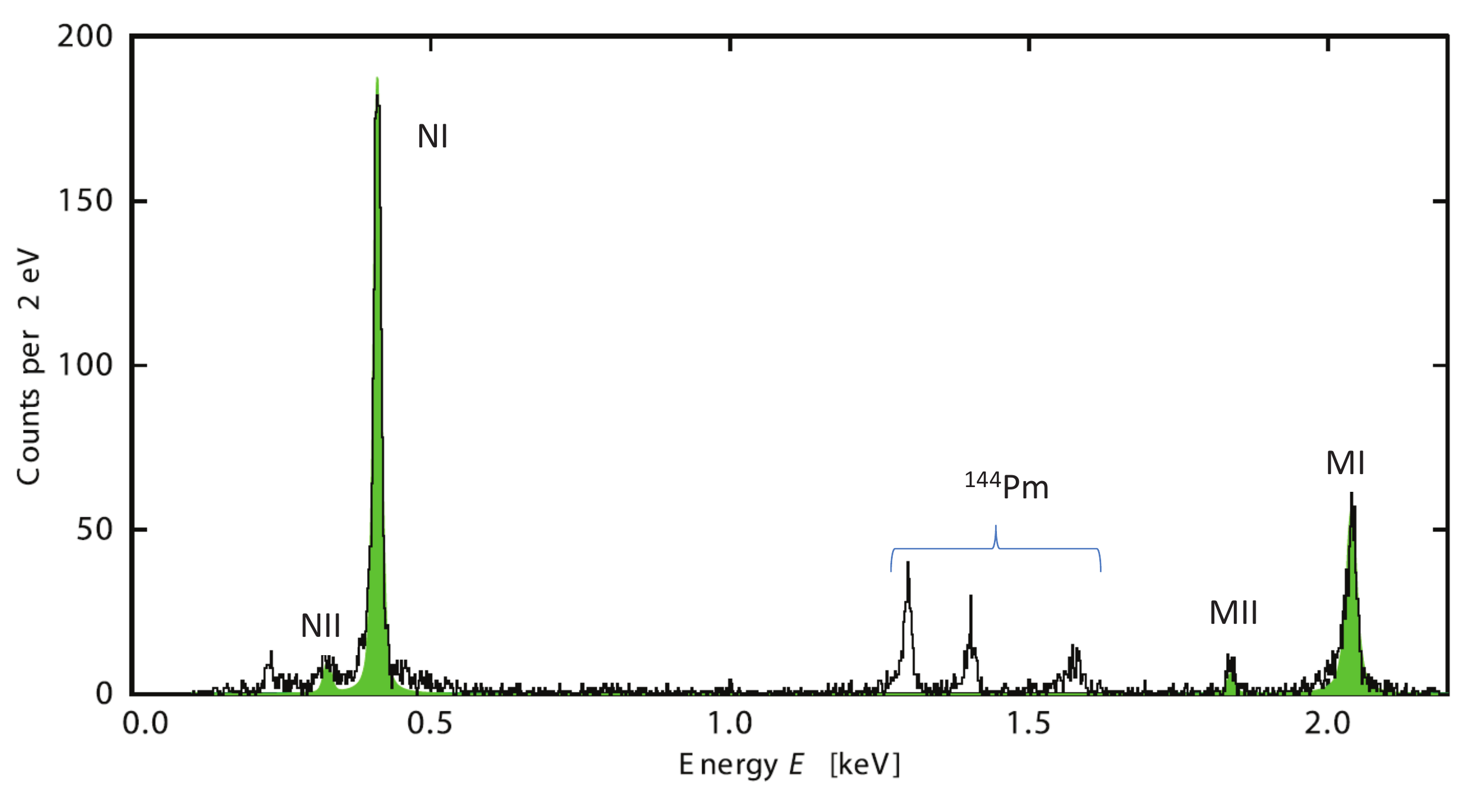}
  \caption{ Calorimetric EC spectrum of $^{163}$Ho as measured (black histogram) and fitted (green
area) The calorimetric lines MI, MII, NI, NII of the $^{163}$Ho EC spectrum can be perfectly seen. The additional lines present in the measured spectrum derive from the EC decay of $^{144}$Pm which was implanted as PmF$^+$ ionized molecules together with the $^{163}$Ho ions. Reprinted from [7]  \vspace{-5mm}}
  \label{163Hospectrum}
\end{figure*}

The results achieved with the first prototype of MMC featuring an absorber with implanted $^{163}$Ho ions and the possibility to improve the energy resolution of the detectors, as discussed in [5] indicate that MMCs meet all the requirements to be used for the high precision measurement of the $^{163}$Ho EC spectrum to investigate the electron neutrino mass. The aim of the ECHo collaboration is to develop MMCs with $10^{11}-10^{13}$ $^{163}$Ho ion in each absorber, corresponding to an activity per pixel of 1 to 100 Bq, having an energy resolution below 3 eV and a pulse rise-time below 100 ns.
A very important aspect of the optimization of the MMCs is the production of the absorber containing the $^{163}$Ho ions. In particular different materials with different concentrations of $^{163}$Ho will be tested as well as different methods to insert the $^{163}$Ho ions in the energy absorbers of MMCs will be investigated as for example ion-implantation and thermo-reduction of the $^{163}$Ho source followed by the preparation of an alloy.

In order to reach the $^{163}$Ho activity to perform the experiment the total number of detectors ranges from $10^4$ to $10^6$. The read-out of such a large number of MMCs can not be done by using the single channel readout since it would produce an enormous heat load on the experimental platform. Therefore it is important to develop a suitable method to multiplex MMCs detectors. The requirements that this method has to fulfill are: low dissipation on the experimental platform, relatively large bandwidth for each pixel of the array and 
ideally no additional noise compared to single pixel read-out.
All these requirements are met by the microwave SQUID multiplexing [8]. 
In this multiplexing scheme every detector is coupled to a non-hysteretic, un-shunted rf-SQUID which is coupled to a superconducting microwave resonator with high internal quality factor and unique resonance frequency. A change of magnetic flux inside the SQUID caused by an event in the detector leads to a change of the effective SQUID inductance and therefore, due to the mutual interaction, to a change of the resonance frequency of the corresponding microwave resonator. It is possible to measure the signal of each detector simultaneously by capacitively coupling the corresponding number of resonators to a common transmission line, injecting a microwave frequency comb driving each resonator at resonance and monitoring either amplitude or phase of each frequency component of the transmitted signal.

Based on experimental results obtained with the first prototype SQUID multiplexer and numerical simulations, the current multiplexer design concerning rf-SQUID layout, SQUID-to-resonator coupling and fabrication of the Nb/Al-AlOx/Nb Josephson junctions [9] have been optimized. Presently new chips consisting of 64 pixels which are read-out using the microwave multiplexing scheme have been developed. After their characterization, these chips will be equipped with absorbers containing $^{163}$Ho and will be used for a small scale experiment.

\subsubsection{$^{163}$Ho source: production and purification}
The production of the proper amount of $^{163}$Ho atoms and the methods to purify this source from the target material and from other isotopes, which are produced within the same process, are of paramount importance for the success of the ECHo experiment. Preliminary studies of different approaches to produce $^{163}$Ho have already been performed by the ECHo collaboration. The production methods for $^{163}$Ho can mainly be divided into two branches: i) charged particle activation of suitable targets in a direct way, that is the optimization of the $^{163}$Ho production, or in an indirect way, that is the optimization of the production of the $^{163}$Ho precursor $^{163}$Er which has a half-life $T_{1/2} \,=\,75\,$min and ii) thermal neutron activation of enriched $^{162}$Er targets.
The reaction that is typically used for the direct activation with proton beams is $^{\mathrm{nat}}$Dy$(p,xn)^{163}$Ho. Another possible direct reaction uses deuteron projectiles $^{163}$Dy$(d,2n)^{163}$Ho. Examples for an indirect way of $^{163}$Ho production are $^{\mathrm{nat}}$Dy$(\alpha,xn)^{163}$Er($\epsilon$)$^{163}$Ho and $^{159}$Tb$(^7$Li$,3n)^{163}$Er($\epsilon$)$^{163}$Ho. For more details on the $^{163}$Ho production, please refer to the related sessions in this paper (\S\ref{Susanta}, \S\ref{Ulli}, \S\ref{Engle}).

In the ECHo experiment all the described methods for production of $^{163}$Ho through charged particle activation processes as well as with neutron irradiation of a $^{162}$Er target will be extensively investigated as well as possible new methods. Moreover methods for the separation of the $^{163}$Ho needs to be optimized in order to have in the final source only traces of the target material and in order to remove the radioactive contaminants to the level in which their contribution to the background of the calorimetric measurement is smaller then the intrinsic pile-up background. Few processes have already been tested. A liquid-liquid extraction technique to separate erbium nuclides from the dysprosium target was performed on $\alpha$-irradiated natural dysprosium target (\S\ref{Susanta}). The developed chemical method is fast enough to complete the entire chemical process within one half-life of $^{163}$Er. On a $^{162}$Er enriched erbium sample which was irradiated for 11 da
 ys with thermal neutrons at the BER II reactor at the Helmholtzzentrum in Berlin (the thermal neutron flux is $\Phi = 1.3 \times 10^{14}\,$s$^{-1}$cm$^{-2}$) an ion-chromatography process using $\alpha$-hydroxyisobutyric acid was used to separate the holmium ions from the erbium target ions. This source after the purification step contains about $10^{16}$ $^{163}$Ho ions and will partially be used for the characterization of the contaminants as well as for the optimization of the separation and purification method. Moreover a large fraction of this source will be used for future detector tests as well for first experiments to determine the $Q_{\mathrm{EC}}$ value with Penning Traps.

In the next future more tests will be performed both at accelerator facilities as well at reactor facilities. The aim of these tests is to investigate and quantify the production of radioactive contaminants and to improve the purification methods in order to reach the required purity.

\subsubsection{$Q_{\mathrm{EC}}$ determination}
\label{QEC}
The total energy available for the decay of $^{163}$Ho to $^{163}$Dy is a fundamental parameter to reach the sub-eV sensitivity to the electron neutrino mass by the analysis of the high energy part of the $^{163}$Ho EC spectrum. The recommended value is $Q_{\mathrm{EC}}\,=\,(2.55\,\pm\,0.016\,)$keV as can be found in [10], but other measurements give values that range from about 2.3 keV [11] to about 2.8 keV obtained by two calorimetric measurements performed using low temperature detectors [12] and [7]. It is to notice that the $Q_{\mathrm{EC}}$ value measured in the cited experiments, has been derived by the analysis of the EC spectrum of $^{163}$Ho.

A very strong method to determine the energy available to the $^{163}$Ho which is also independent from the decay process is the determination of the mass difference between mother and daughter atoms. Within the ECHo experiment the $Q_{\mathrm{EC}}$ of $^{163}$Ho will be measured by means of Penning traps [13].
The superiority of Penning-trap mass spectrometry over the other mass-measurement
methods lies in a determination of the mass $M$ of the nuclide of interest via the direct
measurement of the cyclotron frequency $\nu_{\mathrm{c}}$ of its ionic state with the electric charge $q$ in
a strong static homogeneous magnetic field $B$:
\begin{equation}
\nu_{\mathrm{c}}=\frac{1}{2\pi}\times \frac{q}{M} \times B
\end{equation}
Nevertheless, to perform such a measurement the ion must be confined to a well-localized
volume within the homogeneous magnetic field for at least some seconds. This is achieved by
a superposition of a static three-dimensional quadrupole electric field on the magnetic field,
such that an electrostatic potential well along the magnetic field lines is created. In such a
configuration of the fields, called the Penning trap, the magnetic field confines the motion of the ion
to the plane perpendicular to the magnetic field lines and the electrostatic field does not allow
the ion to escape along the magnetic field lines.
The presence of the electrostatic quadrupole field modifies the pure cyclotron motion of
the ion in the magnetic field to three independent trap motions: two radial motions which are the modified
cyclotron and the magnetron motions with the frequencies $\nu_-$ and $\nu_+$, respectively, and the axial
motion with the frequency $\nu_{\mathrm{z}}$. None of these frequencies are simple functions of the ion's
mass, but the sum of the radial frequencies is equal to the cyclotron frequency $\nu_{\mathrm{c}}$ on the level
of the required accuracy for traps employed in high-precision Penning-trap mass spectrometry.
Thus, a measurement of the radial eigenfrequencies results in a determination of the
cyclotron frequency of the ion. There are two methods to measure the cyclotron frequency. (1)
The cyclotron frequency is measured via detection of the image current induced by the ion's
motion in the resonant tank circuit attached to the trap [14]. (2) In the so-called time-of-flight
ion-cyclotron-resonance technique (ToF-ICR) the cyclotron frequency is determined from the
measurement of the time of flight of the ion through the strong gradient of the magnetic field [15].

The aim of the ECHo collaboration is to reach a precision on the $Q_{\mathrm{EC}}$ of 1 eV or better. This task is planned to be accomplished in the near future by the novel Penning-trap mass spectrometer PENTATRAP [16,17]. The uniqueness and complexity of this Penning-trap mass spectrometer is conditioned by the unprecedented relative accuracy of a few parts in $10^{12}$ with which the $Q_{\mathrm{EC}}$ must be measured. This requires a careful stabilization of the magnetic field, screening the magnet from stray electrical and magnetic fields and a temperature-stabilized experimental room with a vibration-free concrete cushion for the magnet. It is necessary to produce ions of heavy nuclides in very high charge states, which is done with an external electron-beam ion-trap source. Furthermore, the cyclotron frequencies of the decay mother and daughter nuclides have to be measured simultaneously. For this, five cylindrical Penning traps will be employed and the novel cyclotron
 -frequency measurement technique described in Ref. [18] will be applied.

\subsubsection{Conclusions}
The complexity of the ECHo experiment requires that the efforts of several working groups are combined to reach the aimed sensitivity in the sub-eV range for the electron neutrino mass. In the next future the first detector chip with integrated multiplexed read-out will be produced as well as new methods for the production and purification of the $^{163}$Ho source will be analyzed. 
The next goal of the ECHo collaboration is to perform an experiment of reduced size, about 100-1000 pixels read-out as few arrays. The uncertainty for the $Q_{\mathrm{EC}}$ value will be reduced to few tens of eV. With this first small scale experiment it will be possible to set an upper value on the neutrino mass in the few eV range which corresponds to an improvement of about two orders of magnitude compared to the present accepted limit of 225 eV [19].

\begin{center}
{\it References}
\end{center}
\begin{description}
\footnotesize	
\item[1] M. Galeazzi {\it et al.}, http://arxiv.org/abs/1202.4763
\item[2] C. Enss, Topics in Applied Physics, {\bf 99} (2005)
\item[3] A. Fleischmann {\it et al}, AIP Conf. Proc. vol. {\bf 1185} (2009) 571
\item[4] C. Pies {\it et al}, Journal of Low Temperature Physics {\bf 167} 3-4 (2012) 269
\item[5] L. Gastaldo {\it et al.}, Nuclear Inst. and Methods in Physics Research, A {\bf 711} (2013) 150
\item[6] E. Kugler, Hyperfine Interact. {\bf 129} (2000) 23
\item[7] P. C.-O. Ranitzsch {\it et al.}, Journal of Low Temperature Physics, {\bf 167} (2012) 1004
\item[8] J. A. B. Mates {\it et al.}, Applied Physics Letters {\bf 92} (2) [2008) 023514 
\item[9] S. Kempf {\it et al.}, Supercond. Science And Technology {\bf 26} (2013) 065012
\item[10] A. Wapstra, G. Audi, C. Thibault, Nucl. Phys. A {\bf 729} (1), (2003) 129
          G. Audi, A. Wapstra, C. Thibault, Nucl. Phys. A {\bf 729} (1), (2003) 337
          
\item[11] J. U. Andersen {\it et al.}, Phys. Lett. B {\bf 113} (1982) 72
\item[12] F. Gatti {\it et al.}, Physics Letters B {\bf 398} (1997) 415
\item[13] K. Blaum,  Phys. Rep. {\bf 425} (2006) 1
\item[14] H.G. Dehmelt and F.L. Walls, Phys. Rev. Lett. {\bf 21} (1968) 3
\item[15] G. Graeff, H. Kalinowsky and J. Traut, Z. Phys. A {\bf 297} (1980) 35
\item[16] J. Repp {\it et al.}, Applied Physics B {\bf 107} (2012) 983
\item[17] C Roux {\it et al.}, Applied Physics B {\bf 107} (2012) 997
\item[18] S. Eliseev {\it et al.}, Physical Review Letters {\bf 110} (2013) 082501
\item[19] P. T. Springer {\it et al.}, Phys. Rev. A {\bf 35} (1987) 679

\end{description}

 \newpage\subsection{J. Formaggio, for the Project 8 Collaboration ``The Project 8 Neutrino Mass Experiment"}
\begin{description}
\it\small 
\setlength{\parskip}{-1mm}
\item[\small\it D.D.:]Dip. di Fisica ``G. Occhialini'', Universit\`a di Milano-Bicocca, Milano, Italy
\item[\small\it M.M.:]INFN Sezione di Milano-Bicocca, Milano, Italy
\end{description}
%
\subsubsection{Scientific Motivation}

Ever since Enrico Fermi's original proposal~[1], it has been known that the neutrino mass has an effect on the kinematics of beta decay.   Measurements have always
suggested that this mass was very small, with successive generations of experiments giving upper limits~[2][3], most recently ${m_\nu}_\beta < 2.3$ eV.  The
upcoming KATRIN experiment~[4][5] anticipates having a sensitivity of 0.20 eV at 90\% confidence.  If the neutrino mass is much below 0.20 eV, it is difficult to envision any classical spectrometer being able to access it.  Oscillation experiments, however, tell us with great confidence that the tritium beta decay neutrinos are an admixture of all three mass states, at least two of which have a nonzero mass, such that the effective mass must satisfy ${m_\nu}_\beta > 0.005$ eV under the normal hierarchy or  ${m_\nu}_\beta > 0.05$ eV in the inverted hierarchy.  These bounds provide a strong motivation to find new, more sensitive ways to measure the tritium beta decay spectrum if the question of the neutrino hierarchy is ever to be resolved.

Our current knowledge of the neutrino mass scale and the neutrino hierarchy is a powerful reminder that our standard model of nuclear and particle physics remains
incomplete. Direct measurements of the neutrino mass can provide direction as to how to extend that model.  Its implications are  not just limited to the field of
nuclear physics, but extend equally to particle physics and cosmology. There are many theories beyond the Standard Model that explore the origins of neutrino masses
and mixing. In these theories, which often work within the framework of supersymmetry, neutrinos naturally acquire small but finite masses. Several models use the
so-called see-saw effect to generate neutrino masses. Other classes of theories are based on completely different possible origins of neutrino masses, such as
radiative corrections arising from an extended Higgs sector. As neutrino masses are much smaller than the masses of the other fermions, the knowledge of the
absolute values of neutrino masses is crucial for our understanding of the fermion masses in general. Recently it has been pointed out  that the absolute mass scale
of neutrinos may be even more significant and straightforward for the fundamental theory of fermion masses than the determination of the neutrino mixing angles and
CP-violating phases~[6]. It will likely be the absolute mass scale of neutrinos which will determine the scale of new physics.  For a gauge of the impact that neutrino masses can have on particle physics and cosmology, see Table~\ref{tab:knowledge}.

\begin{table*}[htdp]
\caption{Impact of neutrino mass sensitivity level as obtained from beta decay measurements on nuclear physics and cosmology.}
\begin{center}
\begin{tabular}{|l|l|l|}
\hline
Neutrino Mass Sensitivity & Scale & Impact \\
\hline
$m_\nu > 2$ eV & eV & Neutrinos ruled out as primary dark matter \\
\hline
$m_\nu > 0.2$ eV &  Degeneracy &  Cosmology, $0\nu\beta\beta$ reach \\
\hline
$m_\nu > 0.05$ eV &  Inverted Hierarchy & Resolve hierarchy if null result \\
\hline
$m_\nu > 0.01$ eV &  Normal Hierarchy& Oscillation limit, possible relic neutrino sensitivity\\
\hline
\end{tabular}
\end{center}
\label{tab:knowledge}
\end{table*}%

\subsubsection{Tritium Beta Decay and Detection of Electrons via Cyclotron Emission}

The most sensitive direct searches for the electron neutrino mass
up to now are based on the investigation of the electron spectrum
of tritium $\beta$-decay. As both the matrix elements and Coulomb correction are independent
of $m_\nu$, the dependence of the spectral shape on $m_\nu$ is
given by the phase space factor only. In addition, the bound on
the neutrino mass from tritium $\beta$-decay is independent of whether
the electron neutrino is a Majorana or a Dirac particle.

To make advances toward lower and lower masses, it is important to develop techniques that allow for extremely precise spectroscopy of low energy electrons.  Current electromagnetic techniques, such as those employed by the KATRIN experiment, can achieve of order $10^{-5}$ in precision, but are at the limit of their sensitivity.  Therefore, a new technique must be pursued in order to approach the inverted or even the hierarchical neutrino mass scale implied by current oscillation measurements.  The technique proposed here relies on the principle that the frequency of cyclotron radiation emitted by the particle depends inversely on its energy, independent of the electron's direction when emitted.  As the technique inherently involves the measurement of a {\em frequency} in a non-destructive manner, it can, in principle, achieve a high degree of resolution and accuracy.  The combination of these two features makes the technique attractive within the context of neutrino mass measurements, as well as other venues, as will be discussed below.

Imagine a charged particle, such as an electron created from the decay of tritium or from neutrino capture, traveling in a uniform magnetic field $B$.  In the absence of any electric fields, the particle will travel along the magnetic field lines undergoing simple cyclotron motion.  The characteristic frequency $\omega$ at which it precesses is given by

\begin{equation}
\label{eq:cyfreq}
\omega = \frac{e B}{\gamma m_e} = \frac{\omega_c}{\gamma} = \frac{\omega_c}{1+\frac{T}{m_e c^2}},
\end{equation}

\noindent where $\omega_c$ is the cyclotron frequency, $T$ and $m_e$ are the electron kinetic energy and mass, respectively, and $\gamma$ is the relativistic boost factor. The cyclotron frequency, therefore, is shifted according to the kinetic energy of the particle and, consequently, any measurement of this frequency stands as a measurement of the electron energy.  Electrons from tritium decay have a kinetic energy of 18.6 keV or, equivalently, a boost factor $\gamma \simeq 1.0364$.

A charged particle undergoing cyclotron motion will also emit cyclotron radiation as it travels through a magnetic field.  Since the relativistic boost for the
energies being considered is close to unity, the radiation emitted is relatively coherent.  For a magnetic field strength of 1 Tesla, the emitted radiation has a
frequency of 27 GHz.  This frequency band is well within the range of most commercially available radio-frequency antennas and detectors. It is conceivable,
therefore, to make use of radio-frequency (RF) detection techniques in order to achieve precision spectroscopy of electrons.  Furthermore, the typical power emitted
by these electrons is sufficiently high to enable single-electron detection. A more in-depth description of the technique, including a discussion of its potential
sensitivity, can be found in Ref~[7].

\subsubsection{Project 8: A Multi-Phase Approach}

The technique described above represents a potentially novel and effective approach for measuring the energy of electrons. Some of the advantages are enumerated below: 

\begin{enumerate}
\item{\it Source = Detector:}  Since the energy measurement of the electron is non-destructive, it takes place anywhere along the path of the electron.  This feature, in combination with the transparency of the gas to microwave photons, removes the necessity of extracting the electron  from the  source in order to measure its energy .  The combination of the source and detector region as one allows for a more favorable scaling of the experiment.
\item{\it Frequency Measurement:} Frequency techniques number among the most precise and accurate types of measurement that can be made.  The linearity offered by frequency techniques allows for exquisite calibration of these measurements.  The level of precision envisioned for our measurements (of order part per million) can be achieved with standard, commercially available technology.
\item{\it Full Spectrum Sampling:} Unlike previous techniques used in beta decay experiments, the beta decay spectrum is available within a single measurement.  No scanning or integrating of the spectrum is necessary for the measurement.  This counting provides a large increase in the statistical efficiency of the experiment.
\end{enumerate}

\begin{figure}[t]
\begin{center}
\includegraphics[width=0.35\columnwidth]{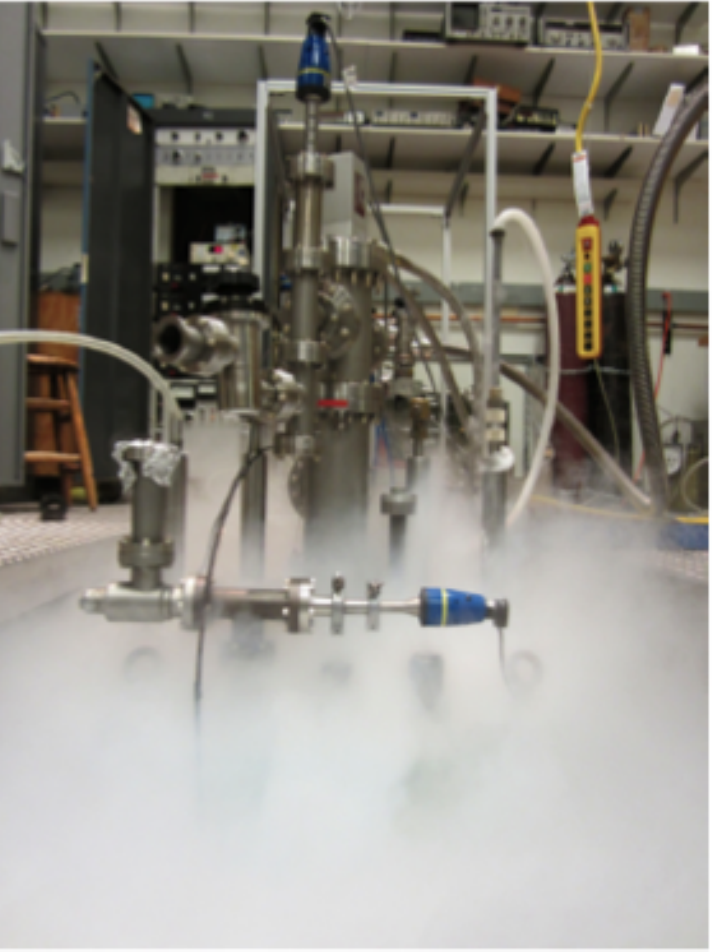} \\
\end{center}
\caption{ A photograph of the top of the Project 8 prototype magnet during cool down.}
\label{fig:cooldown}
\end{figure} 

For all the advantages offered by the above technique, the Project 8 collaboration realizes that there also exist a number of significant challenges in order to realize the technique into a competitive measurement of the neutrino mass.  The Project 8 collaboration is thus moving forward with a multiple-phased approach;  each stage providing both the necessary R\&D and key physics measurements of interest to the physics community.  Phase I establishes a proof-of-principle measurement for the cyclotron emission of energetic electrons by using$~^{83m}$Kr as its electron source.  The prototype incorporates all the main features of the envisioned full-scale experiment: a gaseous electron source, a magnetic trapping region, and the RF detection and amplification scheme.  It is currently assembled at the University of Washington.  Preliminary data analysis is currently underway.

Future phases will shift the physics goals from proof-of-principle to a competitive neutrino mass measurement, with a final goal of 50 meV in sensitivity. Reaching this ultimate sensitivity could address the question of neutrino hierarchy: if the observable (i.e., electron flavor) mass is less than this limit, then the hierarchy is normal and hence resolved.


\begin{center}
{\it References}
\end{center}
\begin{description}
\footnotesize	

\item[1] E. Fermi, Ricerca Scient. {\bf 2}, 12 (1933).
\item[2] C. Weinheimer {\it et al.}, Phys.\ Lett. {\bf B460}, 219 (1999).
\item[3] V. Lobashev {\it et al.}, Nuclear\ Physics-Section B-PS-Proceedings Supplements {\bf 91}, 280 (2001).
\item[4] J. Angrik {\it et al.} (2005), FZKA-7090.
\item[5] A. Osipowicz {\it et al.} (2001).
\item[6] Y. Farzan and A. Y. Smirnov, Phys.\ Lett. {\bf B557}, 224 (2002).
\item[7] B. Monreal and J. A. Formaggio, Phys.\ Rev. {\bf D80}, 051301 (2009), 0904.2860.
\end{description}

 \newpage\subsection{M. Yoshimura ``Neutrino mass spectroscopy using atoms/molecules''}
\begin{description}
\it\small 
\setlength{\parskip}{-1mm}
\item[\small\it M.Y.:] Okayama University, Okayama, Japan
\end{description}
%
With the expected small neutrino mass of a fraction of eV
the energy mismatch becomes a serious problem
in planned neutrino experiments using the conventional nuclear target 
where several MeV energy is released.
On the other hand, 
with the advent of remarkable technological innovations,
manipulation  of 
atoms and molecules may contribute greatly to
fundamental physics.
Neutrino physics may also be one of these areas.
Atoms and molecules are target candidates of precision
neutrino mass spectroscopy, as recently emphasized in
[1]
due to closeness of the energy released 
in their transition to expected neutrino masses.
The relevant process of our interest is cooperative
(and coherent, called macro-coherent subsequently) atomic de-excitation;
$|e \rangle \rightarrow |g\rangle + \gamma + \nu_i\nu_j$
where $ \nu_{i(j)}, i,j = 1,2,3$ is one of neutrino mass eigenstates.
Measured quantities are spectrum rates and
polarization at different photon energies.
This experiment can resolve neutrino mass eigen-states.

To obtain a measurable rate of the process,
it is crucial to develop the macro-coherence [2],[1],
a new kind of coherence that involves both atoms and fields.
The macro-coherent emission of radiative neutrino pairs is stimulated by
two trigger irradiation of frequencies $\omega, \omega'$
constrained by $\omega + \omega' = \epsilon_{eg}/\hbar
\,, \omega < \omega'$.
The measured photon energy in the de-excitation is given by the smaller frequency $\omega$.
The macro-coherence assures that the three-body process, 
$|e \rangle \rightarrow |g\rangle + \gamma + \nu_i\nu_j$,
conserves both the energy and the momentum.
Assuming that $|e\rangle, |g\rangle $ can be taken infinitely heavy and
the atomic recoil may be ignored,
there exist threshold photon energies [3]
at
\begin{eqnarray}
&&
\omega_{ij} = \frac{\epsilon_{eg}}{2} - \frac{(m_i +m_j )^2}{2\epsilon_{eg}}
\,,
\label{threshold location}
\end{eqnarray}
(with
$\epsilon_{eg} = \epsilon_e - \epsilon_g$ the atomic energy difference
of initial and final states).
Each time the measured photon energy decreases below
a fixed threshold energy of $\omega_{ij}$, 
a new continuous spectrum is opened, hence there are six energy thresholds
$\omega_{ij}\,, i,j = 1,2,3$ separated by finite photon energies.
Determination of the threshold location given by eq.(\ref{threshold location}), 
hence decomposition into six mass thresholds,
is made possible by precision of irradiated laser frequencies at 
$\omega \approx \omega_{ij}$ and
not by resolution of detected photon energy.

The macro-coherently amplified radiative emission of
neutrino pair has been coined 
RENP (radiative emission of neutrino pair) [1],
which is the core idea of our neutrino mass spectroscopy
that may determine all unknown neutrino parameters.
This method can  ultimately determine all three masses,
the nature of neutrino
masses (Dirac vs Majorana distinction), and
the new Majorana source of CPV phases [3], [4].
Emission of two identical Majorana particles suffers from
the Fermi exclusion principle, hence rate of Majorana RENP
differs from the Dirac rate where anti-particles are
distinct from particles, with or without interference terms
of identical fermions.
This is how Majorana vs Dirac distinction can be made.

RENP rate for Xe atom is of order 1 mHz for the excited target number density of $10^{21}$cm$^{-3}$ (rate $\propto$ the number density$^3$),
assuming a maximum coherence development.
This atomic target has an excellent potential of discovering RENP process and
can determine the absolute neutrino mass scale.
It can also determine the normal vs inverted mass hierarchical mass pattern.
Majorana vs Dirac distinction of the neutrino mass type 
and CPV phase determination is however difficult
due to a large energy released in de-excitation of Xe, $\sim 8.4$eV:
it is easier for smaller atomic energy available [4].

I shall summarize status of ongoing experiments in our group.
We believe that prior to RENP it is important to
verify the principle of macro-coherence development in
QED process. For this purpose it is best to experimentally
demonstrate the macro-coherent two photon emission 
$|e \rangle \rightarrow |g\rangle + \gamma + \gamma'$,
whose good candidate target is the
vibrational transition of pH$_2$ molecule.
The macro-coherent two photon emission 
is called paired super-radiance (PSR)
in which two back-to-back photons are emitted
with equal energies [2].
If the product of initial coherence between the
initial ($|e\rangle$) and the final ($|g\rangle$) vibrational levels and the excited
atom density in $|e\rangle$ is large enough,
unambiguous signal of  PSR may be obtatined,
in which a significant portion
of  energy stored in $|e\rangle$
are released as a short pulse in a few nano-seconds, 
an event that never occurs in ordinary circumstances due to
a small spontaneous two photon decay rate of order $10^{-16}$sec.$^{-1}$.
Numerous simulations based on
1(time) + 1(space) dimensional Maxwell-Bloch
equation have been done [2], [1] to confirm that
large signal PSR events occur.
Simulations have  taken into account
relaxation, both the phase de-coherence and
population decay. If the relaxation time is larger than
order a few ns and the target length is large enough,
visible PSR can occur.

We almost finished fabrication of three lasers of good quality
for this purpose,
two for pulse excitation to $v=1$ level and one CW laser
for the trigger of $\omega=\omega' = \epsilon_{eg}/2$.
PSR signal is identified by increase of correlated pulse
of frequency $\omega$ (different from
excitation pulses) with the excitation pulse.
We shall start actual PSR experiment soon.

In the next stage towards RENP it is important to
control PSR event which may become backgrounds against
RENP.
Target atoms for RENP must have two levels bridged by a transition of
a character,  M1$\times$E1 
which may have a weak PSR (PSR is stronger for E1$\times$E1
related levels).
The idea for controled PSR comes from the expected
formation of static condensed fields supported by
macroscopic atomic polarization, objects called 
solitons [1].
By soliton formation PSR emission at target ends
is suppressed since fields are confined within
the target, while RENP occurs inside the target
and escapes from target ends.

Theoretical ideas related to this experimental project
are explained in detail in [1], along with
our experimental efforts.

\begin{center}
{\it References}
\end{center}
\begin{description}
\footnotesize	
\item[1]
A. Fukumi et al.,
{\it Progr. Theor. Exp. Phys.}{\bf 2012, 04D002};
arXiv1211.4904v1[hep-ph](2012)
and references cited therein.

\item[2]
M. Yoshimura, N. Sasao, and M. Tanaka,
{\it Phys. Rev}
{\bf A86},013812(2012),
and
{\it Dynamics of paired superradiance},
arXiv:1203.5394[quan-ph] (2012).

\item[3]
M. Yoshimura, {\it  Phys. Lett.}{\bf B699},123(2011) and references therein.

\item[4]
D.N. Dinh, S. Petcov, N. Sasao, M. Tanaka,
and M. Yoshimura,
{\it  Phys. Lett.}{\bf B719},154(2012), and  
arXiv1209.4808v1[hep-ph].

\end{description}

 \newpage\subsection{T. Ota: ``Collider-testable neutrino mass generation mechanisms''}
\begin{description}
\it\small 
\setlength{\parskip}{-1mm}
\item[\small\it T.O.:]
Department of Physics, 
Saitama University, Shimo-Okubo 255, 
338-8570 Saitama-Sakura, Japan
\end{description}
%
If the Standard Model (SM) is an low-energy effective model of 
a fundamental theory that is realized at high energy scales, 
the full Lagrangian that describes physics at the low energy scale 
(the electroweak scale $\Lambda_{\rm EW}$) should contain the series 
of higher-dimensional operators whose mass
dimensions are higher than four. They are suppressed by the inverse power 
of new physics scale $\Lambda_{\rm NP}$.
The lowest higher-dimensional operator, 
which is suppressed by $1/\Lambda_{\rm NP}$,
is the famous Weinberg operator $\overline{L^{c}} L H H$,
which brings Majorana mass to neutrinos after the electroweak
symmetry breaking. The origin of the operator has been extensively 
studied. In the famous seesaw mechanism (type I), which is
well-motivated by supersymmetric Grand Unified Theories (GUT), 
the operator is induced through the tree-level diagram mediated by heavy
right-handed neutrinos.
In that framework,  
$\Lambda_{\rm NP}$ must lie around the GUT scale so as to suppress 
the resulting neutrino mass at the scale of $\mathcal{O}(1)$ eV.
There have been also many attempts to build high-energy models
in which Weinberg operator is provided by loop diagrams.
In such a framework, the resulting neutrino mass obtains an additional 
loop-suppression factor $1/(4\pi)^{2}$. Thank to that factor, the typical scale 
of $\Lambda_{\rm NP}$ can be lowered to the collider-testable energy scale. 

In this talk, we would like to introduce an alternative idea to serve 
neutrino mass from new physics based on the collider-testable 
energy scale~[1]. If the mass dimension five $d=5$ Weinberg operator 
is forbidden for some reason (e.g., by some symmetry), 
the next lowest higher dimensional operator that can bring Majorana mass
to neutrinos is the dimension seven $d=7$ operator $\overline{L^{c}} L HHHH$,
whose resulting neutrino mass is more suppressed by the additional
suppression factor $(\Lambda_{\rm EW}/\Lambda_{\rm NP})^{2}$ than
that from Weinberg operator.
With this new suppression factor, 
we can lower $\Lambda_{\rm NP}$ to TeV scale that is now explored
by the LHC.

\begin{figure}[b]
\unitlength=1cm
\begin{center}
\begin{picture}(18,3)
\thicklines
\put(0,0.5){\includegraphics[width=4cm]{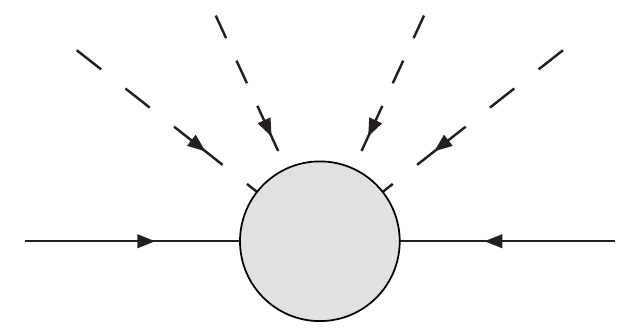}}
\put(4,1.5){\vector(1,0){1}}
\put(5.1,0.5){\includegraphics[width=3cm]{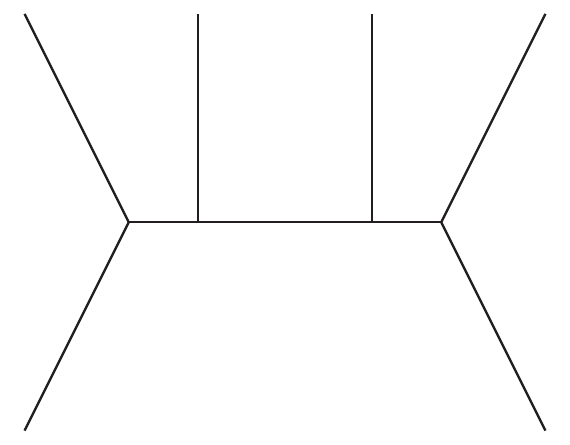}}
\put(8.1,0.5){\includegraphics[width=3cm]{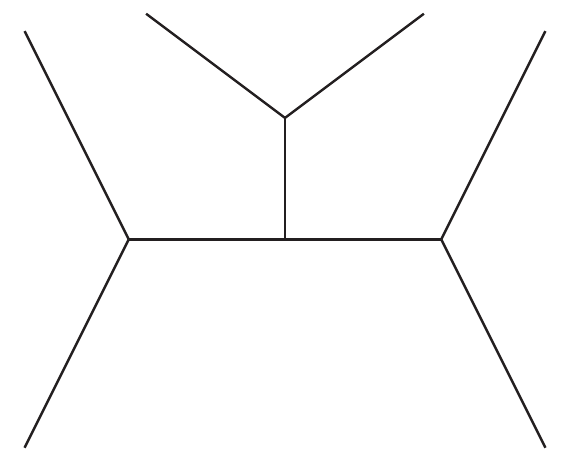}}
\put(11,0.5){\includegraphics[width=3cm]{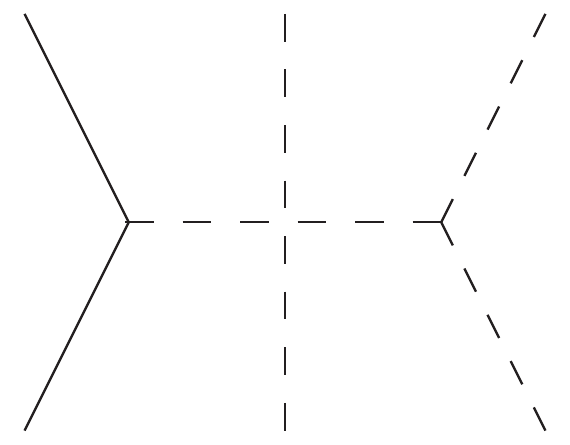}}
\put(14,0.5){\includegraphics[width=3cm]{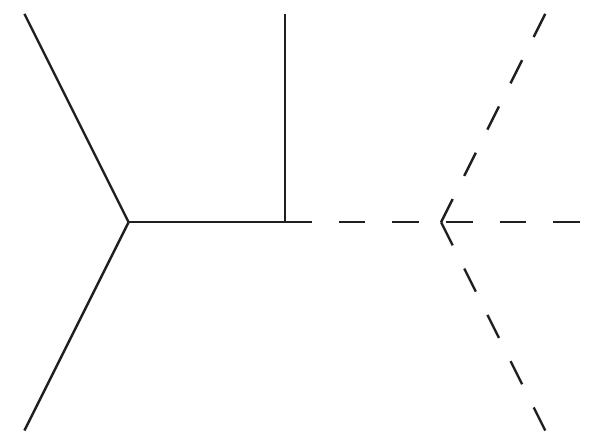}}
\put(0.05,0.7){$L$}
\put(3.7,0.7){$L$}
\put(0.2,2.35){$H_{u}$}
\put(1.2,2.55){$H_{u}$}
\put(2.5,2.55){$H_{d}$}
\put(3.5,2.35){$H_{u}$}
\put(1.65,1){$\mathcal{O}_{d=7}$}
\put(5.7,0){\bf Topology 1}
\put(8.7,0){\bf Topology 2}
\put(11.6,0){\bf Topology 3}
\put(14.6,0){\bf Topology 4}
\end{picture}
\end{center}
\caption{Topologies of tree-level decomposition 
of dimension seven neutrino mass operator.
Dashed lines always indicate scalar fields (Higgs doublets or scalar
 mediators) and solid lines can be interpreted as both scalar fields and 
fermion fields (lepton doublets or fermion mediators).}
\label{Fig:d7}
\end{figure}
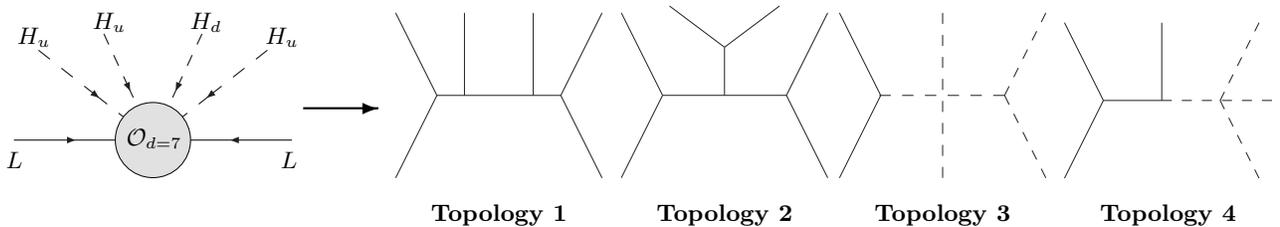

In order to forbid $d=5$ Weinberg operator and allow the $d=7$ operator,
we extend the particle content so as to contain the second Higgs doublet 
(i.e., we employ Two Higgs Doublet Model (THDM))
and introduce $Z_{5}$ matter parity.
Assigning the charges of $Z_{5}$ to the matter fields and Higgs doublets
appropriately, we can have the $d=7$ operator $\overline{L^{c}} L H_{u}
H_{u} H_{u} H_{d}$ without Weinberg operator,
i.e., at the low energy scale, our model is described as
\begin{align}
\mathcal{L}_{\text{eff}} = \mathcal{L}_{\rm THDM} +
 \frac{1}{\Lambda_{\rm NP}^{3}}
 \overline{L^{c}}L H_{u}H_{u}H_{u}H_{d}.
\label{eq:Leff}
\end{align}
Now, our interests lead us to consideration of the high-energy 
completions of this low-energy effective theory.
We survey all the possibilities that can bring the 
low-energy effective model described as Eq.~\eqref{eq:Leff}, 
with the exhaustive bottom-up approach, in which 
we decompose the $d=7$ operator to all the possible tree-level diagrams
in order to list the necessary mediation fields and interactions.
The topologies of the tree diagrams are shown in Fig.~\ref{Fig:d7}.
The list of the high energy models is given in Ref.~[1], and
some extensions of this approach have been discussed in Refs.~[2-4].

We also study the effective $d=9$ operators that can contribute 
to the neutrinoless double beta decay ($0\nu2\beta$) process~[5].
As discussed in this workshop, many experiments are going to 
measure various aspects of neutrino mass, and 
it is expected that we will have numerous information; 
the mass squared differences from oscillation,
the effective Majorana mass from $0\nu2\beta$,
the kinetic mass from single beta decay, 
and cosmological properties.
In future, if some of them will make a conflict in the standard
framework of three-generation neutrino, we will be able to find 
a clue to new physics.
For example, if $0\nu2\beta$ experiment will observe
the signal at the parameter region disfavoured by 
cosmological observations,
it might suggest the existence of new physics that contributes 
to $0\nu2\beta$ measurement but does not affect 
the cosmological observations. 
The $d=9$ operators 
$\overline{u} \overline{u} dd \overline{e} \overline{e}$ 
are a typical way to parameterize such new physics.
With the exhaustive bottom-up approach,
we decompose these $d=9$ operators into all the possible ways 
at the tree level.
As Senjanovic and Rodejohann also pointed out in this workshop, 
an interesting remark is that the typical energy scale of new physics, 
which can make the signal at the sensitivity range of 
the next generation $0\nu2\beta$ experiments, 
is TeV scale, i.e., $0\nu2\beta$ experiments 
are sensitive not only to neutrino mass with
$\mathcal{O}(0.1)$ eV scale but also to new physics at TeV scale.
In Ref.~[5], we list all the necessary particles to mediate 
the $d=9$ operators and find out that all the decompositions except for one 
possibility contain exotic coloured particles, which are now extensively 
searched for at the LHC.
If the LHC will not find the coloured particles, the possibility will
be focused on the one left. We also show that such a possibility 
will be tested by a linear collider, 
because it must contain an exotic interaction with electron.

\begin{center}
{\it References}
\end{center}
\begin{description}
\footnotesize	
\item[1] F. Bonnet {\it et al.}, JHEP {\bf 0910} (2009) 076 
\item[2] S. Kanemura and T. Ota, Phys. Lett. {\bf B694} (2010) 233
\item[3] M. B. Krauss {\it et al.}, Phys.Rev. {\bf D84} (2011) 115023 
\item[4] M. B. Krauss {\it et al.}, arXiv:1301.4221
\item[5] F. Bonnet {\it et al.}, JHEP {\bf 1303} (2013) 055.
\end{description}

 \newpage\subsection{P.Gorla: ``CUORE-0 prototype on the way to CUORE''}
\begin{description}
\it\small 
\setlength{\parskip}{-1mm}
\item[\small\it ] on behalf of the CUORE collaboration
\item[\small\it ] Laboratori Nazionali del Gran Sasso - INFN, Italy
\end{description}
%
CUORE is a bolometric experiment searching for the neutrinoless double beta decay of $^{130}$Te [1]. The 0$\nu$DBD Q-value for this nucleus is at 2527 keV [2]. In the bolometric approach, the energy deposited by particle interacting in an absorber is measured as an increase of temperature in the absorber itself. In CUORE each bolometer will use a cubic 5$\times$5$\times$5 cm$^{3}$ TeO$_2$ crystal with a mass of about 750 g as crystal absorber. In this way the detector contains the source of the decay, achieving high efficiency ($\sim$87\%) and good energy resolution. The temperature variations of the crystal are detected by a thermistor as resistance variations. The thermistor is a neutron transmutation doped (NTD) Ge semiconductor, that exhibits an exponential response to T variations of the resistance at low temperatures. The crystals are held and thermally coupled to a heat sink, that is a copper structure cooled down to about 10 mK, by means of eight PTFE supports. Typical sensitivity of the thermistor is $\sim$100 $\mu$V/MeV of deposited energy. The CUORE detector will be composed by 988 natural TeO$_2$ bolometers for a total mass of 741 kg. As the natural abundance of $^{130}$Te is  $\sim$34\%, the total amount of 0$\nu$DBD active mass is 206 kg. The detectors will be arranged in 19 towers of 13 layers each. The towers will be placed in a roughly cylindrical compact configuration in a new low radioactivity custom dilution refrigerator, to be commissioned and installed in Labotatori Nazionali del Gran Sasso (LNGS), in Italy.
A demonstrator experiment, CUORICINO, was operated in the same laboratory from 2003 to 2008. The CUORICINO detector was composed by 62 TeO$_2$ bolometers, for a total mass of 40.7 kg. The acquired statistics was 19.75 kg($^{130}$Te)$\cdot$ y, and no 0$\nu$DBD signal was found. The background level in the 0$\nu$DBD energy region was 0.169 $\pm$ 0.006 counts/keV/kg/y and the corresponding lower limit on the 0$\nu$DBD half-life of $^{130}$Te is 2.8 $\cdot$ 10$^{24}$ y (90\% C.L.) [2]. This limit translates into an upper limit on the neutrino effective Majorana mass ranging from 300 to 710 meV, depending on the nuclear matrix elements considered in the computation [3].

\subsubsection{The CUORE experimental challenges}
The aim of future neutrinoless experiments, as CUORE, is to probe the inverted hierarchy region of neutrino masses. To achieve the necessary sensitivity the key experimental parameters are i) a large sample of the candidate nuclei to be studied, ii) good detector energy resolution, iii) very low radioactive backgrounds and iv) long live time.
The CUORE collaboration already demonstrated that the bolometric technique is able to provide very good energy resolutions. In CUORICINO the average energy resolution at 2615 keV ($^{208}$Tl line) was 6.3 $\pm$ 2.5 keV. After CUORICINO, the detector performances of the R\&D bolometers for CUORE have been improved thanks to a new layout of the detectors that uses new copper frames and new PTFE crystal supports, and the target energy resolution in the CUORE is 5 keV.
A challenge related to the detector performances is the uniformity of their behaviour. In CUORICINO, we observed a large spread in the pulse shape among the detectors. This leads to a complication in the data analysis that could be more challenging in case of an experiment, like CUORE, with about 1000 single detectors. To improve the detector response uniformity special care has been devoted in the realisation of a new detector assembly system. The background level is expected to be as good as 0.01 counts/keV/kg/y due to: improvements in the radio-purity of the copper and crystal surfaces as well as in the assembly environment;  thicker shields using low-activity ancient Roman lead; and the fact that the 19-tower array affords superior self-shielding and better anti-coincidence coverage (discussion of the different contribution can be found in [5] [6] [7]). 

\begin{figure}[h]
\begin{center}$
\begin{array}{c}
\includegraphics[%
  width=0.45\linewidth,keepaspectratio]{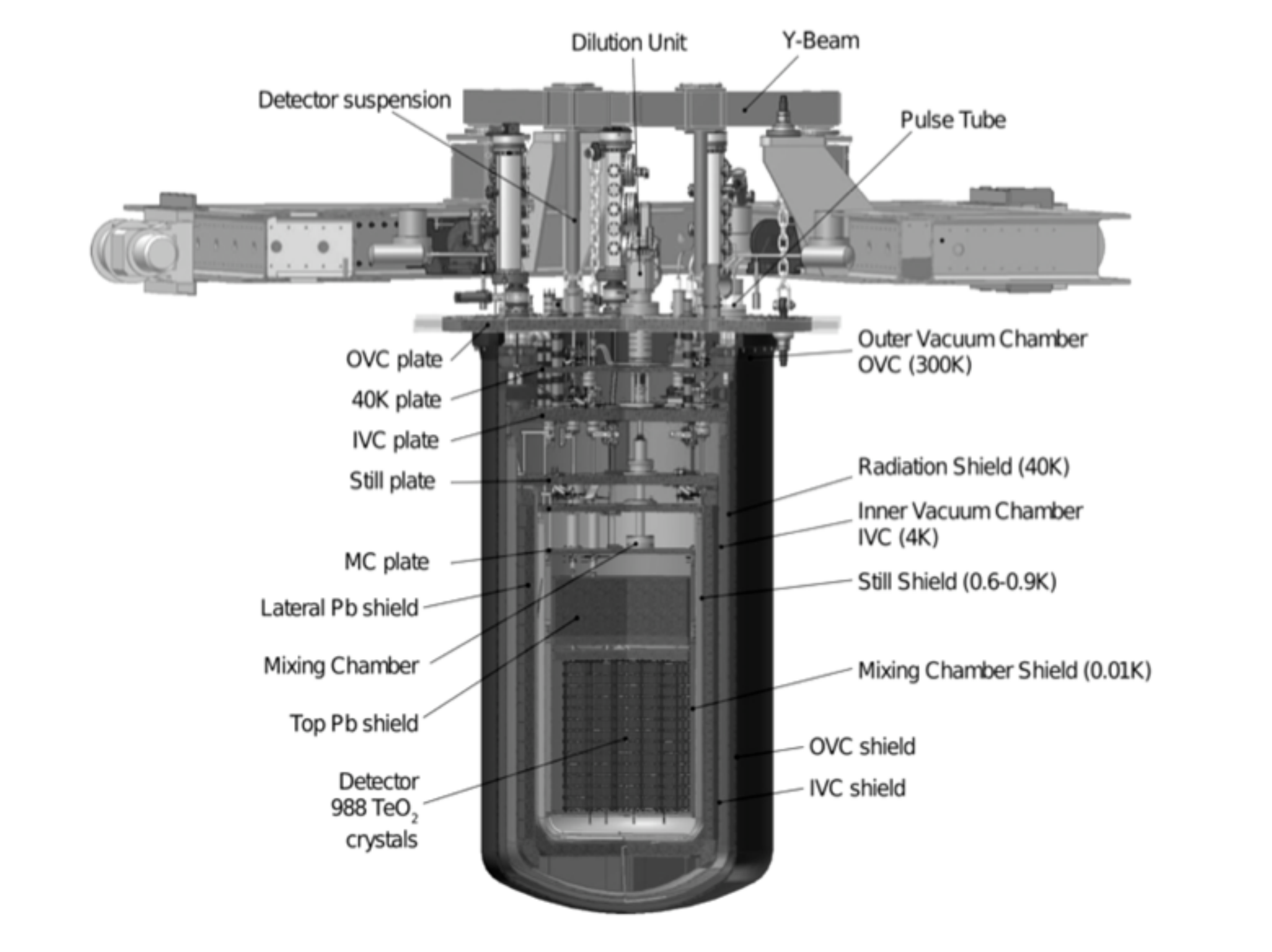}  
\includegraphics[%
  width=0.23\linewidth,keepaspectratio]{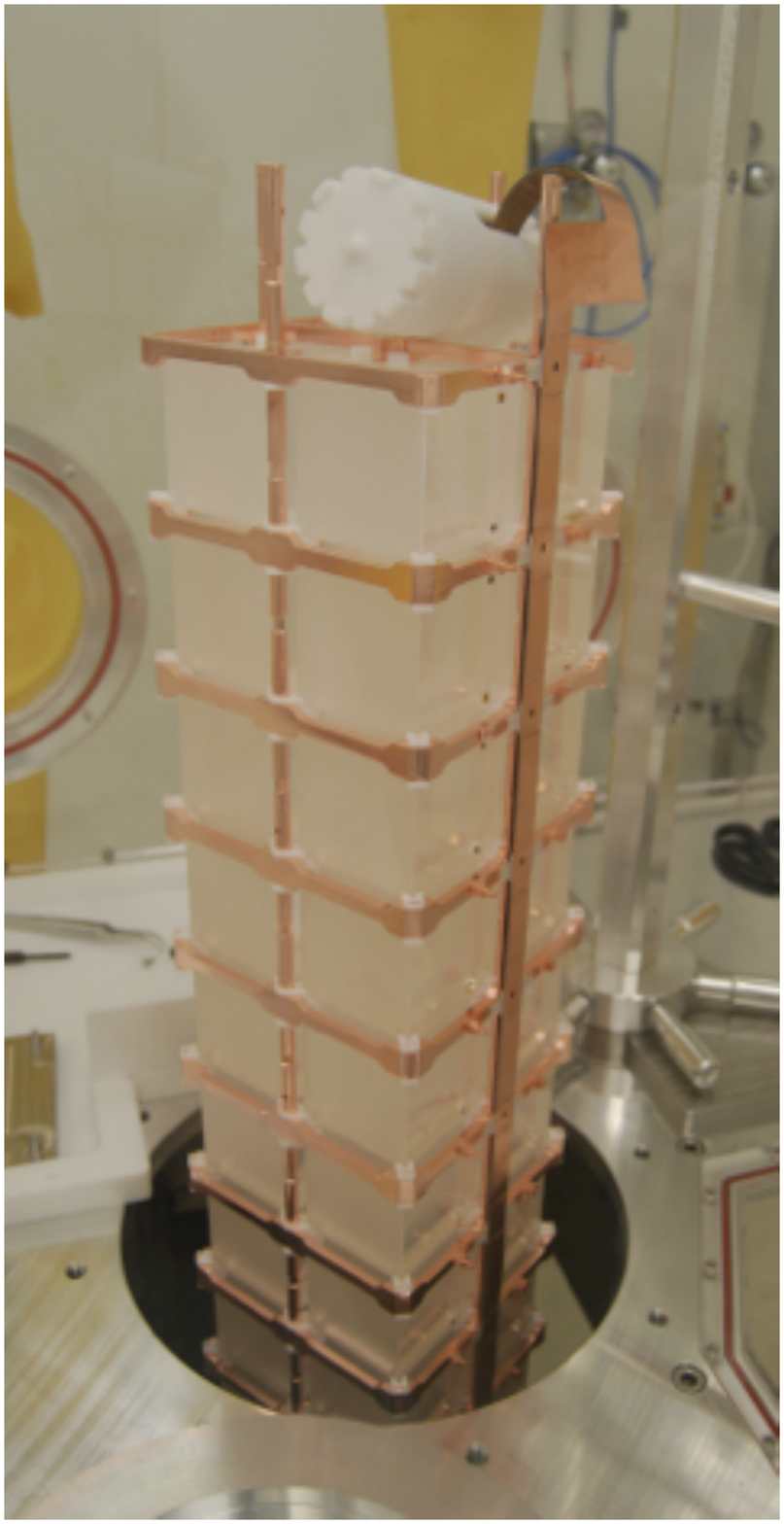}
\end{array}$
\end{center}
\caption{Left: artistic view of the CUORE detector inside the cryostat; the different thermal shields of the cryostat and the internal Roman Pb shield are also visible. Right: CUORE-0 tower during its construction phase.}
\label{fig:CUORE}
\end{figure}

The CUORE detector (see Fig.\,\ref{fig:CUORE}) will be operated by cooling crystals of TeO$_2$ to $\sim$ 10 mK inside a dilution-refrigerator cryostat. The response of each crystal --namely, its energy scale and gain-- is monitored via monthly calibrations with a $^{232}$Th gamma source and by injecting controlled amounts of energy into the crystals at 5-minute intervals using Joule heaters glued to the crystal surface. 

The challenges in constructing CUORE are due largely to its size and the complexity of the engineering. Namely, it is difficult to prepare and
maintain the cleanliness of such a large amount of radio-pure material, and the apparatus involves many interconnected systems occupying the same space under unique conditions. Currently all the assembling lines for the production of the 988 CUORE detectors and their packing into 19 towers are installed and tested in the underground location at LNGS. All the assembling phases, including automatic gluing of the NTD thermistors on the TeO$_2$ crystals, assembling of the crystals in a copper and teflon structure, cabling of the towers, and bonding of the thermistors, are performed under nitrogen atmosphere in dedicated glove boxes to prevent Rn contaminations from air. For the same reason all the component of CUORE are stored in multiple layers radio-clean plastic foils under nitrogen atmosphere. All the assembling lines are installed in the CUORE clean room in the LNGS underground location.

\subsubsection{CUORE-0}

CUORE-0 consists of a CUORE-like tower detector  (see Fig.\,\ref{fig:CUORE}) assembled with the same assembly line and procedure of CUORE.  The CUORE-0 detector is operated in the former CUORICINO dilution refrigerator. The main goal of this project is to perform a full test and debug of the hardware and procedures developed for the CUORE detector. It will also allow to test the improved uniformity of the detectors, and check the improvement of the radioactive background. It will also serve as a test of the analysis tools prepared for CUORE. Moreover, it will be a stand-alone 0$\nu$DBD experiment able to improve the limits achieved by CUORICINO.
The two main steps of the CUORE-0 detector realisation are the sensor-to-crystal connection and the assembly of the tower. These operations have been performed inside the CUORE clean room in the CUORE Hut.

The main difference between CUORE-0 and CUORE is the shielding from external radiation, which is limited by the dimensions of the CUORICINO cryostat to 2.5 cm of roman lead internal shield. Moreover the CUORE detector compact layout will guarantee that the inner TeO$_2$ crystals are shielded by the outer once form contaminations generated inside the roman lead shield, i.e. on the cryostat copper shields. 

CUORE-0 started its cold run in August 2012 and is currently in data taking phase. 
\\

\begin{center}
{\it References}
\end{center}
\begin{description}
\footnotesize	
\item[1] C. Arnaboldi, {\it et al.} [CUORE Collaboration], \emph{\it Nucl. Instrum. Methods A} \textbf{518}, 775 (2004).
\item[2] E.Andreotti,  {\it et al.} [Cuoricino Collaboration], \emph{\it Astropart. Phys.} \textbf{34}, 822 (2011).
\item[3] M. Redshaw, B. J. Mount, E. G. Myers and F. T. Avignone, \emph{\it Phys. Rev. Lett.} \textbf{102}, 212502 (2009);  N. D. Scielzo et al., \emph{\it Phys. Rev. C} \textbf{80}, 025501 (2009); S. Rahaman et al., \emph{\it Phys. Lett. B} \textbf{703}, 412 (2011)
\item[4] F.Simkovic {\it et al.}, \emph{\it Phys.\ Rev.\  C\/} \textbf{77}, 045503, (2008); O.Civitarese {\it et al.},  \emph{\it J. Phys. Conf. Ser. } \textbf{173}, 012012 (2009) ; J. MenŽndez  {\it et al.},  \emph{\it Nucl. Phys. A } \textbf{818}, 139 (2009); J. Barea Phys. {\it et al.},  \emph{\it Rev. C } \textbf{79}, 044301 (2009)
\item[5] F.Bellini {\it et al.},{\it Astropart. Phys.}, \textbf{33}, 169 (2010); C.Arnaboldi  {\it et al.}, {\it Phys.Rev.} \textbf{C78}, 035502 (2008) 
\item[6] F.~Alessandria, {\it et al.}, {\it Astropart. Phys.}, \textbf{35}, 839 (2012)
\item[7]  F.~Alessandria, {\it et al.}, 
 http://arxiv.org/abs/1109.0494v2.
\end{description}

  \newpage\subsection{M. Sisti: ``The future of neutrinoless double beta decay searches with thermal detectors''}
\begin{description}
\it\small 
\setlength{\parskip}{-1mm}
\item[\small\it M.S.:] Dip. di Fisica ``G. Occhialini'', Universit\`a di Milano-Bicocca and INFN Sezione di Milano-Bicocca, Milano, Italy
\end{description}

Neutrinoless double beta decay (\bbn) is a unique tool for investigating neutrino properties [1]. Particularly interesting is the proposed mechanism of virtual exchange of a massive Majoarana neutrino: in this case the decay half life is directly related to the effective Majorana mass \mee. In the next decade, the achievable sensitivity by the so-called \emph{second generation} experiments is expected to approach the inverted hierarchy region of the mass spectrum (10\,meV$\lesssim |m_{ee}| \lesssim 50$\,meV), but none of the proposed projects will be able to fully exclude this region.
Improving the \mee\ sensitivity of future experiments requires to increase the isotope mass (i.e. increasing detector mass and isotopic abundance)
and to drastically reduce the background level. \\
Low temperature detectors (LTDs) are successfully used for \bbn\ since 
long time [2], because they are true calorimeters with very high energy resolution and can be made out of many different materials containing 
$\beta\beta$ active isotopes. CUORE [3],  a ton scale experiment using LTDs made with \teo\ crystals to search for \Te\ \bbn,  is going to be the most sensitive upcoming experiment: its data taking is expected to start at the end of 2014. Still, the advantages of the low temperature technique have not been completely exploited, and LTDs can continue playing a primary role in the \bbn\ field, especially concerning background reduction. First of all, LTDs can be used to search for \bbn\ of isotopes with transition energy \Qbb\ above natural $\gamma$ ($\leq 2.615$\,MeV) and $\beta$ ($\leq 3.27$\,MeV) radioactivity. Furthermore, with scintillating crystal LTDs [4], e/$\gamma$ vs. $\alpha$ particle identification can be achieved by simultaneous light emission measurement or, for some scintillating crystals, by pulse shape analysis, thus eliminating the main background source for isotopes with transition energy above 3\,MeV [5,6,7]. There are also other tools for particle identification available to LTD experiments, like the detection of Cherenkov light in non-scintillating crystals [8,9]. A pioneering scintillating LTD \bbn\ search is presently completing its R\&D phase [10].

It is interesting to estimate the ultimate sensitivity of LTD based \bbn\ searches by assuming a set of very aggressive hypothesis on a future
experiment. First of all we assume that all external and internal sources of radioactive background are made negligible by adopting passive (e.g. shielding, material selection and purification) and active (e.g. detector granularity and e/$\gamma$ vs. $\alpha$ particle identification) measures.
The only background left is then the intrinsic one due to the allowed two-neutrino decay (\bb): while the LTD high energy resolution $\Delta E$ guarantees that practically no \bb\ decay will leak into the region of interest (i.e. about \Qbb$\pm\Delta E$), that is not true for the random coincidences between two \bb\ decays [11].
Table~\ref{tab:ihe} reports the 90\% C.L. sensitivity achievable for four \bbn\ candidates for which good LTDs can be used: \Se, \Cd, \Mo\ and \Te.
The sensitivity is calculated assuming a 90\% isotopic enrichment, a detector mass of  1\,ton, an energy resolution of  $\Delta E = 5$\,keV, a 99.9\% $\alpha$ rejection capability, and a measuring time of  10\,years. For the detector efficiency an average value of 0.8 has been assumed. The background in the region of interest due to random coincidences between \bb\ has been calculated according to~[11]
for a detector size of $5\times5\times5$\,cm$^3$ and a time resolution of  1\,ms. The \mee\ values are calculated using the same selection of nuclear matrix elements adopted for CUORE results [3].
Table~\ref{tab:ihe} shows that even under these extremely challenging hypothesis the sensitivity achievable with next generation LTD experiments may be
not enough for fully probing the inverted hierarchy region of the mass spectrum.
\begin{table}[h!]
\caption{\label{tab:ihe}} Sensitivity of hypothetical future \bbn\ searches (see text for details). The spread in \mee\ values is due to uncertainties in nuclear matrix element evaluations [1].
\begin{center}
\begin{tabular}{ccccc}
\hline 
crystal & isotope & \bb\ bckg & $\tau_{0\nu}$ & \mee \\
& & [counts/ton/year] & [years] & [meV] \\
\hline
ZnSe & \Se & 2.7$\times 10^{-2}$ & 6.5$\times 10^{27}$ & $6\div 18$ \\
CdWO$_4$  & $^{116}$Cd & 7.4$\times 10^{-2}$ & 3.0$\times 10^{27}$ & $9\div 17$ \\
ZnMoO$_4$ & $^{100}$Mo  & 1.5 & 1.4$\times 10^{27}$ & $9\div 25$ \\
TeO$_2$ & $^{130}$Te & 5.0$\times 10^{-4}$ & 6.8$\times 10^{27}$ & $6\div 15$ \\
\hline
\end{tabular}
\end{center}
\end{table}

\begin{center}
{\it References}
\end{center}
\begin{description}
\footnotesize	
\item[1] S. Bilenky, C. Giunti, Mod. Phys. Lett. A27 (2012) 1230015
\item[2] E. Andreotti et al., Astropart. Phys. 34 (2011) 822
\item[3] F. Alessandria et al., "Sensitivity of CUORE to Neutrinoless Double Beta Decay", submitted to Astrop. Phys. (2013)
\item[4] A. Alessandrello et al., Phys. Lett. B 420 (1998) 109
\item[5] C. Arnaboldi et al., Astropart. Phys. 34 (2010) 143
\item[6] J.W. Beeman et al., Eur. Phys. J. C72 (2012) 2142
\item[7] C. Arnaboldi et al., Astropart. Phys. 34 (2011) 344
\item[8] T. Tabarelli de Fatis, Eur. Phys. J. C65 (2010) 359
\item[9] J. Beeman et al., Astropart. Phys. 35 (2012) 558
\item[10] F. Ferroni, Nuovo Cimento C 033 (2010) 27
\item[11] D. Chernyak et al., Eur. Phys. J. C72 (2012) 1.
\end{description}

 \newpage\subsection{M. Sorel (on behalf of the NEXT Collaboration): ``The NEXT Experiment''}
\begin{description}
\it\small 
\setlength{\parskip}{-1mm}
\item Instituto de F\'isica Corpuscular (IFIC), CSIC \& Univ.\ de Valencia, Valencia, Spain
\end{description}
\subsubsection{\label{subsubsec:sorel_xenon_experiments}Xenon-based $\beta\beta0\nu$ experiments}
The field of neutrinoless double beta decay ($\beta\beta0\nu$, in the following) has witnessed in recent years the development of new experimental techniques to search for this hypothetical process. Xenon-based experiments have demonstrated to be an attractive and practical alternative to germanium diodes and tellurium bolometers, the customary technology choice in the field thanks to their excellent energy resolution. The best known examples of current-generation $\beta\beta0\nu$ xenon experiments are the KamLAND-Zen liquid scintillator detector [1] and the EXO-200 time projection chamber (TPC, [2]). Both experiments use the $^{136}$Xe isotope, a $\beta\beta$ emitter with a Q-value of 2.458 MeV. In the former case, the $\beta\beta$ source is provided by an inner balloon filled with 300 kg of $^{136}$Xe dissolved in liquid scintillator, placed in the center of the detector. In the latter case, 90 kg of $^{136}$Xe in its liquid phase fill the TPC, simultaneously acting as source and detector material. The KamLAND-Zen and EXO-200 experiments have not found any evidence for $\beta\beta0\nu$ decay in $^{136}$Xe so far. When combined, the two constraints provide the most stringent half-life lower limit among all $\beta\beta$ isotopes, with $T_{1/2}^{0\nu}(^{136}\mbox{Xe})>3.4\times 10^{25}$ yr at 90\% confidence level. This translates into an upper limit on the effective neutrino Majorana mass of 120--250 meV, for a range of plausible nuclear matrix element values [1]. These results demonstrated that the choice of xenon is already a mature technology for $\beta\beta0\nu$ searches. Aditionally, among all leading $\beta\beta$ candidate isotopes, $^{136}$Xe appears to be the easiest to procure in large quantities, with an indicative cost per unit $\beta\beta$ isotope mass that is about an order of magnitude lower than $^{76}$Ge [3]. The excellent results already provided by current-generation experiments for $\beta\beta0\nu$ searches, together with their relatively easy scalability, make xenon-based experiments a natural choice for future ton-scale $\beta\beta0\nu$ experiments.

In this presentation, a third detector technology using $^{136}$Xe has been discussed, the one to be adopted by the NEXT experiment. This approach is complementary to the KamLAND-Zen and EXO ones, and allows for potentially lower backgrounds, thanks to two unique features: near-intrinsic energy resolution (0.5-1\% FWHM at $Q_{\beta\beta}$), and information on the $\beta\beta$ decay topology via tracking and dE/dx measurements. The NEXT project started in 2009 with an R\&D phase now nearing completion, and construction of the NEXT-100 detector at the Laboratorio Subterraneo de Canfranc (LSC, Spain) has started in 2013. NEXT is an international collaboration with about 80 collaboration members from 14 different institutions from Spain, USA, Portugal, Russia and Colombia.

\subsubsection{\label{subsubsec:sorel_detection_concept}The NEXT detection concept}
Before describing the NEXT R\&D results and the NEXT-100 detector, it is useful to illustrate the detection concepts common to both phases of the project. The NEXT-100 detector is based on the following ideas, which have all been extensively tested via prototypes. First, the detector is a room-temperature TPC filled with xenon gas at about 15 bar pressure. Second, the ionization signal created by $\beta\beta$ decay electrons is amplified using electroluminescence (EL), in order to maintain the ionization yield fluctuations at a level that is close to the intrinsic limit provided by the Fano factor of xenon. Third, once amplified and converted into light, the ionization signal is read by two read-out planes, separately optimized to provide accurate calorimetric and tracking information, respectively. Fourth, the same photo-detectors used for the energy measurement are also used to detect the prompt primary scintillation of xenon following a $\beta\beta$ event, thereby providing the decay time that is needed to reconstruct the drift spatial coordinate. Fifth, the VUV (170 nm) xenon light produced both by primary scintillation and by electroluminescence is shifted to longer wavelengths. This provides better light collection efficiency, a more uniform detector response and permits the use of non-VUV-sensitive photo-detectors.

\subsubsection{\label{subsubsec:sorel_next_randd}Results from the NEXT R\&D phase}
An intense R\&D activity geared toward the NEXT-100 construction started in 2009, in order to acquire the needed technological know-how (2009-2010), to choose the detector concept (2011) and to finalize its technical design (2012). In particular, the main goals of the R\&D phase were: (i) to demonstrate a near-intrinsic energy resolution in a large active volume; (ii) to demonstrate the topological signature of electrons in high-pressure xenon gas (reconstruction of electron tracks and dE/dx along track) and the performance of MPPCs as tracking read-out elements; (iii) to test long drift lengths and high (up to 50 kV) voltages; (iv) to understand gas recirculation and purification in a large volume, including operation stability and robustness against gas leaks; (v) to understand the light collection efficiency, with and without wavelengths shifters. To this end, several small-scale (up to 2.5 kg of natural xenon) prototypes have been built and operated at NEXT collaborating institutions. Two EL-based prototypes (NEXT-DBDM at LBNL, and NEXT-DEMO at IFIC Valencia) are fully operational since 2011, and selected results from those are presented in the following.

\begin{figure}[t!b!]
\begin{center}
\includegraphics[scale=.50]{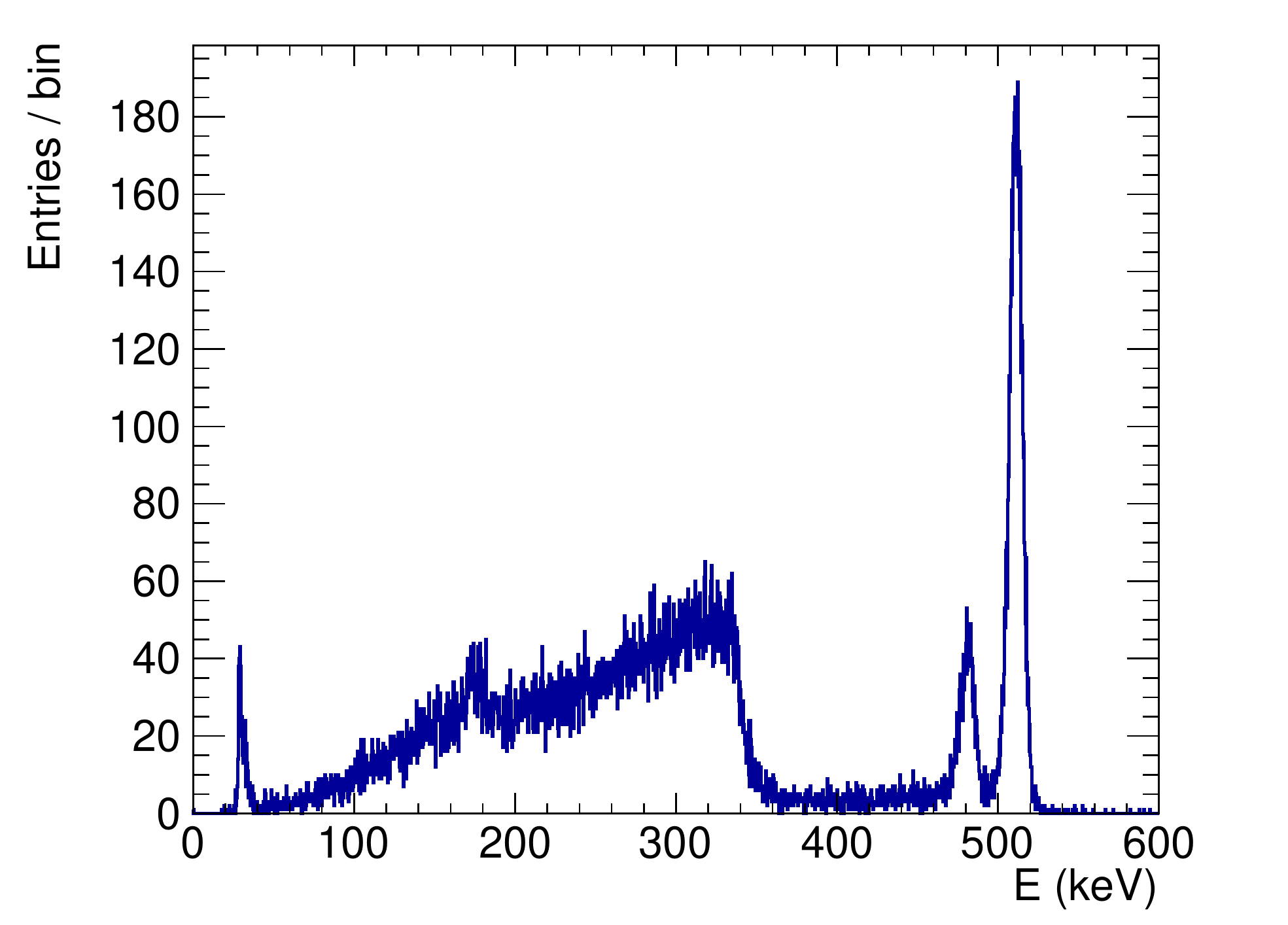}
\end{center}
\caption{Energy spectrum for 511 keV gammas interacting in NEXT-DEMO. From low to high energies, the X-ray peak ($\sim$30 keV), the Compton continuum (100--340 keV), the X-ray escape peak ($\sim$480 keV) and the photo-electric peak (full energy) are clearly visible.}\label{fig:sorel_spectrum}
\end{figure}

Concerning energy resolution, NEXT-DBDM has accomplished a 1\% FWHM resolution for full-energy (photo-electric) events from 662 keV gamma interactions occurring within the central ($\simeq$1 cm radius) region of the detector [4]. If assumed to follow a $E^{-1/2}$ dependence, this resolution extrapolates to about 0.6\% FWHM at $Q_{\beta\beta}$=2.458 MeV. A good resolution was also achieved in the larger (3.5-5 cm radius) fiducial volume of the NEXT-DEMO prototype, see Fig.~\ref{fig:sorel_spectrum} for a typical spectrum. In this case, an energy resolution of about 1.8\% FWHM for 511 keV electrons was obtained, extrapolating to about 0.8\% FWHM [5]. In the NEXT-DEMO (and NEXT-100) case, a mapping of the energy response of the detector in the (x,y) plane perpendicular to the drift field is necessary, to correct for the $\sim$10\% level variations in the detector energy response with spatial position.

Regarding the topological signature, work has concentrated on two fronts. On the one hand, the MPPC-based read-out planes in NEXT-DBDM and NEXT-DEMO have been used to demonstrate tracking capabilities. Straight cosmic ray muon tracks, $\sim$500 keV electrons tracks dominated by multiple Coulomb scattering, and isolated X-ray energy depositions of about 30 keV have been reconstructed [4]. On the other hand, the tracking plane information can be combined with the energy (PMT) plane information in order to identify the number of Bragg peaks signaling the number of electrons ranging out in the detector. This is useful for $\beta\beta0\nu$ searches, given that the ``blob'' (\textit{ie}, a track segment with higher energy deposition) multiplicity per event is expected to provide an additional background suppression factor. Signal events tend to yield two ``blobs'' from two electrons emanating from a common vertex. In the case of background, dominated by gamma interactions, only one ``blob'' per event is typically expected. Energy blobs where electron tracks range out have been clearly identified in NEXT-DEMO using the energy plane information, by projecting the electron tracks' dE/dx pattern along the drift direction [5].

In addition to demonstrating the R\&D goals toward NEXT-100 construction, the NEXT prototypes are ideal tools to study the xenon gas detection properties in a broader context. Ionization electron transport properties (drift velocity, longitudinal diffusion) have been accurately measured using alpha particles and X-rays [4,6]. In addition, electron-ion recombination studies have been carried out. In xenon gas, recombination is important only for highly ionizing particles such as alpha particles, not for electrons. In particular, the first observation of correlated fluctuations between ionization and scintillation in xenon gas was obtained using the NEXT-DEMO prototype [6].  

\subsubsection{\label{subsubsec:sorel_next100}NEXT-100}
The NEXT-100 detector will be a TPC filled with about 100 kg of xenon gas at 15 bar pressure. The xenon will be depleted in the $\beta\beta$ isotope $^{136}$Xe during a first phase (for commissioning, background studies), and isotopically enriched to a 90\% fraction during the second phase (physics run). The experiment is now under construction at the Laboratorio Subterraneo de Canfranc (LSC, Spain). The main features of the NEXT-100 design are given below [7].

The pressure vessel is a 140 cm diameter, 230 cm length cylinder made of low-radioactivity stainless steel (alloy 316Ti), capable of withstanding a 15 bar pressure. In order to screen radioactive contaminants in the vessel material, the inner surface of the vessel is covered with a radiopure (5-10 $\mu$Bq/kg)  inner copper shield of 12 cm thickness. The TPC field shaping rings are mounted on the inside of a 2.5 cm thick cylindrical shell made of high-density polyethylene, in contact with the inner copper shield. The cathode, gate and anode wire meshes defining the drift (cathode to gate) and EL (gate to anode) regions are similar to the NEXT-DEMO ones. We expect to operate the detector at a 0.3 kV/cm drift field and a 3 kV/(cm$\cdot$bar) EL field, corresponding to an optical gain of about $2.5\times 10^3$ EL photons per ionization electron. Reflector panels coated with tetraphenyl butadiene (TPB), again similar in concept to the NEXT-DEMO panels, cover the entire barrel surface. The TPB material shifts the VUV (170 nm) light into blue (430 nm) light. The energy plane on one end-cap is made of 60 low-radioactivity PMTs (Hamamatsu R11410-10) housed in pressure-resistant enclosures made of radiopure copper, providing 30\% photo-cathode coverage. The PMTs have a 35\% (30\%) quantum efficiency for 170 nm (430 nm) light. The tracking plane on the other end-cap consists of 7,000 MPPCs (Hamamatsu S10362-11-050P) mounted on TPB-coated Cuflon boards, at 1 cm pitch. These MPPCs have a photon detection efficiency of about 50\% for 430 nm photons, and are not sensitive to VUV light. The tracking plane electrical lines for power supply, signal and ground are extracted through the vessel via three custom-made, pressure-tight, flat cable connectors. The detector is shielded against gamma rays from LSC laboratory walls via a 20 cm thick lead castle structure. 

A careful selection of all detector components that are either massive or inside the vessel has been made. This material screening campaign has been carried out mainly via germanium gamma-ray spectrometry measurements at the LSC [8]. Background rates expected in NEXT-100 have been estimated from these material screening measurements and from the background rejection factors obtained from simulations. The main backgrounds around $Q_{\beta\beta}$ are expected to be de-excitation gammas from $^{208}$Tl and $^{214}$Bi decay daughters. For such backgrounds, $\lesssim 2\times 10^{-7}$ rejection factors are expected from simple (and preliminary) reconstruction and event selection algorithms. Overall, a background rate at the level of $8\times 10^{-4}$/(keV$\cdot$kg$\cdot$yr) or better is expected by summing all $^{208}$Tl and $^{214}$Bi contributions. The detector inner elements (two read-out planes, and possibly the field cage) are expected to dominate the background budget. As a result, NEXT-100 should be sensitive to Majorana masses as small as 100 meV after five years of operation. This figure is comparable to the projected sensitivities of the most promising current-generation experiments, including EXO-200, KamLAND-Zen, GERDA-2 and CUORE [9].

\subsubsection{\label{subsubsec:sorel_next_ton}Ideas for a ton-scale detector}
If successful at the 100 kg scale with NEXT-100, the NEXT approach could be further pursued with a more sensitive detector. This second-generation detector should be capable of fully exploring the inverted hierarchy of neutrino masses, that is of reaching an effective Majorana mass sensitivity as low as 15--20 meV. In order to fulfill this goal, an exposure in excess of 10 ton$\cdot$yr is required, implying a detector at the ton or (more likely) multi-ton scale. In addition, to fully explore this exposure increase compared to NEXT-100, the background rate of such future detector should not exceed 0.1--1 background counts per ton$\cdot$yr. Even though still at a very early stage of development, several ideas on how such a future detector might look like are being discussed within the NEXT Collaboration already. This detector would likely be based on the same NEXT detection concepts discussed above (Sec.~\ref{subsubsec:sorel_detection_concept}). However, ideas for possible improvements over the NEXT-100 design are being considered, for example: (i) a mild increase in operating pressure; (ii) the use of gas additives to shift the 170 nm xenon light; (iii) the possibility of a higher light collection and a better PMT shielding by guiding the light to the PMTs through wavelength-shifting plastic panels; (iv) a different tracking readout system for a more powerful identification of the $\beta\beta$ topology, especially if operating in higher pressure conditions; (v) the addition of an active outer veto system (such as an instrumented  water or liquid scintillator tank) against vessel and external backgrounds. These ideas will need to mature over the next few years, learning from the NEXT-100 experience and from dedicated R\&D setups. 

\subsubsection*{\label{subsubsec:sorel_acknowledgments}Acknowledgments}
I would like to thank the workshop organizers for arranging this lively and inter-disciplinary forum for discussion on the topic of neutrino mass. This work was supported by the Ministerio de Economia y Competitividad of Spain under grants CONSOLIDER-Ingenio 2010 CSD2008-0037 (CUP) and FIS2012-37947-C04-01.

\begin{center}
{\it References}
\end{center}
\begin{description}
\footnotesize	
\item[1] A.~Gando {\it et al.}  [KamLAND-Zen Collaboration], Phys.\  Rev.\  Lett.\  {\bf 110} (2013) 062502 [arXiv:1211.3863 [hep-ex]].
\item[2] M.~Auger {\it et al.}  [EXO Collaboration], Phys.\ Rev.\ Lett.\  {\bf 109} (2012) 032505 [arXiv:1205.5608 [hep-ex]].
\item[3] A.~Giuliani and A.~Poves, Adv.\ High Energy Phys.\  {\bf 2012} (2012) 857016.
\item[4] V.~Alvarez {\it et al.}  [NEXT Collaboration], Nucl.\ Instrum.\ Meth.\ A {\bf 708} (2013) 101 [arXiv:1211.4474 [physics.ins-det]].
\item[5] V.~Alvarez {\it et al.}  [NEXT Collaboration], JINST {\bf 8} (2013) P04002 [arXiv:1211.4838 [physics.ins-det]].
\item[6] V.~Alvarez {\it et al.}  [NEXT Collaboration], arXiv:1211.4508 [physics.ins-det].
\item[7] V.~Alvarez {\it et al.}  [NEXT Collaboration], JINST {\bf 7} (2012) T06001 [arXiv:1202.0721 [physics.ins-det]].
\item[8] V.~Alvarez, I.~Bandac, A.~Bettini, F.~I.~G.~M.~Borges, S.~Carcel, J.~Castel, S.~Cebrian and A.~Cervera {\it et al.}, JINST {\bf 8} (2013) T01002 [arXiv:1211.3961 [physics.ins-det]].
\item[9] J.~J.~Gomez-Cadenas, J.~Martin-Albo, M.~Mezzetto, F.~Monrabal and M.~Sorel, Riv.\ Nuovo Cim.\  {\bf 35} (2012) 29 [arXiv:1109.5515 [hep-ex]].
\end{description}

 \newpage\subsection{M. Faverzani, C. Arnaboldi, G. Ceruti, E. Ferri, F. Gatti, A. Giachero, C. Gotti, C. Kilbourne, S. Kraft-Bermuth, M. Maino, A. Nucciotti, G. Pessina, D, Schaeffer, M. Sisti: ``Status of the Mare-1 in Milano experiment''}
\begin{description}
\it\small 
\setlength{\parskip}{-1mm}
\item[\small\it M.F., C.A., G.C., E.F., A.G., C.G., M.M., A.N., G.P., M.S.:]Dip. di Fisica ``G. Occhialini'', Universit\`a di Milano-Bicocca and INFN Sezione di Milano-Bicocca, Milano, Italy
\item[\small\it F.G.:]Universit\`a di Genova and INFN Sezione di Genova, Genova, Italy
\item[\small\it C.K.:]Goddard Space Flight Center, Greenbelt, Maryland, USA
\item[\small\it S.K.:]Institut f\"{u}r Physik, Johannes-Gutenberg-Universit\"{a}t Mainz, Germany
\item[\small\it D.S.:]ABB AB, Corporate Research, Vster\aa{}s, SE
\end{description}
%
\subsubsection{Abstract}
The international project MARE (Microcalorimeter Array for a Rhenium Experiment) aims at the direct and calorimetric measurement of the electron neutrino mass with sub-eV sensitivity. Although the baseline of the MARE project consists in a large array of rhenium based thermal detectors, a different option for the isotope -$^{163}$Ho- is also being considered. The potential of using $^{187}$Re for a calorimetric neutrino mass experiment has been already demonstrated, while no calorimetric spectrum of $^{163}$Ho has been measured so far with the precision required to set a useful limit on the neutrino mass. The first phase of the project (MARE-1) is a collection of activities aimed to sort out both the best isotope and the most suited detector technology to be used for the final experiment. One of the MARE-1 activities is carried out in Milan by the group of Milano-Bicocca in collaboration with NASA/GSFC and Wisconsin groups. The Milan MARE-1 arrays are based on semiconductor thermistors, provided by the NASA/GSFC group, with dielectric silver perrhenate absorbers, AgReO$_{4}$. The experiment is designed to host up to 8 arrays. With 288 detectors, a sensitivity of 3 eV at 90\% CL on the neutrino mass can be reached within 3 years.

\subsubsection{Introduction}
Nowadays it is known that neutrinos are massive particles. The experiments based on kinematic analysis of electrons emitted in single $\beta$-decay are the only ones dedicated to effective electron-neutrino mass determination. The method consists in searching for a tiny deformation caused by a non-zero neutrino mass to the spectrum near its end point. The most stringent results come from electrostatic spectrometers on tritium decay (E$_0$ =18.6 keV). The Troitsk experiment has set an upper limit on neutrino mass of 2.5 eV/c$^2$ [1] , while the Mainz collaboration has reached m$_{\nu}$ $\leq$ 2.3 eV/c$^2$ [2]. KATRIN, the next generation experiment, is designed to reach a sensitivity of 0.2 eV/c$^2$ in five years [3],[4]. Calorimetry is an alternative approach to spectrometry, where the $\beta$-source is embedded in the detector, so that all the energy emitted in the decay is measured, except for the one carried away by the neutrino. In this way the measurement is not affected by systematics due to energy loss in the source and to undetected excited final states. The residual systematics may be due to energy lost in metastable states living longer than the detector response time. Differently from the spectrometric approach, the full beta spectrum is acquired. Therefore, the source activity has to be limited to avoid pile-up effects which would deform the shape of beta spectrum. As a consequence the statistics near the end-point is limited as well. This limitation might be partially balanced by using $\beta$-emitting isotopes with an end-point energy as low as possible.

MARE is a new large scale experiment aimed to measure directly the neutrino mass with the calorimetric technique. The MARE project has a staged approach. The goal of the last phase (MARE-2) is to achieve a sub-eV sensitivity on neutrino mass, while the first phase (MARE-1) is a collection of activities aimed to sort out both the best isotope and the most suited detector technology to be used for the final experiment. The two competing isotopes are $^{187}$Re and $^{163}$Ho. Rhenium is in principle suited for fabricating thermal detectors: metallic rhenium crystals, or dielectric compounds, allow to reach high sensitivity thanks to their low thermal capacity. Up to now, only two $\beta$ decay experiments have been carried out with thermal detectors: MANU [5],[6] and MIBETA [7],[8] experiments. MANU used metallic rhenium single crystal as absorber, while MIBETA used AgReO$_4$ crystals. Collecting a statistic of about 10$^7$ events, they achieved an upper limit on neutrino mass of about 26 eV/c$^2$ at 95\% CL and 15 eV/c$^2$ at 90\% CL, respectively. In these experiments, the systematic uncertainties are still small compared to the statistical errors. The main sources of systematics are the background, the theoretical shape of the $^{187}$Re $\beta$ spectrum and the detector response function. In order to have an alternative to the rhenium as $\beta$ source, the MARE collaboration is considering the possibility to use $^{163}$Ho electron capture (EC) [9]. $^{163}$Ho EC decay has been the subject of many experimental investigations aimed at determining the neutrino mass thanks to its low transition energy ($\sim$ 2.5 keV). The EC may be only detected through the mostly non radiative atom de-excitation of the daughter atom ($^{163}$Dy) and from the Inner Bremsstrahlung (IB) radiation. So far, there has been no experiment attempting to study the end-point of the total absorption spectrum, as proposed by De Rujula and Lusignoli [10]. The total spectrum is composed by peaks with Breit-Wigner shapes and it ends at E$_0$ - m$_{\nu}$, in analogy to what happens for $\beta$ spectrum.

More MARE-1 activities are devoted to the design of the single detector for the final MARE large scale experiment and to the optimization of the coupling between rhenium crystals - or ${163}$Ho implanted absorbers - and sensitive sensors, like Transition Edge Sensor (TES) [11], Metallic Magnetic Calorimeters (MMC) [12] or Microwave Kinetic Inductance Detectors (MKIDs) [13].

\subsubsection{MARE 1 in Milano}
The goal of MARE-1 in Milano is to achieve a sensitivity on the neutrino mass of $\sim$2eV, through 8 arrays provided by the NASA/GSFC group, composed by 6x6 detectors each. The detectors consist of implanted Si:P thermistors (300 x 300 x 1.5 $\mu$m$^3$) coupled with dielectric silver perrhenate absorbers, AgReO$_4$. The AgReO$_4$ absorber are grown by Mateck GmbH in Germany. Mateck has developed a procedure to grow large single crystal with high purity and to cut them as precisely as possible. Our crystals are cut in regular shape of 600 x 600 x 250 $\mu$m$^3$, corresponding to a mass around 500 $\mu$g, giving 0.27  dec/s. To adapt the thermistors developed by NASA to our purpose, silicon spacers of 300 x 300 x 10 $\mu$m$^3$ are glued between the thermistor and the larger absorber. With such detectors it is possible to achieve an energy and time resolution of 25 eV at 2.6 keV and 250 $\mu$s, respectively. With 288 detectors and such performances, a sensitivity on neutrino mass of 3 eV at 90 \% CL can be achieved in 3 years [14]. The performance of such bolometers, which have by high impedance at low temperature (around 4 M$\Omega$ at 85 mK), depends not only on the thermistors and the quality of the crystals but also on the read-out electronics. For that reason a cold buffer stage, based on JFETs at 140 K, is installed as close as possible to the detectors to shift down the thermistor impedance. This stage is followed by a room temperature amplifier which subtracts the ground signal from the signal present at the output of the cold buffer. The presence of only one ground signal for all channels, which corresponds to the detector holder, cancels the ground loop interference. The output signal is filtered with an active Bessel low pass antialiasing filter, placed close to the DAQ system [15]. In the original design [16], the electrical contacts between the cold buffer stage (140 K) and the detectors (85 mK) were made by two different stages of microbridges. Microbridges were low thermal conductance wires produced by Memsrad/FBK in Trento, Italy. The first microbridge stage, made of Titanium, provided the thermal decoupling between the detectors and the JFET holder at 4 K. The second microbridge stage, in Aluminum, were between the JFETs and their box. During the assembly of the setup there was an unexpected failure of the two microbridge stages. For that reason an R\&D work has been dedicated to determine the best solution to replace the Ti and Al microbridges. Our studies have shown that Stainless Steel and Al/Si 1\% wires are acceptable replacements for Ti and Al microbridges, respectively. The length of Stainless Steel wires is around 2 cm and the diameter is 12.5 $\mu$m, while the length of Al/Si 1\% wires is around 1.3 cm with diameter of 17.5 $\mu$m. Stainless Steel and Al/Si 1\% wires can not guarantee the  mechanical stability of the setup. Therefore, material with low thermal conductivity are used as mechanical support, namely Kevlar and Vespel. The mechanical support of the PCB, where the Stainless Steel wires are soldered, consists of 3 Kevlar crosses; two thin Vespel rods are used as mechanical support for the JFET PCB. The energy calibration system, located between the detector holder and the JFETs boxes, consists of fluorescence sources with $^{55}$Fe, with an activity of 10mCi, as a primary source movable in and out of a Roman lead shield [16]. The fluorescence targets, made of Al, Si, NaCl and CaCO$_3$, are at the Mixing Chamber temperature and they allow a precise energy calibration around the end-point of ${187}$Re with the K$\alpha$ and K$\beta$ X-rays. A new thermal shield at 35mK has been added in order to shield the detectors from the thermal radiation. The cryogenic set-up of MARE-1 in Milan is mounted in a Kelvinox KX400 dilution refrigerator, located in the cryogenic laboratory of the University of Milano-Bicocca. It can host up to eight arrays but only two of them with their electronics have been funded so far. Therefore, the read-out electronics was installed only for 80 channels. At the beginning of the year all the available detectors (31 of 36; 5 pixel were broken) of one array have been glued with perrhenate absorbers. Since it is useful to study the thermal coupling between thermistors and silicon spacer, two different kinds of epoxy resins are tested: 6 silicon spacers are attached with Araldite Normal, 10 with Araldite Rapid and the other 15 with ST1266 epoxy. ST2850 epoxy is used to glue all the AgReO$_4$ absorbers on the silicon spacers. A measure run aimed to test the performance of this setup is scheduled for the next few months, after which the absorbers will be glued also on the second array. With two arrays, a sensitivity of 4.5 eV at 90\% C.L. is expected in three years running time.

Another MARE-1 activity consists in developing superconducting microwave microresonators for the electron-capture decay end-point measurement of the neutrino mass using Holmium [13]. These low temperature detectors are compatible with large-scale multiplexed frequency domain readout. The devices, developed in collaboration with Fondazione Bruno Kessler -FBK- in Trento, are sensitive to the variation of the number of quasi-particles in a superconductor by measuring a change of microresonator characteristic parameters. The characterization and optimization of these microresonators is still in progress. So far, the gap parameter of superconducting films made of TiN and multilayer of Ti/TiN have been measured, and they resulted to be, respectively, 0.8meV and 0.26meV. Besides, the devices were tested with test sources emitting X-rays of 1.5keV (Al fluorescence) and 6keV ($^{55}$Fe), collecting pulses which allowed to build an energy spectrum. At the present time, due to interaction of X-rays in the Si substrate (
and consequently creation of phonons), it was not possible to determine the energy resolution of the detector. In order to avoid this effect, different solution will be tested.
Subsequently the optimization work, the devices will be ready to be implanted with Ho nuclei.

\subsubsection{Acknowledgments}

The activity concerning the development of microresonators is supported by Fondazione Cariplo through the project "Development of Microresonator Detectors for Neutrino Physics (grant 2010-2351).

\begin{center}
{\it References}
\end{center}
\begin{description}
\footnotesize
\item[1] V. M. Lobashev, Nucl. Phys. B (Proc Suppl.) 91 280-286 (2001)
\item[2] Ch. Kraus, Eur. Phys. J. C 40 447-468 (2005)
\item[3] Osipowic A {\it et al.}, Letter of intent hep-ex/0109033 (2001)
\item[4] Angrik J. {\it et al.}, Design Report 7090 http://bibliothek.fzk.de/zb/berichte/FZKA7090.pdf (2004)
\item[5] F. Gatti {\it et al.}, Nucl. Phys. B 91 293 (2001)
\item[6] M. Galeazzi {\it et al.}, Phys. Rev. C 63 (2001)
\item[7] M. Sisti {\it et al.}, NIM A 520 125 (2004)
\item[8] C. Arnaboldi {\it et al.}, Phys. Rev. Lett. 91 (2003)
\item[9] F. Gatti {\it et al.}, J. Low Temp. Phys. 151 603 (2008)
\item[10] A. De Rujula and M. Lusignoli, Phys. Lett. B 118 72 (1982)
\item[11] M. Ribeiro-Gomes {\it et al.}, AIP Conf. Proc. 1185 (2009)
\item[12] J.-P. Porst {\it et al.}, J. Low Temp. Phys. 167 (2012) 254
\item[13] M. Faverzani, P. Day, E. Ferri, A. Giachero, C. Giordano, B. Marghesin and A. Nucciotti, Nucl. Instrum. Methods Phys. Res. A (2012)
\item[14] M. Ribeiro-Gomes et al, AIP Conf. Proc. 1185 (2009)
\item[15] C. Gotti {\it et al.}, J. Low Temp. Phys. 167 (2012) 620
\item[16] D. Schaeffer {\it et al.}, J. Low Temp. Phys. 151, 623–628 (2008)
%
\end{description}

 \newpage\subsection{Ulli K\"oster: ``Production and Separation of $^{163}$Ho''}
\label{Ulli}
\begin{description}
\it\small 
\setlength{\parskip}{-1mm}
\item[\small\it ]Institut Laue Langevin, Grenoble, France
\end{description}
%
``All roads lead to Rome'' and many different nuclear reactions lead to $^{163}$Ho.
Among these, two ``highways'' leading to rather pure samples of $^{163}$Ho are discussed in the following:
spallation production coupled with on-/off-line isotope separation and reactor irradiations of enriched 
$^{162}$Er samples. Other reactions induced by protons, alphas or $^7$Li beams on Dy, Er or Tb targets 
are discussed in [1,2].

\subsubsection{Isotope separation on-line of spallation products}
The on-line isotope separator ISOLDE at CERN uses 1.4 GeV proton beams to induce spallation, fragmentation and fission 
in a variety of targets and provides mass-separated radioactive ion beams of over 1000 different radioisotopes from 
over 70 elements [3].
Spallation of a tantalum foil target combined with a hot tungsten surface ionizer provides $^{163}$Ho with cumulative yields (including the precursors Er, Tm, Yb) up to $10^{10}$ ions/$\mu$C [4].
The mass separation leads naturally to rather pure samples since $^{163}$Ho is the only long-lived radioisotope on mass 163 and neighboring masses are suppressed by orders of magnitude.
Nevertheless, ``quasi-isobars'' may occur due to molecular ions being separated at the same mass.
Typically oxide sidebands and, to a lesser degree, fluoride sidebands are observed for Ta foil targets.
Hence long-lived contaminants such as $^{147}$Eu$^{16}$O$^+$, $^{147}$Pm$^{16}$O$^+$, $^{145}$Sm$^{18}$O$^+$
and $^{144}$Pm$^{19}$F$^+$ may occur together with $^{163}$Ho.
Complete elimination of such background can be achieved by an additional wet chemical separation of the holmium
fraction. In the 1980s this path was followed with samples collected at ISOLDE to produce extremely pure $^{163}$Ho samples 
for precision decay spectroscopy [5,6].
Today, with the use of the resonance ionization laser ion source (RILIS) [7], one could still slightly improve 
the cumulative $^{163}$Ho yield and the $^{163}$Ho-to-contaminants ratio respectively. The lasers could either be
tuned to resonantly ionize Ho or to ionize Yb since this isobar has the highest spallation cross-section 
on mass 163 and contributes most to the cumulative yield of $^{163}$Ho [4]. 
$^{163}$Yb has already been successfully laser ionized at ISOLDE [8].
Under optimum conditions one could expect to collect about $2 \cdot 10^{15}$ atoms (10 kBq) of $^{163}$Ho per day
of on-line operation.
However, $^{163}$Ho samples with ultimate purity, i.e. without molecular sidebands, would probably still require 
an additional radiochemical step.

\subsubsection{Reactor irradiations of $^{162}$Er}

An alternative production path uses thermal neutron capture on enriched targets of $^{162}$Er [9,10,11].
The produced $^{163}$Er decays quickly ($T_{1/2} = 75$~min) to $^{163}$Ho and can be radiochemically
separated from remaining erbium to provide $^{163}$Ho in non-carrier-added quality.
$^{162}$Er has a natural abundance of only 0.14\%, but material enriched up to 40\% 
is commercially available.

Long irradiations in a very high neutron flux are favorable for an efficient transmutation of the 
valuable target material into $^{163}$Ho.
The V4 irradiation position of ILL's high flux reactor provides thermal neutron fluxes up to $1.5 \cdot 10^{15}$ cm$^{-2}$s$^{-1}$ [12] and irradiations of 50 days or longer are possible.
A 50 day irradiation of a 10~mg target enriched to 35\% $^{162}$Er in a thermal neutron flux of $10^{15}$ cm$^{-2}$s$^{-1}$
should produce about $10^{18}$ atoms (5 MBq) of $^{163}$Ho, provided that the, yet unknown, neutron capture
cross-section of $^{163}$Ho is not too big ($\ll 100$~b). Else the product burnup $^{163}$Ho(n,$\gamma$) would limit the achievable activity of $^{163}$Ho. 

Radiochemical separation can efficiently remove stable isotopes and radioisotopes of other elements (Er, Tm, etc.) but
it cannot eliminate disturbing holmium isotopes.
Traces of stable holmium present in the target material would therefore lead to a reduced specific activity and to radionuclidic impurity of $^{166m}$Ho. A 0.1\% Ho impurity leads in the above-mentioned irradiation to a
contamination of about 500~Bq $^{166m}$Ho.
Here long irradiations in high neutron flux are beneficial since the $^{166m}$Ho activity is limited by massive 
product burnup due to the high neutron capture cross-section of $^{166m}$Ho (3100 b). At lower flux or shorter 
irradiations the ratio $^{166m}$Ho/$^{163}$Ho gets less favorable.

\begin{figure}[h]
 \centering
 \includegraphics[width=10cm]{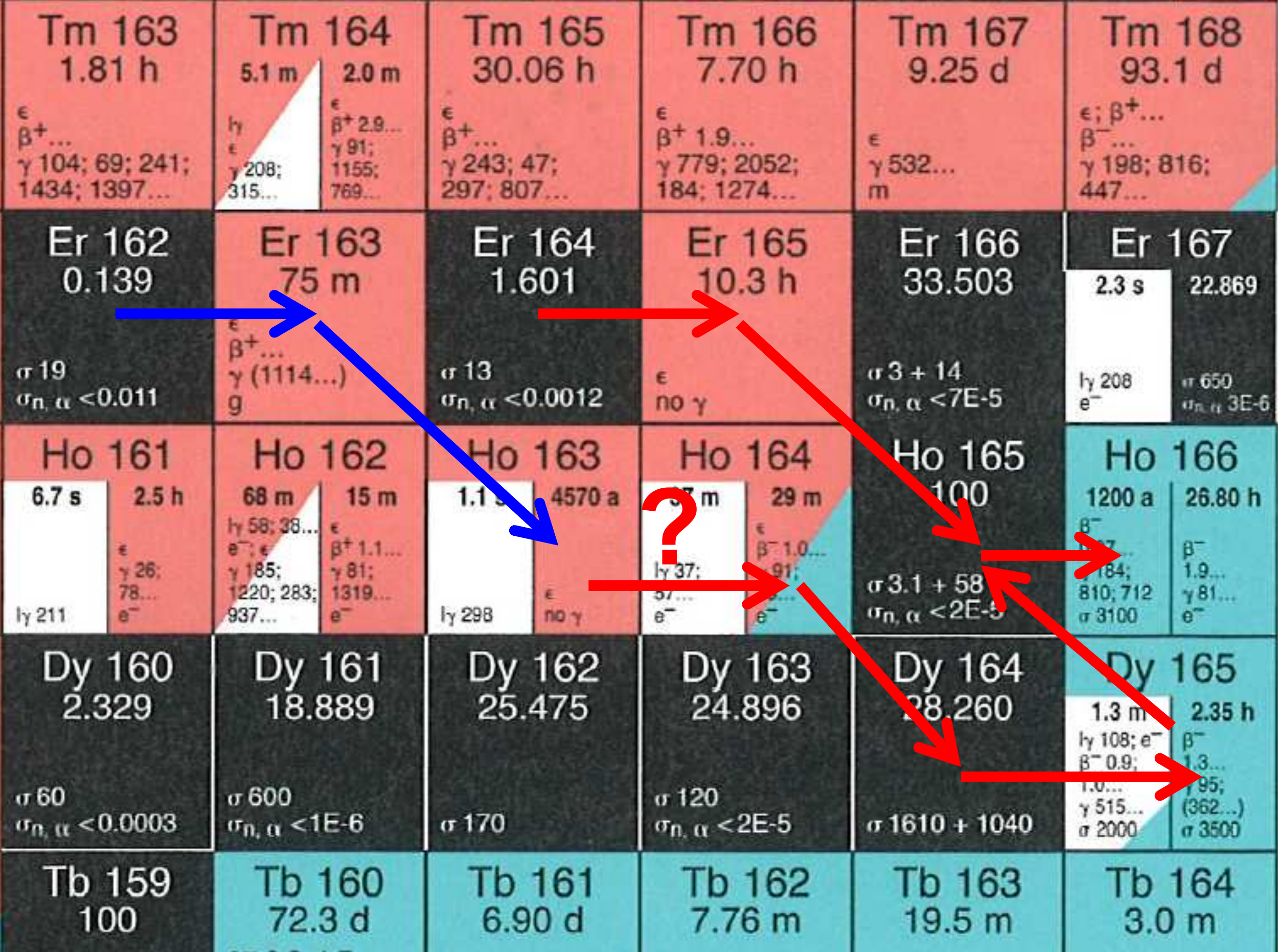}
\vspace{-3mm}
\caption{Chart of the nuclides in the vicinity of $^{163}$Ho. Blue arrows show the desired production path of $^{163}$Ho, red arrows indicate reactions leading to undesired background of stable $^{165}$Ho and long-lived $^{166m}$Ho respectively.} 
\label{chart}
\end{figure}

Also traces of dysprosium present in the target material may transmute via $^{164}$Dy(n,$\gamma$)$^{165}$Dy$(\beta^-) ^{165}$Ho(n,$\gamma)^{166m}$Ho, see Fig. \ref{chart}. A 0.1\% Dy impurity leads in the above-mentioned irradiation 
to a contamination of about 200~Bq $^{166m}$Ho.

Finally, also $^{164}$Er is transmuted to $^{166m}$Ho via $^{164}$Er(n,$\gamma$)$^{165}$Er(EC)$^{165}$Ho(n,$\gamma)^{166m}$Ho.
Every percent of $^{164}$Er produces in the mentioned irradiation about 300 Bq of $^{166m}$Ho.
The achievable $^{163}$Ho/$^{166m}$Ho ratio scales with the $^{162}$Er/$^{164}$Er ratio of the target
material.
Although the natural abundance of $^{164}$Er is only 1.6\%, enrichment of $^{162}$Er leads also to an enrichment of $^{164}$Er.
Commercially available enriched $^{162}$Er contains 3 to 8\% of $^{164}$Er, hence for high flux irradiations 
the total $^{166m}$Ho impurity is dominantly produced from $^{164}$Er and not from the chemical impurities Ho or Dy.

If, for certain applications, such $^{166m}$Ho impurities at a level of few $10^{-4}$~Bq/Bq relative 
to $^{163}$Ho are inacceptable, it could be removed by an additional off-line mass separation after the chemical separation.
Again resonant laser ionization can help to achieve good ionization efficiencies ($> 20$\% for Ho [13]).

\subsubsection{Conclusions}

On-line collections at ISOLDE can provide up to $\approx 10^{16}$ $^{163}$Ho atoms in form of a low-energy (30-60 keV) 
ion beam, suitable for direct ion-implantation into detectors. The presence of molecular sidebands limits the achievable
beam purity.
Larger quantities of $^{163}$Ho can be produced by irradiating enriched $^{162}$Er targets in a high flux reactor.
Radiochemical purification removes radionuclide impurities efficiently, except for $^{166m}$Ho.
If the remaining traces of $^{166m}$Ho are not acceptable and/or if the activity needs to be ion-implanted
into the detector, then ISOLDE could be used off-line for mass separation and ion implantation. 

\begin{center}
{\it References}
\end{center}
\begin{description}
\footnotesize
\item[1] J. Engle, these proceedings.
\item[2] S. Lahiri, these proceedings.
\item[3] U. K\"oster for the ISOLDE Collaboration, Radiochim. Acta 89 (2001) 749.
\item[4] T. Bj{\o}rnstad et al., Nucl. Instr. Meth. B26 (1987) 174.
\item[5] J.U. Andersen et al., Phys. Lett. 113B (1982) 72.
\item[6] H.L. Ravn et al., AIP Conf. Proc. 99 (1983) 1.
\item[7] U. K\"oster, V.N. Fedosseev, V.I. Mishin, Spectrochimica Acta B58 (2003) 1047
\item[8] U. K\"oster et al., Nucl. Instr. Meth. B204 (2003) 347.
\item[9] R.A. Naumann, M.C. Michel, J.L. Power, J. Inorganic Nucl. Chem. 15 (1960) 195.
\item[10] P.A. Baisden et al., Phys. Rev. C28 (1983) 337.
\item[11] A. Laegsgaard et al., Proc. 7th Int. Conf. Atomic Masses Fundamental Constants, 3-7 Sep 1984, Darmstadt-Seeheim.
\item[12] U. K\"oster et al., Radiotherapy and Oncology 102, Suppl. 1 (2012) S170.
\item[13] G.D. Alkhazov et al., Preprint 1309, Leningrad Nucl. Phys. Inst., Leningrad (1987).
\end{description}

 \newpage\subsection{Susanta Lahiri, Moumita Maiti, Zoltan Szucs, Sandor Takacs: ``Alternative Production routes and new separation methods for $^{163}$Ho''}
\label{Susanta}
\begin{description}
\it\small 
\setlength{\parskip}{-1mm}
\item[\small\it S.L.:]Chemical Sciences Division, Saha Institute of Nuclear Physics, 1/AF Bidhannagar, Kolkata - 700064, India
\item[\small\it M.M.:]Department of Physics, Indian Institute of Technology Roorkee, Roorkee-247667, India
\item[\small\it Z.S.:]Institute of Nuclear Research, Hungarian Academy of Sciences, Debrecen, Hungary
\item[\small\it S.T.:]Institute of Nuclear Research, Hungarian Academy of Sciences, Debrecen, Hungary
\end{description}

Investigation of neutrino mass in the sub-eV energy range is one of the challenging problems in particle physics. In view of this, several international collaborations have been established among which KATRIN and MARE are important to mention. Recently a new collaboration called ECHO (Electron Capture of Holmium) has also been formed with the same goal. In the ECHO experiment, the decay of $^{163}$Ho will be measured by micro-calorimeter.The importance of $^{163}$Ho source in neutrino mass experiments is to measure the low Q value of $\approx$ 2.8 keV for the Electron Capture process. As per the present estimation about 10$^{6}$Bq activity of $^{163}$Ho is required to perform the experiment. However, difficulty lies in the artificial production of long-lived $^{163}$Ho(4570 yr), since it is not abundant in nature.

$^{163}$Ho can be produced either by charged particle reaction using accelerator or by neutron activation in a nuclear reactor. 
In eighties attempts were made to produce high purity $^{163}$Ho. In one attempt, high energy proton beam of $\approx$ 2.4 $\mu$A current from CERN PS was shoot on a tantalum target and mass fraction 163 was separated by physical separator. Later $^{163}$Ho was chemically separated from the other isobars. However, the chemical separation of adjacent lanthanide is not an easy task; moreover handling of $^{163}$Ho is difficult as its nuclear properties cannot be exploited for detection. In other case, production possibility of $^{163}$Ho was investigated by bombarding Dy targets with protons. 

It has been found that chemistry plays an important role in solving challenging and forerunner problems in physics. In this paper we propose alternative production routes of $^{163}$Ho and the corresponding analytical method for chemical purification of $^{163}$Ho towards solving the present issue.

Based on the theoretical study we propose to produce $^{163}$Ho by numbers of direct and indirect reactions based on charged particle activation and the chemical procedure to separate the desired product from the corresponding target matrix. Among direct reactions $^{163}$Dy(p,n)$^{163}$Ho offers $\approx$ 300 mb at 10 MeV whereas about four times higher cross section ($\approx$ 1250 mb at 19 MeV) is expected from $^{164}$Dy(p,2n)$^{163}$Ho reaction. The $^{163}$Dy(d,2n)$^{163}$Ho reaction is also found to be efficient to provide five times higher cross section compared to $^{163}$Dy(p,n)$^{163}$Ho reaction. Although it is hard to obtain 100\% enriched $^{163}$Dy or $^{164}$Dy as their natural abundances are only 24.9\% and 28.2\%, respectively, use of enriched Dy targets ($^{163}$Dy/$^{164}$Dy) has advantage over natDy as they do not lead to the production of long lived radioactive contaminants such as $^{157}$Tb (99 yr), $^{158}$Tb (180 yr), $^{159}$Dy (144.4 d) below 25 MeV projectile energy. The main disadvantage of using direct reactions is that the activity of 163Ho will be very low because of its long half life, no way to monitor $^{163}$Ho during its chemical separation from the target dysprosium. Moreover, difficulty lies in separation of adjacent lanthanide while Dy is in macro-quantity and the amount product $^{163}$Ho will be in the order of 10$^{14}$ atoms only. 

The indirect way of production essentially means the production of $^{163}$Ho from the decay of its short lived precursor $^{163}$Er (75 min). We proposed for the first time to produce $^{163}$Ho indirectly [1-2], with the help of the reactions (i) $^{nat}$Dy($\alpha$, xn)$^{163}$Er and (ii) $^{159}$Tb($^{7}$Li, 3n)$^{163}$Er,
short-lived $^{163}$Er will be produced which will eventually decay to $^{163}$Ho via electron capture process. The reactions $^{162}$Dy($\alpha$ ,3n)$^{163}$Er and $^{161}$Dy($\alpha$,2n)$^{163}$Er offer cross sections of $\approx$ 1 b and $\approx$ 800 mb at $\alpha$-particle bombarding energy around 40 MeV and 26 MeV respectively. The benefit of using indirect reaction over the direct is twofold: (i) production of $^{163}$Er will be higher compared to the direct production of $^{163}$Ho (ii) chemical separation between $^{163}$Er-Dy pair might be easier compared to $^{163}$Ho-Dy pair. Like direct reaction use of enriched $^{162}$Dy/$^{161}$Dy target is equally important in these cases. Pure $^{163}$Ho can also be produced using naturally abundant $^{159}$Tb via $^{159}$Tb($^{7}$Li,3n)$^{163}$Er reaction which has $\approx$ 300 mb cross section $\approx$ 32 MeV projectile energy. In this particular case, product will be free of radioactive contaminants but one has to compromise with cross section. 

Encouraged by the theoretical investigation we simulated the separation of trace amount of Er from bulk Dy matrix by liquid-liquid extraction (LLX) where HNO$_{3}$ was used as aqueous phase and liquid cation exchanger, di-(2-ethylhexyl)phosphoric acid (HDEHP) dissolved in cyclohexane was used as organic phase. The fate of Er and Dy were monitored by Inductively Coupled Plasma Optical Emission Spectrometry (ICPOES). After successful simulation, a natural $^{nat}$Dy$_{2}$O$_{3}$ target was irradiated by 40 MeV $\alpha$-particle with 3 $\mu$A current and 11 h duration at the Variable Energy Cyclotron Centre, Kolkata, India. Chemical separation method was developed to separate $^{163}$Er produced in the $^{nat}$Dy$_{2}$O$_{3}$ target, which was found in accordance with the earlier simulation. About 50\% separation of Er was achieved with less than 1\% contamination from bulk Dy. The developed method is fast enough to complete the entire chemical process within one half-life of $^{163}$Er.

\begin{center}
{\it References}
\end{center}
\begin{description}
\footnotesize	
\item[1] S. Lahiri, M. Maiti, L. Gastaldo, 14$^{th}$ International Workshop on Low Temperature Detectors (LTD-14), August 1-5, 2011, Heidelberg University, Germany, pp 96.
\item[2] L. Gastaldo {\it et al.},  14$^{th}$ International Workshop on Low Temperature Detectors (LTD-14), August 1-5, 2011, Heidelberg University, Germany, pp100.
\end{description}

 \newpage\subsection{ J. W. Engle, E. R. Birnbaum, H. R. Trellue, K. D. John, M. W. Rabin, and F. M. Nortier: ``Producing $^{163}$Ho for Micro-Calorimetric Electron Capture Spectroscopy to Measure the Mass of the Neutrino''}
\label{Engle}
\begin{description}
\it\small 
\setlength{\parskip}{-1mm}
\item[\small\it J.W.E., E.R.B., H.R.T., K.D.J., M.W.R., F.M.N.:]Los Alamos National Laboratory, Los Alamos, NM, USA
\end{description}
%
LA-UR-13-22103

\subsubsection{Introduction}
The rare-earth isotope $^{163}$Ho is targeted for electron capture spectroscopy experiments to improve the precision of neutrino mass measurements. Production of this isotope in useful quantities and purities is challenging, not least because the necessary radioisotopic and radiochemical purity requirements are poorly understood. The co-production of $^{166m}$Ho may limit the utility of a given production method, as decay emissions from $^{166m}$Ho potentially have deleterious effects on the electron capture spectrum measured from $^{163}$Ho. In order to contribute to the global effort to refine estimates of neutrino mass, the chosen production technique must lend itself to high-volume production of  $^{163}$Ho. Previous experimenters have shown that even ~10$^{10}$ atoms of  $^{163}$Ho (10$^{-1}$ Bq) are useful for proof of concept experiments in detector design [1], but larger quantities of  $^{163}$Ho are potentially necessitated by troublesome chemistries required to isolate the desired radionuclides or to offset losses incurred during preparation of the sample for incorporation into detector systems. Others have estimated that to achieve 0.1 eV uncertainty in the neutrino mass via calorimetric spectroscopy of $^{163}$Ho decay, experimental statistics will necessitate a minimum of 10$^{17}$ atoms of $^{163}$Ho implanted into absorbers [2][3]. For this end goal, two primary production methods were considered, proton and neutron irradiation of dysprosium and erbium targets, respectively. The necessary quantities of $^{163}$Ho can likely only be produced by the devotion of weeks or even months of irradiation time using these methods examined below. 

\subsubsection{Proton Irradiation of Dysprosium}
Charged particle irradiation (using proton or alpha beams) can produce $^{163}$Ho directly (via a single nuclear transmutation reaction) from dysprosium targets, taking advantage of $^{163}$Ho's relatively long half live compared to other neutron-poor isotopes of the same element. This beneficial characteristic of holmium radioisotopes supports a natural radioisotopic purification process with a very tolerable duration following elemental separation chemistry. The notable foil to this beneficial rule is $^{166m}$Ho (t$_{1/2}$ = 1200 y, 100\% EC) which is expected to remain in target solutions following both radiochemical separation and even lengthy waiting periods spent allowing other radioisotopes to decay. For this reason, only irradiation parameters such as target isotopic enrichment, the intensity, duration, and energy distribution of incident particle flux, and the availability of mass-based separation methods can be expected to mitigate the effect that $^{166m}$Ho may have on spectroscopy experiments. \\
\\
Cross sections, or energy dependent measures of the probability of forming a desired radionuclide from a designated nuclear reaction, for production of $^{163}$Ho using protons on dysprosium are unmeasured. In the absence of these data, which would permit estimates of radionuclide yields across a range of irradiation energies and fluences, computational estimates (see Figure 1) are often used to predict the results of irradiations. Measured data quantifying production yields from thin dysprosium targets are contributed by Yasumi and coauthors, who applied 24 hour, 100 $\mu$A proton currents with 20 MeV incident energy to dysprosium targets braised onto copper cooling block backings, producing $^{163}$Ho masses of approximately 7 $\mu$g (10$^{16}$ atoms) [4] [5] [6]. These measurements confirm the approximate shape and structure of computational predictions within reasonable (20-30\%) uncertainty.
\begin{figure}[h!]
  \caption{TALYS predictions of cross sections for proton induced reactions on dysprosium [7].}
  \centering
    \includegraphics[width=0.4\textwidth]{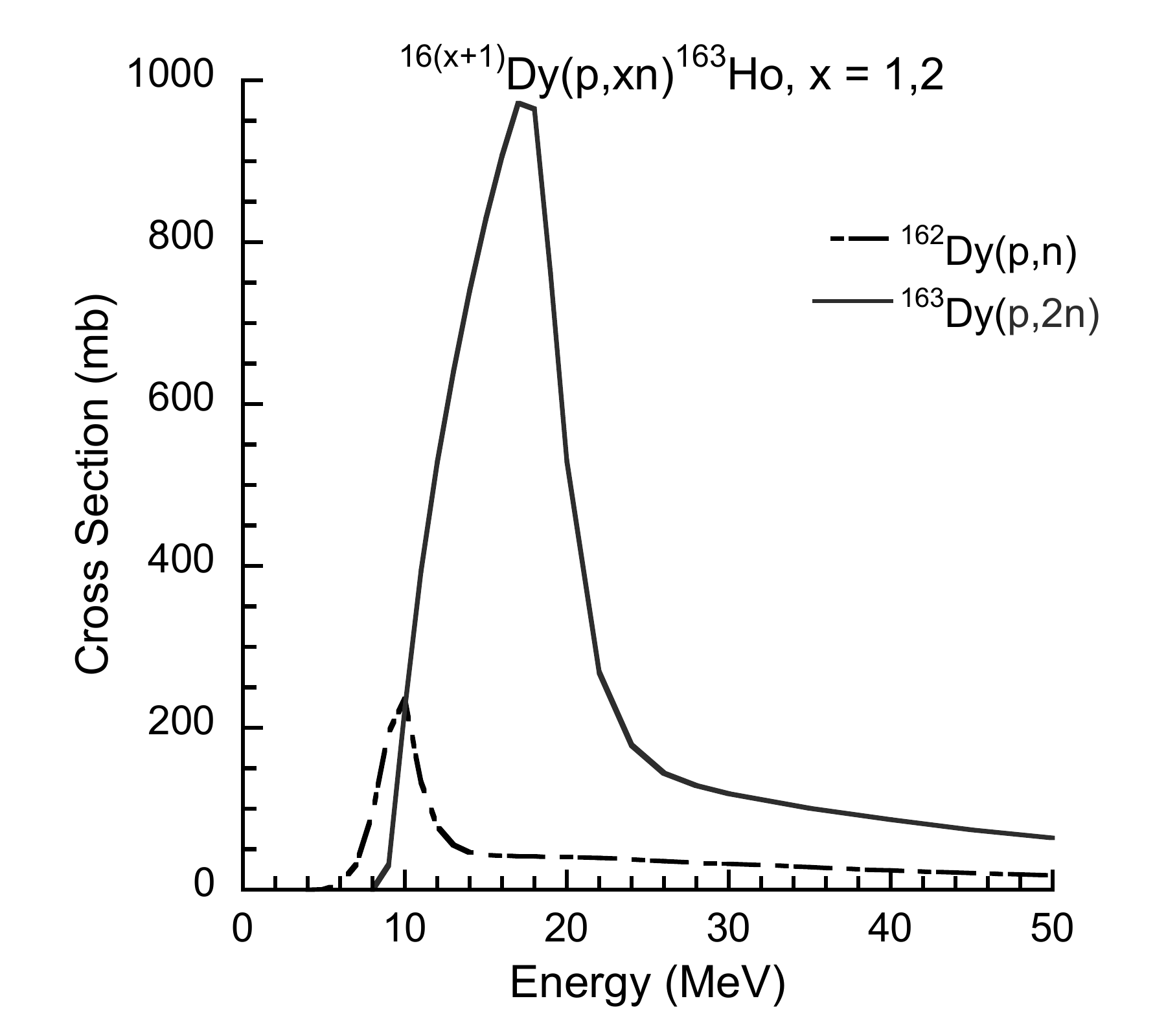}
\end{figure}
Notably, this approach avoids the necessity of using enriched target material. Radioisotopic purity of an isolated final product, which would be achieved by radiochemical separation from the dysprosium target material, is expected to be primarily influenced by the quantity of secondary neutrons (ejected from primary, proton-induced reactions) in a chosen irradiation scheme. These secondary neutrons will certainly initiate $^{165}$Ho(n,$\gamma$)$^{166m}$Ho reactions on any trace quantities of stable holmium present in the target, potentially producing large quantities of $^{166m}$Ho. Furthermore, the two target isotopes of dysprosium germane to a proton-irradiation scheme are $^{163}$Dy and $^{164}$Dy, which have abundances in natural dysprosium of 24.90\% and 28.26\%, respectively. The $^{164}$Dy(n,$\gamma$) reaction produces $^{165}$Dy, which quickly decays to $^{165}$Ho, ultimately contributing $^{166m}$Ho to the final product in addition to that formed from expected stable $^{165}$Ho contamination in the target (Figures 2 and 3). The significance of this contribution will be determined by the mass of $^{164}$Dy exposed to the secondary neutron flux, and the effect of this indirect reaction mechanism is expected to be magnified by the extremely large (3600 barns) neutron capture cross section of $^{164}$Dy. 
\begin{figure}[h!]
  \caption{TALYS predictions of cross sections for neutron capture on the $^{165}$Ho nucleus [7].}
  \centering
    \includegraphics[width=0.4\textwidth]{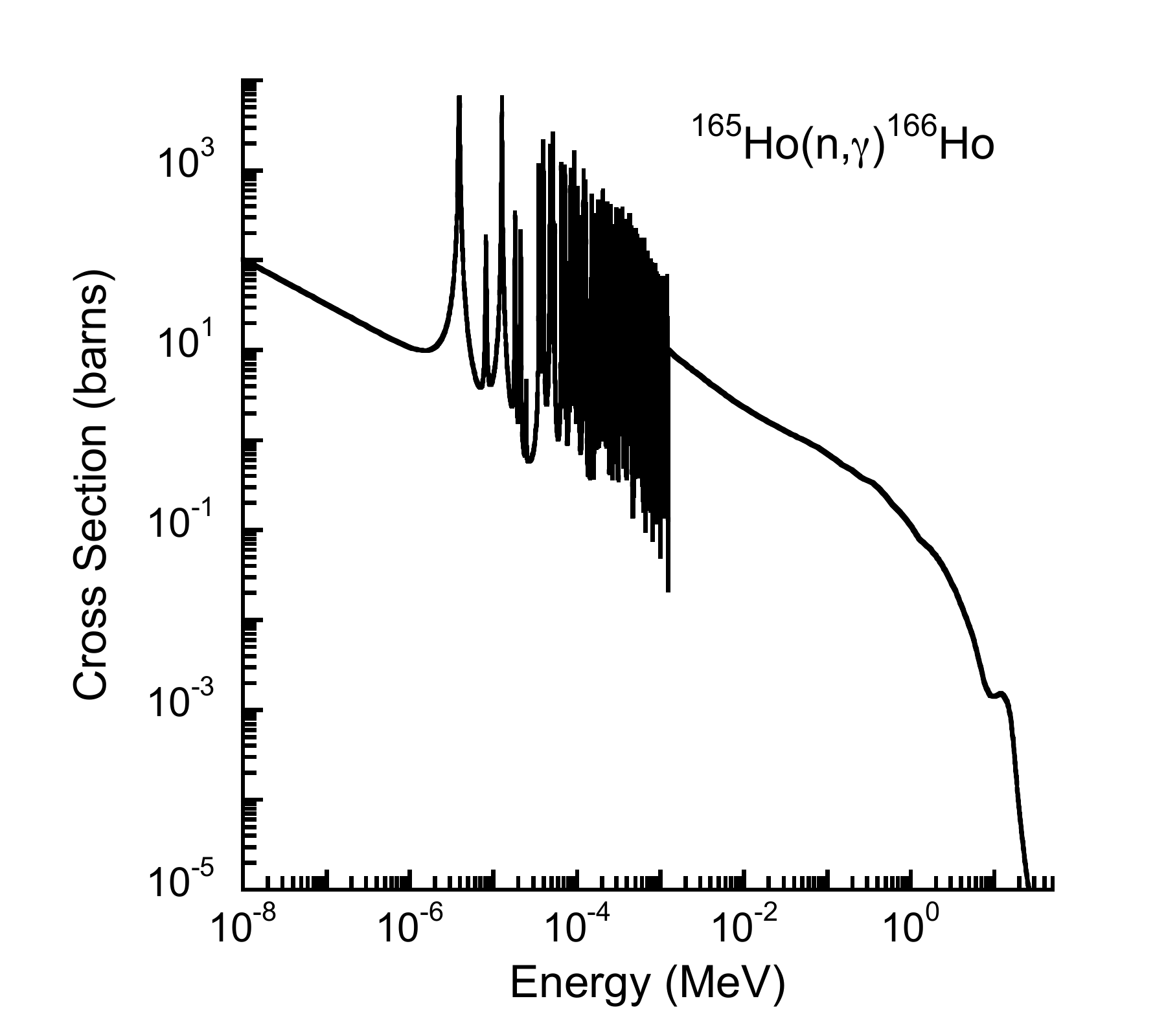}
\end{figure}
\begin{figure}[h!]
  \caption{TALYS predictions of cross sections for neutron capture on the $^{164}$Dy nucleus [7].}
  \centering
    \includegraphics[width=0.4\textwidth]{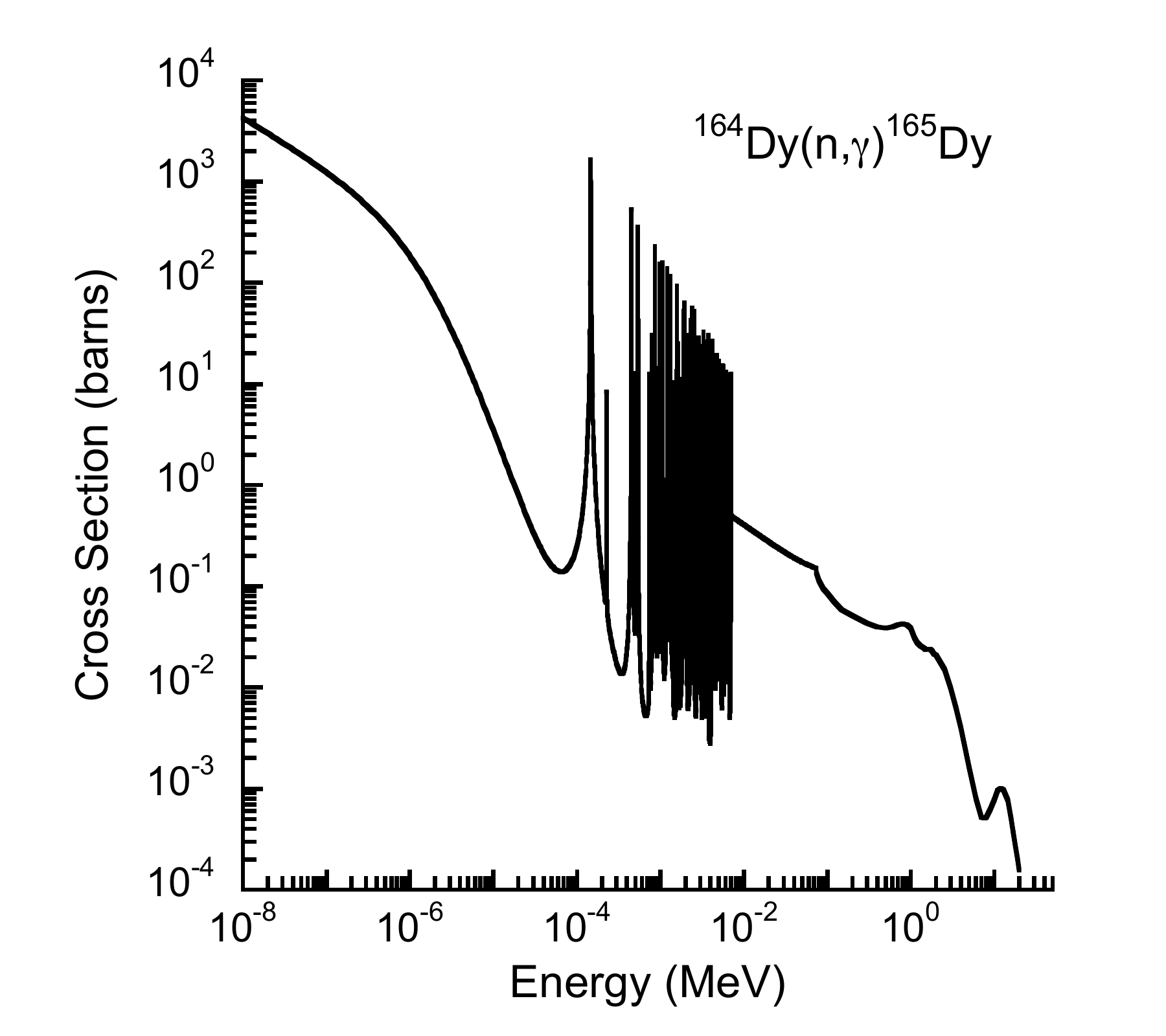}
\end{figure}

\subsubsection{Neutron Irradiations of Erbium}
Neutron irradiation provides an alternative to ÒdirectÓ production, making $^{163}$Ho by first producing its parent, $^{163}$Er, and allowing this isotope to decay. No erbium isotopes have half-lives which exceed that of $^{160}$Er (t$_{1/2}$ = 28.6 h), and so decay of radioisotopic impurities likely co-produced with $^{163}$Er can proceed in a reasonable time frame. The neutron capture cross section of the $^{162}$Er nucleus is estimated to peak near 102 barns [2], though little data exists to support this estimate in any energy region. Production rates corresponding to the medium-intensity flux of 10$^{13}$ thermal neutrons/cm$^{2}$/sec are estimated at 10$^{15}$ atoms/week per mg of$^{162}$Er. Such a production could reasonably support preliminary counting experiments. However, $^{166m}$Ho can be produced even via ÒindirectÓ methods where the objective of a reactor experiment is $^{163}$Er. 

Several considerations are expected to limit the utility of neutron irradiations to the formation of $^{163}$Ho for ECS. First, a $^{nat}$Er target is composed of only 0.139\% $^{162}$Er. The target will therefore require isotopic enrichment lest the expected production rate drop by a factor corresponding to target enrichment. The presence of isotopic impurities as a consequence of unavoidably imperfect enrichment schemes will produce $^{166m}$Ho by various means. Most notably, $^{164}$Er(n,$\gamma$) leads to $^{166m}$Ho by decay of 10.5-hour $^{165}$Er to stable $^{165}$Ho and subsequent neutron capture by the holmium nucleus as described above. Second, stable $^{165}$Ho impurities in the target will be exposed to the same intense neutron flux being used for the production of $^{163}$Er/$^{163}$Ho, possibly mandating additional elemental purification of the enriched target prior to irradiation. Such purification is likely to be accomplished by mass separation, which may ultimately simplify incorporation of $^{163}$Ho into the absorber of choice in microcalorimetry experimental equipment, but carries additional cost and risk of material losses from approximate 10\% throughput efficiencies. Finally, the neutron capture cross section for the $^{163}$Ho nucleus is unknown. If a lengthy reactor irradiation is required in order to produce $^{163}$Ho on the unprecedented scale demanded by multiplexed arrays of 10$^{3}$ individual detector elements, then the quantity of $^{163}$Ho formed during a lengthy irradiation may be reduced by transmutation of $^{163}$Ho to $^{164}$Ho as well. Neutron capture reactions on $^{164}$Dy, which is likely present as trace stable contamination in the target and also formed by $^{163}$Ho(n,$\gamma$)$^{164}$Ho (t$_{1/2}$ = 29 m, 60\% EC to $^{164}$Dy), are expected to contribute additional $^{166m}$Ho via the same neutron capture processes described above for secondary neutron-initiated reactions in proton irradiation schemes. 

\subsubsection{Discussion}
The factors distinguishing proton- and neutron-irradiation products will be the quantity of $^{163}$Ho and the relative quantity of $^{166m}$Ho associated with each method. In 1997, Springer and coauthors published the production of 90 $\pm$ 9 $\mu$g $^{163}$Ho (10$^{17}$ atoms) made from an enriched $^{162}$Er target in the high-flux beam reactor at Brookhaven National Laboratory; the irradiation parameters were not reported. They noted the presence of contaminating $^{166m}$Ho and used mass separation in an attempt to remove the radioactive contaminant [7]. At present, the capability of separating radioactive materials by mass exists primarily at various European facilities. Following mass separation, Springer and coauthors estimate from independent spectroscopy of $^{166m}$Ho sources that the radiocontaminant contributed approximately 20\% of the observed background in the energy range of interest (between 2.5 and 3 keV) in $^{163}$Ho spectra. Published production campaigns which report yields from proton irradiations do not describe problems with $^{166m}$Ho contamination [4], but past measurement techniques may not have been sensitive to this radioisotopic contaminant in acquired spectroscopic samples. \\
\\
Exploratory studies have modeled production schemes using neutron and proton irradiations as described above; alpha-initiated reactions were also modeled, though they are not addressed in detail above. The code Monte Carlo for Neutral Particles (MCNP) was used to describe simulation geometry and irradiation physics [8]. The isotope depletion and production code CINDER90 was used to obtain an understanding of changing target isotopic compositions during irradiation, and for reactor calculations the linkage code Monteburns was run to coordinate iterative interaction between the transport and depletion calculations [9] [10]. Monteburns is beneficial for analyzing irradiation behavior of specific reactors because MCNP uses continuous-energy cross sections to accurately model reactions in energy regions where more generalized codes may lack specific data.  That information can be fed into the depletion code to determine the system- and time-dependent behavior of a reactor.  
The results of these simulations are presented below in Table 1, with particular emphasis on the atom ratio of $^{163}$Ho to $^{166m}$Ho for each irradiation scheme. An attempt is made to consider a variety of probable scenarios germane to each incident particle. For protons, this includes both small, medical proton accelerators and the national-scale Isotope Production Facility at Los Alamos National Laboratory, where encapsulated targets have been designed to dissipate the thermal energy deposited by 240 $\mu$A beam currents. Alpha irradiations were simulated for dysprosium targets of natural isotopic abundance and for targets enriched in $^{161}$Dy at energies and beam intensities presently available at accelerators in the United States (40 MeV, 30 $\mu$A). Neutron irradiations were simulated using $^{162}$Er targets of an enrichment level available from commercial suppliers (40\% $^{162}$Er), and reported results attempt to estimate yields from reactor fluxes between 10$^{13}$ and 10$^{15}$ n/cm$^{2}$/sec in the physical geometry surrounding the erbium target. Results in the form of approximate yields are presented in the table below. 
\begin{center}
    \begin{tabular}{|{l}|{c}|{c}|{c}|}
    \hline
    Incident Particle & Target & $^{163}$Ho Production Rate (atoms/hr) & $^{163}$Ho/$^{166m}$Ho Atom Ratio \\ \hline 
    \hline
    16 MeV p$^{+}$ & $^{nat}$Dy & 10$^{14}$ & 10$^{10}$ \\ \hline
    26 MeV p$^{+}$ at LANL IPF & $^{nat}$Dy & 10$^{15}$ & 10$^{6-8}$ \\ \hline
    40 MeV $\alpha$ & $^{nat}$Dy & 10$^{13}$ & 10$^{5}$ \\ \hline    
    40 MeV $\alpha$ & $^{161}$Dy & 10$^{10}$ & 10$^{7}$ \\ \hline
    Reactor neutrons & $^{162}$Er (40\%) & 10$^{13-16}$ (per mg $^{162}$Er) & unknown \\ \hline
    \hline
    \end{tabular}
\end{center}
Based upon available nuclear data, high-flux reactors loaded with tens to hundreds of milligram quantities of $^{162}$Er will outpace the production rate of charged particle irradiations by orders of magnitude. However, the necessity of obtaining highly-enriched $^{162}$Er to avoid $^{166m}$Ho production highlights the need to better understand the  effect the complex electron and x-ray spectrum emitted by $^{166m}$Ho may have on calorimetric detectors, and prods the researcher towards the relative safety of charged particle irradiations. Uncertainty in the reactor production rate of $^{166m}$Ho is the principle contributor to uncertainty in the radioisotopic purity of reactor-based productions. Additional variance in production rate estimates for $^{163}$Ho derives from irradiation parameters, in particular the reactor-dependent magnitude and energy distribution of the incident neutron flux and a lack of data for other neutron capture reactions expected to contribute smaller quantities of $^{166m}$Ho. Given presently available knowledge of relevant nuclear data and a conservative approach to estimating microcalorimetersÕ tolerance of radioisotopic impurities, proton-induced reactions balance production quality and quantity most effectively. 

\begin{center}
{\it References}
\end{center}
\begin{description}
\footnotesize	
\item[1] L Gastaldo {\it et al.}, Nucl. Inst. Methods in Phys. Res. A (2013) http://dx.doi.org/10.1016/j.nima.2013.01.027i
\item[2] F. Gatti {\it et al.}, Phys. Lett. B preprint
\item[3] A. Nucciotti, Nucl. Phys. B, {\bf 00} (2010) Proceedings Supplement 1
\item[4] S. Yasumi {\it et al.}, Phys. Lett. B {\bf 181} (1986) 169
\item[5] S. Yasumi {\it et al.}, Phys. Lett. B {\bf 334} (1994) 229
\item[6] S. Yasumi {\it et al.}, Phys. Lett. B {\bf 122} (1983) 461
\item[7] A. Koning {\it et al.}, International Conference on Nuclear Data for Science and Technology, Nice (2007) 211
\item[8] T. Goorley {\it et al.}, Nucl. Tech. {\bf 180} (2011) 298
\item[9] J. Galloway {\it et al.}, Proc. of the Int. Cong. on Adv. in Nucl. Power Plants (2012) Curran Assoc., Inc.: NY
\item[10] W. Wilson {\it et al.}, Los Alamos Nat. Lab. Report  (2007) LA-UR-078412
\end{description}

 \newpage\subsection{G.~Pagliaroli, F.~Rossi-Torres, F.~Vissani: ``Core Collapse Supernovae and Neutrino Mass Bound''}
\begin{description}
\it\small 
\setlength{\parskip}{-1mm}
\item[\small\it G.P.:] 
INFN, Laboratori Nazionali del Gran Sasso, L'Aquila, Italy 
\item[\small\it F.R-T.:]
Instituto de F\'isica Te\'orica--Universidade Estadual Paulista, S\~ao Paolo, Brazil
\item[\small\it F.V.:] 
INFN, Laboratori Nazionali del Gran Sasso and Gran Sasso Science Institute, L'Aquila, Italy
\end{description}
%
The difference between the velocity of the neutrinos and $c$ was 
immediately pointed out to the attention of the 
``radioactive ladies and gentlemen'', in the same letter
where the neutrino was introduced  
(note that, nine years earlier, the 21-year-old Pauli authored 
the famous essay {\em Relativit\"atstheorie}).
In 1969, Zatsepin suggested to probe the mass of neutrinos 
by measuring their times of arrival from 
a core-collapse supernova.
At the time, it was believed that most 
supernova neutrinos were emitted in a few ms burst.

All subsequent studies, from Nadyozhin (1978), Bethe \& Wilson (1986), till present ones, drew a different scenario for the neutrino emission during the formation of the neutron star, summarized in the following table:
\centerline{\small\begin{tabular}{|c||c|c|c|}
\hline
Emission & Number of Events & Energy & Duration \\[-.3ex]
phase & [in SK at 10 kpc] & [in $10^{51}$ erg] &  [in sec] \\ \hline 
neutronization & 1-2 (ES) &  $\sim$ 2  &$\sim$  0.003  \\
{accretion} & 500-1,000 (IBD) & $\sim$ 50 & $\sim$  0.5  \\
cooling & 3,000-4,000 (IBD) &  $\sim$ 200 & $\sim$  10  \\ \hline
\end{tabular}}
On top of a brief initial emission due to $ep\to \nu_e n$ 
(`neutronization') most neutrinos are radiated on a thermal time scale of 
$\mathcal{E}_B/(6\, L_{\mbox{\tiny NS}})\sim 10$ sec, where  
$\mathcal{E}_B \sim G_N M_{\mbox{\tiny NS}}^2/R_{\mbox{\tiny NS}}\sim (2-3)\times 10^{53}$ erg 
is the binding energy and 
$L_{\mbox{\tiny NS}}\sim  R_{\mbox{\tiny NS}}^2 T^4_{\mbox{\tiny NS}}\sim (3-5)\times 10^{51}$ erg/sec
the luminosity per neutrino and antineutrino species, written in terms of the neutron star  radius, 
$R_{\mbox{\tiny NS}}\sim 10$ km and of 
the temperature of the region where neutrinos are emitted, 
$T_{\mbox{\tiny NS}}\sim 4$ MeV.
Moreover, and despite some provisional character of the expectations, the first fraction of a second is predicted to be particularly luminous. This phase of emission is called `accretion', since it happens during the rapid accretio of matter onto the nascent neutron star.

It is remarkable that this scenario is not contradicted but rather supported by the neutrinos 
that have been observed by Kamiokande-II, IMB and Baksan from SN1987A: see [1] and 
Astropart.~Phys.~31 (2009) 163.
 

On the bases of these observations, 
many upper bounds on the neutrino mass have been derived, 
see [2] for a review.
Only in a few cases a confidence level is quoted.  Two groups of upper bounds satisfy this criterion; one at about 6 eV, the other in the range 15-20 eV (95\% CL). 
The difference is due to the assumed model of the neutrino emission: the first group uses models of the emission that account for a phase of accretion, while the other does not. Evidently, the existence of a short phase of emission, in this case accretion, is  
 important to constrain the neutrino mass better.
 (Note that the bound $m_\nu<\infty$ given in Astropart.\ Phys.\ 35 (2011) 177  has not bearing with SN1987A).

Let us analyze the bound in some detail.
The time of flight of a neutrino is  
$T=\frac{D}{v} \mbox{ with } v=\frac{p c}{E}\approx 1- \frac{(m c^2)^2}{2 E^2}$,
that it is smaller than $c$. This causes a delay of the neutrinos of
$$
\Delta t\equiv \frac{D}{v}-\frac{D}{c}=\frac{D}{2 c} \left(\frac{m c^2}{E}\right)^{\!\!2} =
{\bf 2.6 \mbox { \bf sec}}\ \frac{D}{50\mbox{ kpc}} \left(\frac{m c^2}{\mbox{10 eV}} \right)^{\!\!2} 
\left(\frac{\mbox{10 MeV}}{E} \right)^{\!\!2} 
$$
where we recall that the distance of SN1987A was $D=52\pm 2$ kpc.
Thus, if we know that there is a time structure of a few seconds in the (anti)neutrino emission, 
we expect to derive a typical bound of 10 eV.  For instance, if we collect $N$ events from an emission
phase lasting $\tau$, we can impose the condition 
$
\Delta t< (\mbox{number of }\sigma)\times \tau/\sqrt{N-1}
$.
Then, using (number of $\sigma$)=2, this gives 
$$
m_\nu<\frac{2 \langle E_\nu\rangle}{(N-1)^{1/4}} \sqrt{\frac{\tau}{D/c}}
$$
that means {5.7 eV} using  $D=50$ kpc, 
$N=5$ (i.e., the initial set collected by Kamiokande-II)
$\tau=0.5$ s and   $ \langle E_\nu\rangle=13$ MeV. 
This is similar to the bounds of 6 eV [1,3] mentioned above and quoted 
in the PDG biannual report.

However, the electromagnetic radiations must propagate  with the shock wave; thus, they emerge from the star some hours after the neutrinos. This limits the possibility of comparing directly the travel time of the neutrinos and the one of the light.
But, as emphasized by Fargion in Lett.~Nuovo Cim.~{31} (1981) 499, also the gravity waves travel at velocity $c$. If, as expected in most models, the collapse is immediately followed by an emission of a burst of gravity waves (GWB), and if this can be detected, the bound on the neutrino mass can be improved
(see [3] and PRL 103 (2009) 031102 for more discussion).  
In the opposite case, the analysis of the data is slightly more complicated.

In fact, the neutrino flux $\Phi_\nu(t)$
regulates the expected counting rate    $\dot{N}[\Phi(t)]$. 
This is different from zero only if the relative emission time is $t>0$. 
The events detected at the time $T_i^d$ were emitted at $T^e_i=T_i^d-D/v_i$, and when this is compared with the earliest possible emission time $T_0^e=T_0^d-D/c$, we get
$$
\stackrel{\mbox{\tiny relative-emission-time}}{t_i}=T^e_i-T^e_0\approx \stackrel{\mbox{\tiny 
relative-detection-time}}{(T^d_i-T^d_1)}\!\!\!\!\! +
\stackrel{\mbox{\tiny delay-of-detection}}{(T^d_1-T^d_0)}- 
\stackrel{\mbox{\tiny delay-due-to-mass}}{\frac{D}{2 c}\left( \frac{m c^2}{E_i}\right)^2}
$$
Thus we need to evaluate 3 terms. The 1st can be obtained 
from   
the data and the 3rd one have to be obtained from the fit.
What about the 2nd term? In principle, 
$T^d_0$ could be measured by the GWB, otherwise, it should be obtained from the fit of the data, just as the 3rd term.
In this second case the fit is more difficult,
since the 2nd and 3rd terms have opposite signs and are subject to a partial cancellation. 
Note that for the analysis of SN1987A
we need to introduce three delay-of-detection times, 
since the clocks of Kamiokande-II and Baksan were not synchronized [1,3].

Next, we consider the ultimate bound that could be possibly obtained from a 
future galactic supernova:\\
1) If  Super-Kamiokande (SK) data are analyzed
knowing $T_0^d$ from the GWB [4,3]:\hfill {\bf 0.5-1.1 eV}\\ 
2) If elastic scattering events from the neutronization are seen by
Hyper-Kamiokande (SK$\times 20$) [3]:\hfill {\bf $\sim$ 0.4 eV}\\ 
3) If we use IceCUBE hypothesizing the existence of few ms bursts of antineutrinos 
during accretion [5]:\hfill {\bf 0.14 eV}\\
Therefore, the ultimate bound depends upon the assumptions.

Two  remarks are in order; 
{\em (1) The distance is not a critical parameter, as long as the time structure of 
the supernova emission is seen}
(in the calculations, a galactic supernova at 10 kpc is typically considered).
This can be understood already from the rough 
bound that one obtains requiring that the emission time is positive, $t_i>0$, i.e., 
$$
m_\nu<\mbox{Min}_i\left(\frac{E_{\nu,i}}{c^2} \sqrt{\frac{T_i^d-T_0^d}{D/(2 c)}}\right)
$$
Indeed, $T_i^d-T_0^d\propto D$, because   
the counting rate grows as $\dot{N}\propto T\ M_{\mbox{\tiny det}}/D^2$.
A similar conclusions stems from the previous formula on $m_\nu$, noting that $N\propto 1/D^2$. 
{\em (2) Other hypothetical sources of high energy neutrinos are unlikely to 
provide a better bound.} 
Assume e.g., $m c^2=2$ eV (the MAINZ + TROITSK bound). A neutrino with
$
\Delta t=20 \mbox{ ms}
$
has 10 MeV if it comes from a galactic SN at 10 kpc; or has $E=200$ MeV if it comes from CEN-A; or has $E=7$ GeV if it comes from 5 Gpc.  For a supernova-like emission, the corresponding events 
in a Mton detector are about  
$$N_{\mbox{\tiny events}}\sim 20\times \frac{M_{\mbox{\tiny det}}}{\mbox{Mton}} \times
\frac{\mathcal{E}_{B}}{3\times 10^{53}\mbox{ erg}} \times
\left(\frac{1\mbox{ Mpc}}{D}\right)^2$$
that limits the search of such an emission
to the Local Group. The neutrinos of higher energy can be detected more easily, since 
the cross section increases with the energy. However, it is difficult to 
imagine good reasons why a burst 
from cosmological distance and comprising neutrinos of the right energy  should be intense enough to be seen 
(different views are occasionally expressed, e.g., 
PRD67, 2003   and 
Nature Phys. 3, 2007.)

Summarizing, 
the analysis of supernova neutrino events allows us 
to put a bound on their mass.
The bound from SN1987A is not competitive with the direct one, but the analysis has 
methodological merits and interesting statistical aspects.
There are good perspectives of improvement with the next galactic supernova.
However, in order to reliably assess  the reach of the method, we need to know the emission spectrum. 
In this connection, it is important to stress that our understanding of the supernova, and of neutrino emission, is still partial and should be improved.
\leftline{\bf Discussion:}
{\sc Question [Gorla]:} What is the most important information we expect to obtain 
from a future supernova?\\
{\sc Answer:} High-statistics electron antineutrino counting rate will provide us with 
a lot of useful information. 
It would be great to 
measure for the first time also 
neutral current events, discussed in the talk of  Biassoni.\\ 
{\sc Question [Cattadori]:} What about neutrino oscillations, and in particular those with sterile neutrinos?\\
{\sc Answer:} Above, astrophysical issues have been emphasized. Ordinary oscillations are discussed  
in arXiv:1008.4726 and J.Phys.Conf.Ser.~309 (2011) 012025;  
 sterile neutrinos are considered in  Nucl.Phys.~B708 (2005) 215.\\
{\sc Question [Weinheimer]:} Is neutrino emission resulting from black hole formation useful to probe the mass?\\
{\sc Answer:} Perhaps, depending on how sharply the emission is terminated and on how intense is the initial emission. 
In the Milky Way we expect a couple of such events each 1,000 years, 
thus we will need time and/or good luck. 

\begin{center}
{\it References}
\end{center}
\begin{description}
\footnotesize	
\item[1] T.~J.~Loredo and D.~Q.~Lamb,
  Phys.\ Rev.\ D {\bf 65} (2002) 063002
  [astro-ph/0107260].
\item[2] F.~Vissani, G.~Pagliaroli and F.~Rossi-Torres,
  Int.\ J.\ Mod.\ Phys.\ D {\bf 20} (2011) 1873, 
  arXiv:1005.3682.
\item[3]  G.~Pagliaroli, F.~Rossi-Torres and F.~Vissani,
  Astropart.\ Phys.\  {\bf 33} (2010) 287.
\item[4] E.~Nardi and J.~I.~Zuluaga,
  Phys.\ Rev.\ D {\bf 69} (2004) 103002.
\item[5] J.~Ellis, H.~-T.~Janka, N.~E.~Mavromatos, A.~S.~Sakharov and E.~K.~G.~Sarkisyan,
  Phys.\ Rev.\ D {\bf 85} (2012) 105028.


\end{description}

 \newpage\subsection{M. Biassoni: ``The CUORE experiment potential as supernova neutrinos observatory''}
\begin{description}
\it\small 
\setlength{\parskip}{-1mm}
\item[\small\it M.B.:]Dip. di Fisica ``G. Occhialini'', Universit\`a di Milano-Bicocca, Milano, Italy and INFN Sezione di Milano-Bicocca, Milano, Italy
\end{description}
%

CUORE (Cryogenic Underground Observatory for Rare Events, [1]) is a 1-ton scale bolometric experiment. Its primary goal is the observation of neutrinoless double beta decay of $^{130}$Te. An array of 988 natural TeO$_{2}$ crystals is under construction and will be kept at $\sim$10mK in a cryostat located in the Hall A of Gran Sasso Underground Laboratories, in Italy. Since neutrinoless double beta decay is an extremely rare (forbidden in the Standard Model) nuclear decay, the detector is heavily shielded against environmental radioactivity and the setup is the result of a careful selection of ultra radiopure materials and carefully studied building protocols to avoid recontamination.

Thanks to the large mass, excellent energy resolution, high detector segmentation, and extremely low background CUORE proves to be sensitive also to other rare phenomena, involving interactions at low energy. 
WIMPs and neutrinos, for example, can be in general detected via the observation of the recoil of target nuclei they scatter on. Through neutrino-nucleus neutral current coherent scattering (a Standard Model phenomenon that is relatively well known from the theoretical point of view, but has never been observed experimentally) a MeV neutrino can transfer a fraction of its energy to the target nucleus generating a recoil with a typical energy of few to tens of keV.
A very low threshold of few keV is mandatory to detect these interactions. In CUORE it can be achieved by means of a trigger algorithm based on the optimum (matched) filter technique which is able to reject non physical events based on the analysis of the shape of the recorded pulses, thus maximising the signal to noise ratio [2].

Once the energy threshold of the detector is as low as the typical recoil energy of a nucleus scattered by a neutrino of some MeV energy, CUORE becomes sensitive to neutrinos emitted by core-collapse supernovae in the galactic environment.
Coherent scattering cross section quadratically depends on the number of neutrons in the target nucleus, thus heavy elements (like Te) are favourite targets and the expected number of events in a 1ton-scale detector is comparable with the yield in a 100ton experiment where neutrinos are detected via inverse beta decay. Moreover, neutrino-nucleus coherent scattering is flavor-blind, that is, all types of neutrinos can be detected with the same efficiency both improving the sensitivity to a supernova explosion (during which neutrinos of all families are produced with similar branching ratios) and enabling CUORE to collect, in case of detection, valuable and model independent information on supernova explosion parameters.

Most supernova theoretical models [3, 4] predict that the largest fraction of the neutrinos emitted by a supernova explosion is produced in the so-called cooling phase. The binding energy of the exploding star is shared among the different neutrino families in equal parts. The average energy of each neutrino flavour, however, depends on the cross section of its interaction with the supernova high density matter: electron neutrinos, whose cross section is larger, must diffuse to lower temperature and density outer regions of the supernova before free-streaming; their average energy is therefore lower compared to muon and tau neutrinos and antineutrinos that are emitted by deeper, hotter regions of the proto-neutron star. In the simple model used in this analysis, the neutrino spectra are Boltzmann shaped, with three different temperatures for electron neutrinos, $\nu_{e}$, electron anti-neutrinos, $\overline{\nu_{e}}$ and all the other neutrinos, $\nu_{x}$.
In order to calculate the expected number of interactions in the detector, the emission spectra must be integrated together with the neutrino-nucleus coherent scattering cross section. This cross section is calculated within the Standard Model as
\begin{equation}\label{eq:xsec}
\frac{\mathrm{d}\sigma_{\mathrm{NCCS}}}{\mathrm{d}\Omega} = \frac{G_{F}^{2}}{4\pi^{2}} E^{2} (1+\cos \theta) \frac{Q_{W}^{2}}{4} F(Q^{2})^{2}
\end{equation}
where $G_{F}$ is the Fermi constant and $\theta$ is the angle between the original and the scattering direction. $Q_{W}$ is the weak charge of the nucleus. Since $Q_{W} = N-(1-\sin^{2}\Theta_{W})Z$
the cross section for a nucleus with $N$ neutrons and $Z$ protons is enhanced by a factor $\sim N^{2}$ compared to the cross section for the standard $\nu$-nucleon NC elastic scattering. Heavy nuclei (like Te, 592kg of which are present in CUORE detectors) are favorite targets. As coherent scattering happens only when the wavelength associated to the momentum transferred in the interaction is comparable with the nuclear dimension, the form factor $F(Q^{2})$ accounts for the probability of having coherent scattering as a function of the nucleus recoil energy ([5, 6, 7]).	

\begin{figure}[!htb]
	\centering
	\captionsetup{labelfont=bf, width=.9\textwidth}
	\caption{Number of signal (for a supernova at a distance of 8.5kPc and an energy of $10^{53}$erg) events as a function of the energy threshold (violet = tellurium recoils, green = oxygen recoils)}
	\vspace{.2cm}
	\includegraphics[width=.9\textwidth]{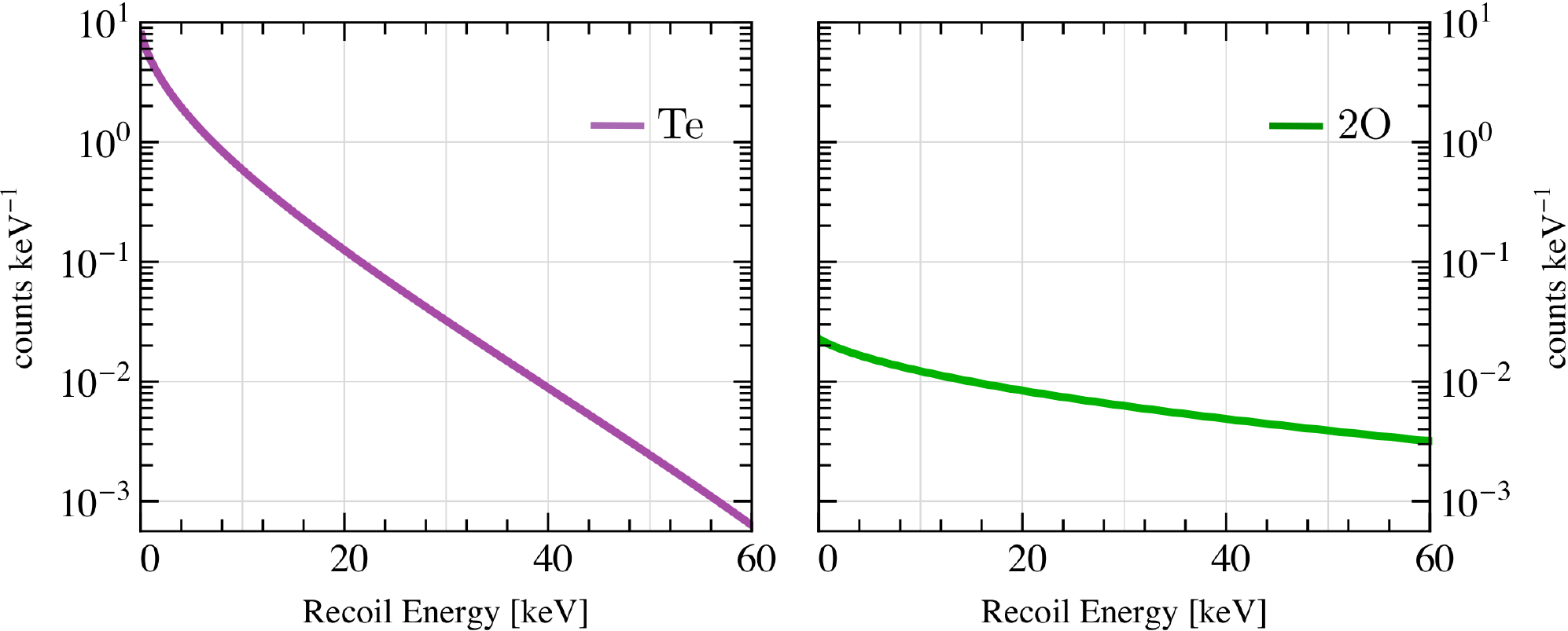}
	\label{fig:yield}
\end{figure}

The integration of the cross section and emission spectra is performed numerically as in [8] and the resulting event yields are shown in the plot of fig.~\ref{fig:yield} for tellurium and oxygen nuclei.

\begin{figure}[!htb]
	\centering
	\captionsetup{labelfont=bf, width=.9\textwidth}
	\caption{Statistical significance of the observation (for a supernova with an energy of $10^{53}$erg): {\bf (left)} as a function of the energy threshold, at 8.5kPc; {\bf (right)} as a function of the distance, with 3keV energy threshold.}
	\vspace{.2cm}
	\includegraphics[width=.9\textwidth]{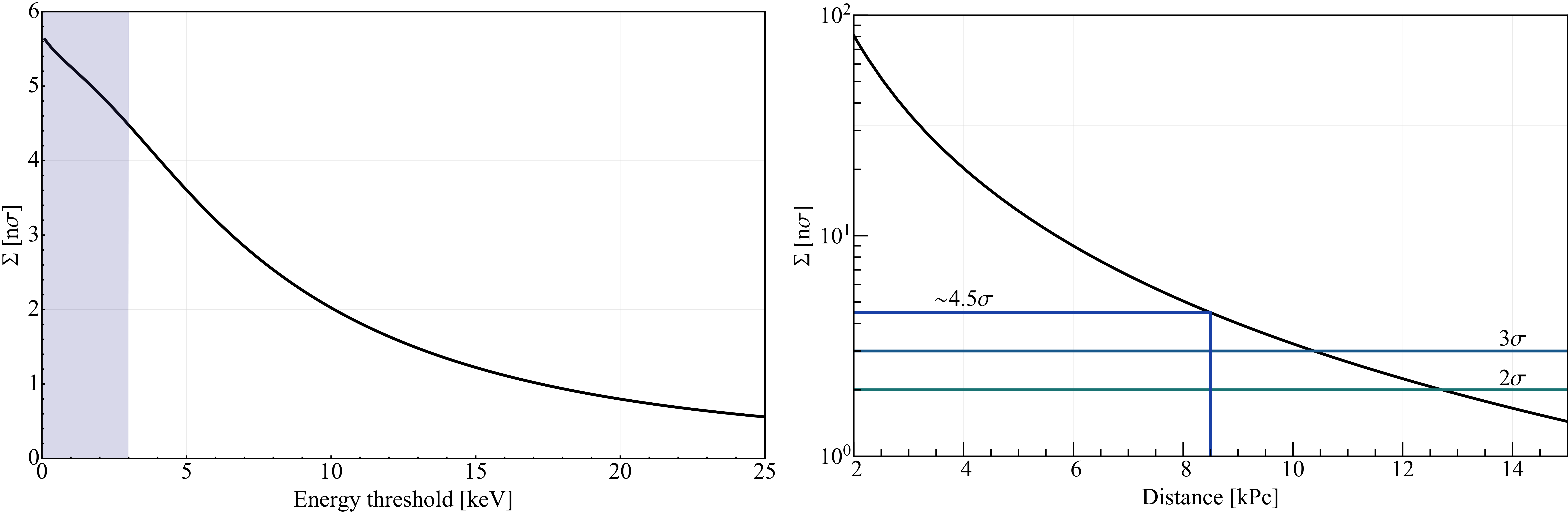}
	\label{fig:significance}
\end{figure}

To quantify the potential of CUORE, the expected signal is compared to the background in the low energy region of the spectrum. The background is extrapolated from CCVR2 and CUORICINO data [9]. The statistical significance of a supernova neutrinos burst observation is defined as the ratio between the expected number of signal events and the statistical fluctuation of the number of background events recorded in the same time interval. The plot of this quantity as a function of the detector energy threshold is reported in fig.~\ref{fig:significance}(left), while in fig.~\ref{fig:significance}(right) the same quantity is plotted as a function of the supernova distance for a fixed threshold of 3keV.

\begin{figure}[!htb]

	\begin{minipage}[b]{0.49\textwidth}
		\captionsetup{format=hang, labelfont=bf}
		\centering
		\includegraphics[width=\textwidth]{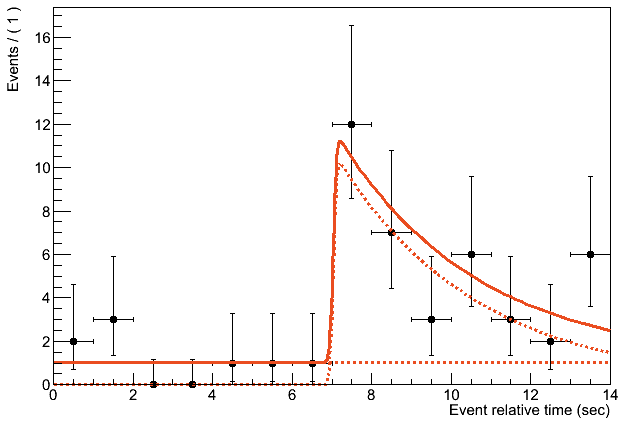}
		\caption{Montecarlo generated data with fit output (distance = 5.5kPc)}
		\vspace{2.2\baselineskip}
		\label{fig:toyMC}
	\end{minipage}
	\hspace{\fill}
	\begin{minipage}[b]{0.49\textwidth}	
		\captionsetup{format=hang, labelfont=bf}
		\centering
		\includegraphics[width=\textwidth]{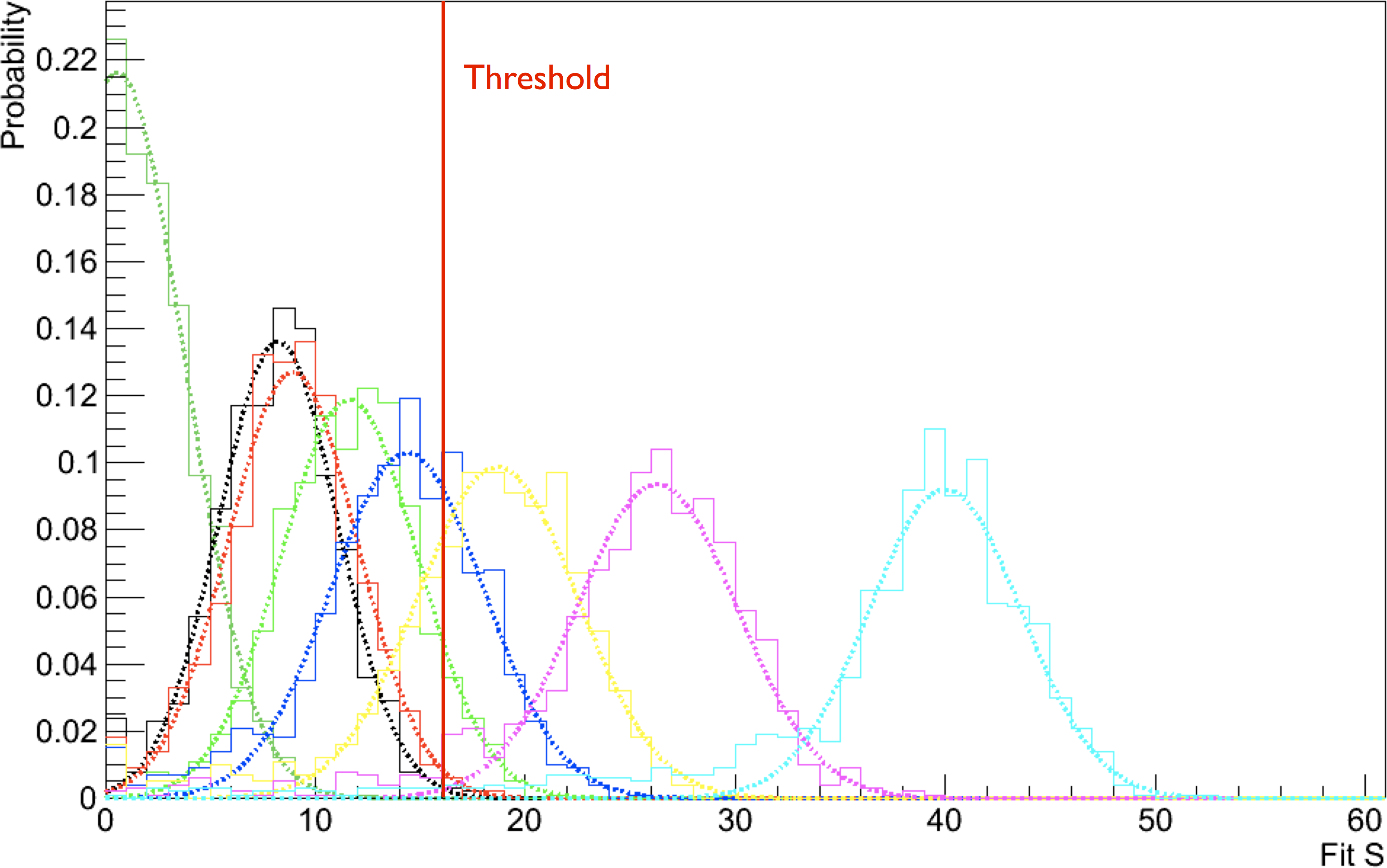}
		\caption{Distributions of the signal amplitude reconstructed by the trigger algorithm for a set of supernova distances (from 10.5 to 4.5kPc, plus zero signal distribution - dark green line)}
		\label{fig:toyMCdist}
	\end{minipage}
	
\end{figure}

In order to understand if CUORE could also discover a supernova explosion without any external information on the timing of the neutrino burst, a supernova trigger algorithm has been developed and tested on toy Montecarlo generated data.
The trigger is based on a maximum likelihood fit of the time distribution of the events recorded in the whole detector at any time during the data acquisition. During the online operation of the DAQ system, for each low energy particle event triggered in any of the CUORE crystals, all the events in a 14 seconds time windows are also considered, and a theoretical model of the time distribution of supernova neutrinos arrival times is fitted to the data. The model is built by summing a flat background (whose magnitude is related to the average background rate) and an exponentially decaying signal whose amplitude and starting points are free parameters of the fit, while the decay time is fixed to 3.5 seconds as predicted in literature by the cooling phase theoretical models.
The fit algorithm is applied to Montecarlo generated time windows (see fig.~\ref{fig:toyMC} for an example) where the magnitude of the signal can be varied to study the performance of the trigger itself as a function of the supernova distance (fig.~\ref{fig:toyMCdist}).

\begin{figure}[!htb]
	\centering
	\captionsetup{labelfont=bf, width=.9\textwidth}
	\caption{Trigger efficiency as a function of the supernova distance: {\bf (left)} for CCVR2 background extrapolated to CUORE mass; {\bf (right)} for a factor 10 improvement in the low energy background with respect to CCVR2.}
	\vspace{.2cm}
	\includegraphics[width=.9\textwidth]{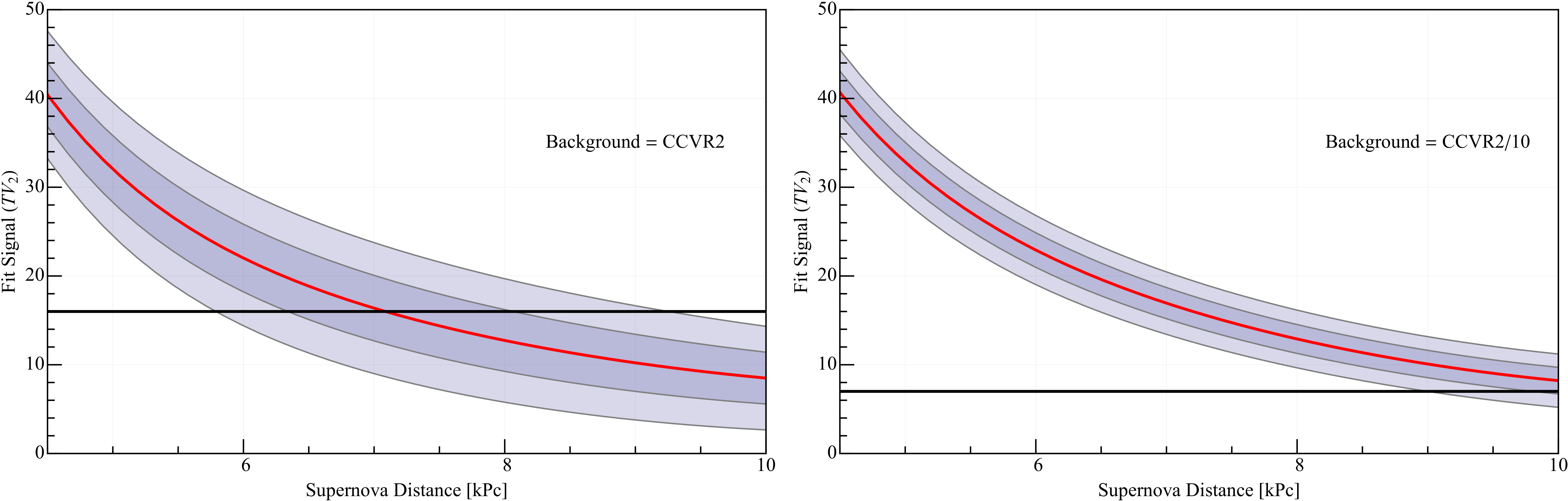}
	\label{fig:discovery}
\end{figure}

The threshold depicted in fig.~\ref{fig:toyMCdist} is defined by requiring that the fraction of zero-signal simulated events that make the algorithm fire due to a background fluctuation is small enough to guarantee that the average rate of false positive triggers is smaller than one per week. This reference value can be modified according to the experimental conditions. Once the threshold is defined, the trigger efficiency (or discovery power) can be calculated as a function of the supernova distance (fig.~\ref{fig:discovery}(left)), also foreseeing a reduction of the low energy background in CUORE with respect to CCVR2 and CUORICINO (fig.~\ref{fig:discovery}(right)).

\begin{figure}[!htb]
	\centering
	\captionsetup{labelfont=bf, width=.9\textwidth}
	\caption{Number of signal events above 3keV energy threshold as a function of the supernova distance for three different values of the $T_{e}$ parameter of the cooling phase emission model}
	\vspace{.2cm}
	\includegraphics[width=.5\textwidth]{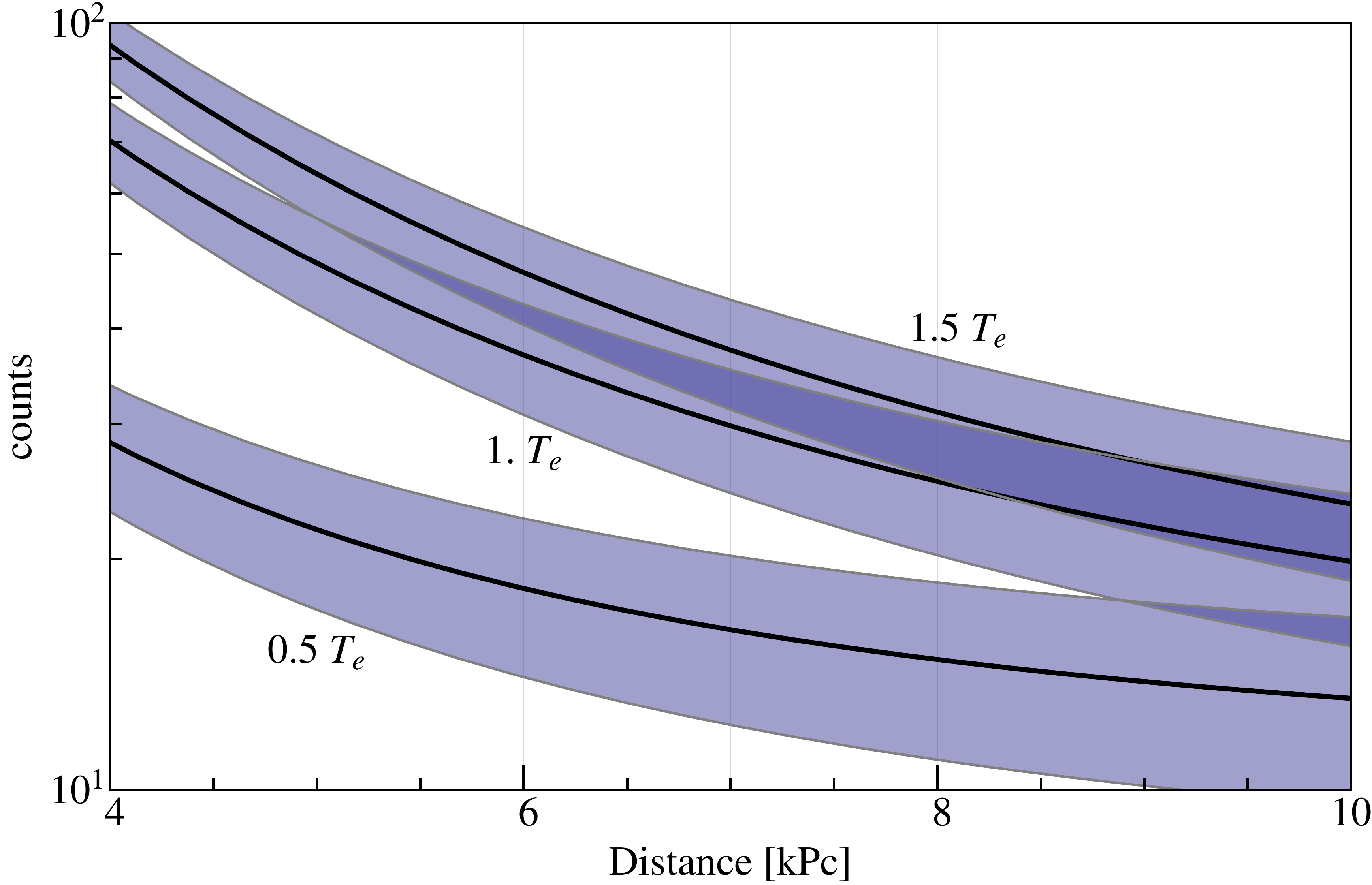}
	\label{fig:models}
\end{figure}

The measurement of the neutrino flux from a nearby supernova with CUORE can potentially provide valuable information on the physics of supernova explosions. As already explained, the expected number of interactions is calculated starting from the three emission fluxes of $\nu_{e}$, $\overline\nu_{e}$ and $\nu_{x}$, each one characterised by a different average energy. The ratio between the average energies is fundamentally related to the ratio between the cross sections of interaction of each neutrino with the supernova matter, but the absolute energy scale depends on the temperature and density of the proto-neutron star during the cooling phase. A variation of the average energy of $\nu_{e}$ generates different emission fluxes, with a consequent variation in the expected number of interactions in the detector. For each of three emission spectra, corresponding to a difference in the average energy of the $\nu_{e}$ flux ($T_{e}$) of 50\%, the total yield (integral of the events associated to the neutrino burst) is represented in fig.~\ref{fig:models} as a function of the supernova distance. The bands associated to each curve are 1$\sigma$ fluctuations of the number of events.

The presented analysis shows that the observation of a galactic supernova via neutrino-nucleus coherent scattering in CUORE is possible, given the expected low energy performance. The observation of neutral current coherent scattering would be an important achievement by itself; nevertheless, in case of a nearby explosion, valuable information on supernova models, like the emission ``temperature'' of the neutrinos, can also be extracted.

\begin{center}
{\it References}
\end{center}
\begin{description}
\footnotesize	
\item[1] M. Pedretti {\it et al.}, Int. J. Modern Phys., {\bf A 23} (2008) 3395
\item[2] S. Di Domizio, F. Orio and M. Vignati, {\bf 6} (2011) P02007
\item[3] J. Gava, J. Kneller, C. Volpe and G.C. McLaughlin, Phys. Rev. Lett. {\bf 103} (2009) 071101
\item[4] G. Pagliaroli, F. Vissani, M. Costantini and A. Ianni, Astropart. Phys. {\bf 0927-6505 31} (2009) 163
\item[5] J. Engel, Phys. Lett. B, {\bf 0370-2693 264} (1991) 114
\item[6] P.S. Amanik and G.C. McLaughlin, J. Phys. G: Nucl. Part. Phys, {\bf 36} (2009) 015105
\item[7] J. Lewin and P. Smith, Astropart. Phys., {\bf 0927-6505 6} (1996) 87
\item[8] M. Biassoni and C. Martinez, Astropart. Phys., {\bf 36} (2012) 151-155
\item[9] F. Alessandria {\it et al.}, JCAP, {\bf 01} (2013) 038

\end{description}
	%

\end{document}